\documentclass[a4paper,11pt]{article}
\pdfoutput=1 
             
\usepackage{jcappub}
\usepackage{amsmath}
\usepackage{amsfonts}
\usepackage{amssymb}
\usepackage{mathtools}
\usepackage{graphics}
\usepackage{tabularx}
\usepackage{adjustbox}
\usepackage{siunitx}
\usepackage{booktabs}
\usepackage{multirow}
\usepackage{citesort}
\usepackage{graphicx}
 \usepackage{url}
\usepackage{soul}
\usepackage{bm}
 \usepackage{xcolor}
\usepackage[utf8]{inputenc}
\usepackage[normalem]{ulem}
\usepackage{hyperref}

\title{Neutrino signal dependence on   gamma-ray burst emission mechanism}

\author[a]{Tetyana Pitik,}
\author[a]{Irene Tamborra,}
\author[b]{and Maria Petropoulou}

\affiliation[a]{Niels Bohr International Academy and DARK, Niels Bohr Institute, University of Copenhagen, Blegdamsvej 17, 2100, Copenhagen}
\affiliation[b]{Department of Physics, National and Kapodistrian University of Athens, Panepistimiopolis, 15783 Zografos, Greece}

\emailAdd{tetyana.pitik@nbi.ku.dk}
\emailAdd{tamborra@nbi.ku.dk}
\emailAdd{mpetropo@phys.uoa.gr}

\abstract{Long duration gamma-ray bursts (GRBs) are among the least understood astrophysical transients powering the high-energy universe.  To date, various mechanisms have been proposed to explain the observed electromagnetic GRB emission. In this work, we show that, although different jet models may be equally successful in fitting the observed electromagnetic spectral energy distributions, the neutrino production strongly depends on the adopted emission and dissipation model. To this purpose, we compute the neutrino production for a benchmark high-luminosity GRB in the  internal shock model, including a dissipative photosphere as well as three emission components,
in the jet model invoking internal-collision-induced magnetic reconnection and turbulence (ICMART), in the case of a magnetic jet with gradual dissipation, and in a jet  with dominant proton synchrotron radiation. We find that the expected neutrino fluence can vary up to three orders of magnitude in amplitude and  peak at  energies ranging from $10^4$ to $10^8$~GeV. For our benchmark input parameters, none of the explored GRB models is  excluded by the targeted searches carried out by  the IceCube and ANTARES  Collaborations. 
However, our work highlights the potential of high-energy neutrinos of pinpointing the underlying GRB emission mechanism and  the importance of relying on different jet models for unbiased stacking searches. }

\begin{document}
\maketitle

\section{Introduction}
\label{sec:introduction}

Gamma-ray bursts (GRBs) are irregular pulses of gamma-rays that have puzzled astronomers for a long time~\cite{Klebesadel:1973iq}. Exhibiting  a non-thermal spectrum, typically peaking in $10$--$10^4$~keV  energy band~\cite{Gruber:2014iza}, bursts lasting for more than $2$~s are named  long-duration  GRBs and  are thought to be harbored within collapsing massive stars~\cite{Woosley:1993wj,MacFadyen:1998vz,Woosley:2006fn}.
They are the brightest explosions in our universe and can release isotropic energies as high as $ 10^{54}$~erg in gamma-rays over  few tens of seconds~\cite{Atteia:2017dcj}.

 The central engine of a long-duration GRB jet can either be a hyper-accreting black hole or a rapidly spinning magnetar.
 Because the central engine cannot be directly observed, its nature can be inferred only indirectly through its impact on the electromagnetic properties of GRBs (see, e.g., Ref.~\cite{2020MNRAS.496.2910P} and references therein).
 A bipolar outflow is continuously powered for a certain time interval, during which gravitational energy~\cite{Chen:2006rra,Lei:2017zro} (for accreting systems) or spin energy~\cite{Blandford:1977ds,Usov:1993qm} (for spinning-down systems) is released in the form of thermal energy or Poynting flux energy, respectively. Subsequently, the outflow propagates through the star and it is strongly collimated by the stellar envelope. Once it succeeds to break out of the stellar surface, it manifests itself as the jet responsible for the GRBs that we observe at Earth. The dynamical evolution of the jet strongly depends on the initial conditions of the central engine. If the magnetic field is negligible, the evolution of the outflow can be well described by the fireball model~\cite{Paczynski:1986px}. If instead the central engine harbours  a strong magnetic field, the jet dynamics is significantly different~\cite{Drenkhahn:2002ug}. 

Gamma-ray bursts are  candidate sources of ultra-high energy cosmic rays and high energy neutrinos~\cite{Meszaros:2017fcs}. In the prompt phase, if the jet contains baryons, protons and nuclei are expected to be accelerated~\cite{Heinze:2020zqb}. If a photon field is also present, photo-hadronic ($ p\gamma $) interactions can lead to a significant flux of neutrinos~\cite{Waxman:1997ti,Guetta:2003wi,Wang:2018xkp}. Another copious source of neutrinos comes from hadronic collisions ($pp$ or $pn$) which, however, are most efficient inside the progenitor star where the baryon density is large~\cite{Razzaque:2003uv,Murase:2013ffa,Metzger:2011xs,Heinze:2020zqb}.
Given the typical GRB parameters,  neutrinos produced in the optically thin region are expected to be emitted in the TeV-PeV energy range~\cite{Waxman:1997ti,Meszaros:2015krr,Waxman:2015ues,Murase:2015ndr}.

The IceCube Neutrino Observatory routinely detects  neutrinos of astrophysical origin in the TeV--PeV energy range~\cite{Ahlers:2018fkn,Ahlers:2015lln,Abbasi:2020jmh,Abbasi:2020zmr}. However, despite the fact that several sources have been proposed as possible candidates to explain the neutrino flux that we observe~\cite{Anchordoqui:2013dnh,Meszaros:2015krr,Murase:2015ndr,Waxman:2015ues,Vitagliano:2019yzm}, we are still lacking clear evidence on the sources producing the observed neutrinos. Among the candidate sources,  high-luminosity GRBs are deemed to be responsible for less than $10\%$ of the observed diffuse emission  in the TeV energy range~\cite{Abbasi:2020jmh,Albert:2020lvs}. On the other hand, over the years, the IceCube and ANTARES Collaborations have searched for high-energy neutrinos emitted in coincidence with GRBs observed by the  Fermi satellite~\cite{Aartsen:2017wea,Aartsen:2016qcr,Albert:2020lvs}, gradually placing  more  stringent upper limits on somewhat optimistic GRB emission models. Recent work suggests that current limits are still not stringent enough to rule out more realistic estimations proposed in the literature~\cite{Li:2012gf,Hummer:2011ms,Heinze:2020zqb,Tamborra:2015qza,Tamborra:2015fzv,Bustamante:2014oka,He:2012tq}. 

Intriguingly, besides the need for increased detection sensitivity, one of the reasons for the non-detection of GRB neutrinos could be connected to the theoretical modeling of the neutrino emission, which is strictly linked to the electromagnetic modeling of the jet. In fact, a comprehensive explanation of the GRB emission and dissipation mechanism is still lacking due to the failure of existing models in addressing all observations in the spectral and temporal domains.

On the other hand, the scarce amount of data on high energy photons and the related statistical challenges allow for a certain flexibility in fitting the same set of data with different input models for GRBs--see, e.g.,~Refs.~\cite{Oganesyan:2019fpa,Zhang:2015bsa,Acuner:2019rif,Acuner:2020zvi,Burgess:2018dhc}. 

Different GRB models may lead to very different predictions for the neutrino emission. The latter  depends on the target photon spectrum and the properties of the accelerated proton distribution (i.e., energy density, power-law slope, and maximum energy), both depending on the emission and dissipation mechanisms as well as the location of the proton acceleration region. 

In this work, we compute the neutrino emission for a benchmark high-luminosity GRB in various jet emission and dissipation scenarios. In particular, we consider an internal shock (IS) model~\cite{Rees:1994nw}, a dissipative photosphere model in the presence of ISs (PH-IS)~\cite{Toma:2010xw}, a three-component model (3-COMP) with emission arising from the photosphere, the IS, and external shock~\cite{Guiriec:2015ppa}, and the internal-collision-induced magnetic reconnection and turbulence model (ICMART)~\cite{Zhang:2010jt}. We also compute, for the first time, the neutrino signal expected in two models where the jet is assumed to be magnetically dominated, namely a magnetized jet model with gradual dissipation (MAG-DISS)~\cite{Beniamini:2017fqh,Gill:2020oon}, and a  proton synchrotron emission model (p-SYNCH)~\cite{Ghisellini:2019lgz}. Our goal is to make a fair comparison among the proposed models for dissipation and electromagnetic emission in GRBs for what concerns  the expected neutrino signal. 

This paper is organized as follows. In Sec.~\ref{sec: Properties of Astro jets}, we outline the basics of the dynamical evolution of the GRB jets considered in this paper. The main model ingredients as well as the proton  energy distributions are reported in Sec.~\ref{sec:ingredients}.  The neutrino production mechanism is discussed in Sec.~\ref{subsec:neutrino production}. The neutrino emission  is presented in Sec.~\ref{sec:emission_models}, first in various scenarios involving ISs, then in the case of magnetized jets, and lastly for the proton synchrotron mechanism. A discussion on our findings, also in the context of detection perspectives as well as uncertainties on the input GRB parameters, and conclusions are reported in Secs.~\ref{sec:discussion} and \ref{sec:conclusions}, respectively. The fitting functions adopted for the photon spectral energy distributions are listed in Appendix~\ref{sec:fitting_models}. A discussion on the dependence of the neutrino emission on the input parameters for the magnetic model with gradual dissipation is reported in Appendix~\ref{MAG-DISS2}. A comparison of the quasi-diffuse neutrino emission with standard input assumptions reported in the literature is provided in  Appendix~\ref{Appendix: Standard shock parmeters}.

\section{Dynamical evolution of gamma-ray burst jets}
\label{sec: Properties of Astro jets}
In this section, we introduce  the main physics describing the jet models considered in this work. We present the models in the context of kinetic dominated jets, then focus on two cases of Poynting flux dominated jets, and the proton synchrotron model. Note that, despite the fact that the proton synchrotron model has a Poynting luminosity larger than the kinetic one (see Ref.~\cite{Florou2021} for a dedicated discussion), we treat it separately from the Poynting flux dominated jets because it does not require knowledge of the jet dynamics.

The general GRB model envisages a relativistic jet propagating with Lorenz factor $ \Gamma$, with respect to the central engine frame, and  half opening angle $ \theta_{j} $. As long as $ \Gamma^{-1} < \theta_{j} $, which is expected to hold during the prompt phase~\cite{Bromberg:2011fg}, the radiating region can be considered spherically symmetric. We therefore use isotropic equivalent quantities throughout the paper.

The reference frames used in our calculation are the observer frame (on Earth), the frame of the central engine (laboratory frame), and the jet comoving frame. A  quantity characteristic of the jet is labeled as $X$, $\tilde{X}$, and $X^{\prime}$, in each of these frames, respectively.  
For example,  energy is transformed through  the following relation: $\tilde{E} = (1 + z) E = (1 + z) \mathcal{D} E^{\prime}$; time instead transforms as $t = (1 + z) \tilde{t} = (1 + z) \mathcal{D}^{-1} t^\prime$, with $\mathcal{D} = [\Gamma (1 - \beta \cos\theta)]^{-1}$ being the Doppler factor, $\beta = v/c$, $\Gamma=1/\sqrt{1-\beta^{2}}$ the Lorentz boost factor and $\theta$ the angle of propagation of an ejecta element with respect to the line of sight. 
A characteristic quantity of the jet is the isotropic-equivalent energy, $\tilde{E}_{\rm{iso}}$, which  represents  the energetic content of the outflow and it is related to the bolometric energy $\tilde{E}_{\rm{bol}}$ through the opening angle  by the following relation:  ${\tilde{E}_{\rm{bol}} = (1 - \cos\theta_{j}) \tilde{E}_{\rm{iso}}} \approx (\theta_j^2/2)\tilde{E}_{\rm{iso}}$, where the approximation holds for small opening angles.

The dominant source of energy in a GRB jet is related to the initial conditions. The jet is powered by accretion onto a newly formed black hole~\cite{Woosley:1993wj} or a rapidly spinning massive  neutron star~\cite{Usov:1993qm}. Two mechanisms  are invoked to extract energy from the central compact object and power the GRB jet:  neutrino annihilation~\cite{1989Natur.340..126E, 1999ApJ...518..356P, Chen:2006rra} or tapping of the spin energy of the central object by means of magnetic fields~\cite{Blandford:1977ds, 1997ApJ...482L..29M}.

 \subsection{Kinetic dominated jets }
 \label{subsubsec:GRB fireball model}

We start with the case of a generic fireball composed of photons, electron/positron pairs, and a small fraction of baryons (primarily protons and neutrons), with negligible magnetic fields~\cite{Goodman:1986az,Piran:1993jm}.
The dynamical evolution of the fireball is  sketched in Fig.~\ref{Fig:sketch fireball} and  consists of three phases, namely acceleration, coasting, and deceleration: 
\begin{figure}[b]
	\centering
	\includegraphics[width=0.7\textwidth]{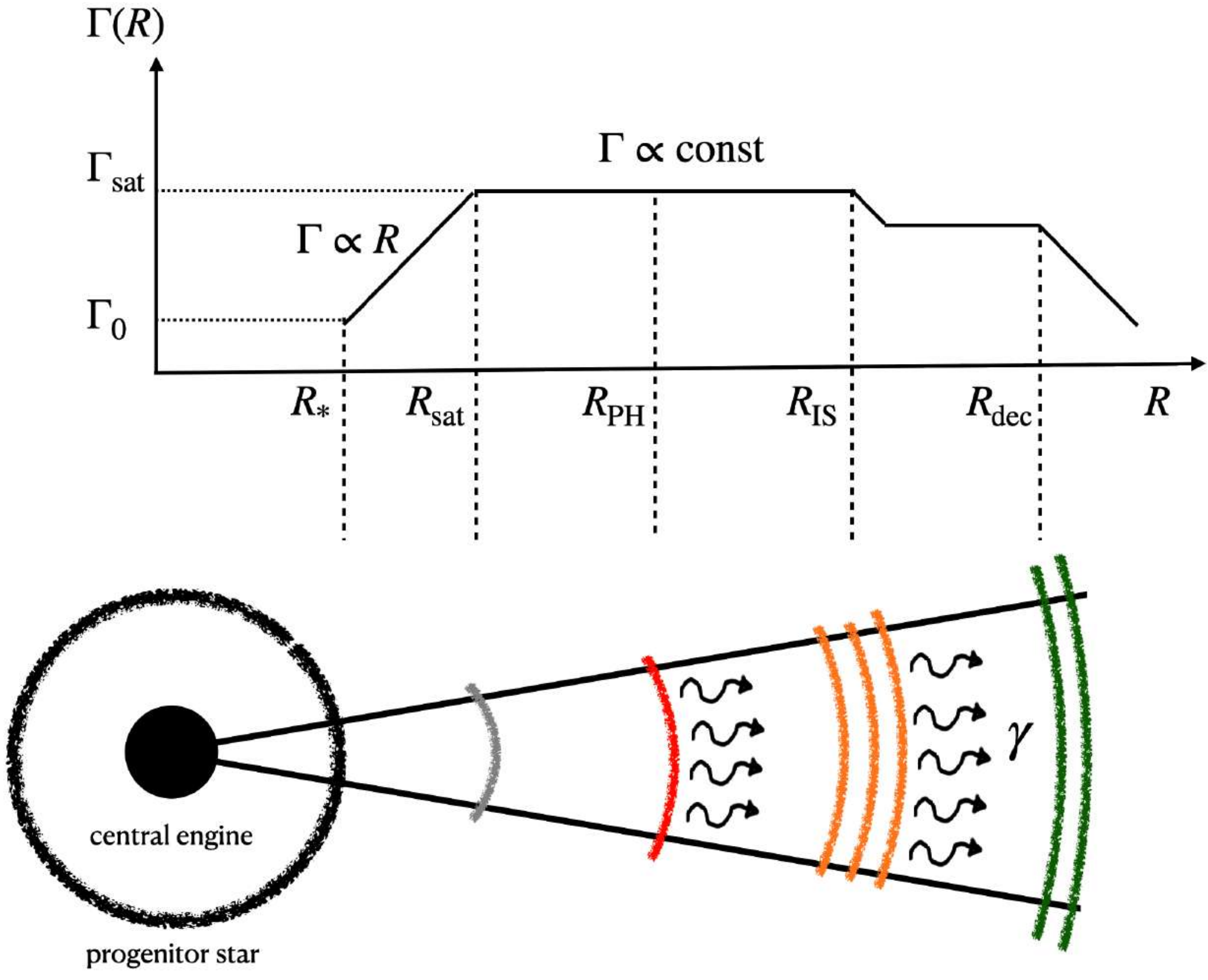}
	\caption{Schematic representation of a GRB jet (not in scale) where energy dissipation takes place through relativistic shocks.
	The Lorentz factor $ \Gamma $ is shown  as a function of the fireball radius for the case in which the photosphere occurs in the coasting phase, so that the photospheric radius ($ R_{\rm PH}$) lies above the saturation radius ($ R_{\rm sat}$).  The photosphere is assumed to produce thermal $\gamma$-rays, the ISs forming at $R_{\rm IS}$ are thought to
	produce  non-thermal $\gamma$-rays, and the external shock, which starts to decelerate
	at $ R_{\rm dec} $, is responsible for the afterglow. When energy dissipation takes place below the photosphere, non-thermal radiation is also expected  from $R_{\rm PH}$.}
	\label{Fig:sketch fireball}
\end{figure}

\begin{enumerate}
	\item \textit{Fireball acceleration}:
	A hot relativistic fireball of isotropic energy $\tilde{E}_{\rm{iso}} = \tilde{L}_{\rm{iso}} \tilde{t}_{\rm{dur}}$ is created and launched at the radius  $R_{0}$ by the central engine emitting energy with  luminosity 
	$\tilde{L}_{\rm{iso}}$ for a time $\tilde{t}_{\rm{dur}}$. Since after the propagation through the envelope of the progenitor star, the fireball can be re-born~\cite{Lazzati:2013ym}, we adopt as size of the jet base 
	$R_{0} = R_{\star}\theta_{j}$, with $R_{\star} \simeq 10^{11}\,\rm{cm}$  being the progenitor star radius. The width of the emitted shell is ${\tilde{\Delta} = c \tilde{t}_{\rm{dur}}}$. As the fireball shell undergoes adiabatic expansion, and while the pair plasma retains relativistic temperatures,  baryons are accelerated by radiation pressure and the bulk Lorenz factor  increases linearly with radius ($\Gamma \propto R$), until it reaches its maximum value. We assume that the latter coincides with the dimensionless entropy per baryon $\eta = \tilde{E}_{\rm{iso}}/M c^2$, where $M$ is the  baryonic mass injected into the outflow. The maximum Lorenz factor is achieved at the saturation radius 
	$R_{\rm{sat}} = \eta R_{0}/\Gamma_{0}$, where $ \Gamma_{0} = \Gamma (R_{\star}) \simeq 1$--$10$~\cite{Mizuta:2013yma} is the  breakout Lorenz factor.
	
	\item \textit{Fireball coasting}:
	Beyond $R_{\rm{sat}}$,  the flow coasts with $\Gamma = \Gamma_{\rm sat} \sim \eta = \rm{const}$. As the fireball shell keeps on expanding, the baryon density, obtained by the mass continuity equation $\dot{M} = 4 \pi R^{2} \Gamma \rho^\prime c = \rm{const}$~\cite{Dermer:2009zz} for a relativistic flow with spherical symmetry, drops as
	\begin{equation}
		\label{eq: baryon_density}
		n^\prime_{b} = \frac{\rho^\prime}{m_{p}} = \frac{\dot{M}}{4 \pi m_{p} R^{2} c  \Gamma} \simeq \frac{\tilde{L}_{\mathrm{iso}}}{4 \pi R^{2} m_{p} c^{3} \eta \Gamma}\ ,
	\end{equation}
	where $ \rho^\prime$ is the baryon density in the comoving frame, $R$ is the distance from the central engine, and $\tilde{L}_{\rm iso}= \eta\dot{M}c^2$. At a certain point,  photons 
	become optically thin to both pair production and Compton scattering off  free leptons associated with baryons entrained in the fireball. Once  the Thomson optical depth ($\tau_{\rm{T}} = n^\prime_{l} \sigma_{\mathrm{T}} R/\Gamma$) drops below $1$, the energy that has not been converted into  kinetic energy is released at the photospheric radius $R_{\rm{PH}}$. Let $\mathcal{R}$ denote the number of leptons per baryon ($n^\prime_{l} = \mathcal{R} n^\prime_{p}$), we can define the critical dimensionless entropy~\cite{Meszaros:1993cc}:
	\begin{equation}
		\label{eq: critical_dimensionless_entropy}
		\eta_{\ast} = \left(\frac{\sigma_{\mathrm{T}} \mathcal{R} \tilde{L}_{\mathrm{iso}} \Gamma_{0}}{8 \pi R_{0} m_{p} c^3}\right)^{1/4}\ ,
	\end{equation}
	where ${\sigma_{T} = 6.65 \times 10^{-25} \mathrm{cm}^{2}}$ is the Thomson cross section. $\eta_{\ast}$ represents the limiting value of the Lorenz  factor which separates  two scenarios: $\eta > \eta_{\ast}$ (the photosphere occurs in the acceleration phase) and  $\eta < \eta_{\ast}$ (the photosphere occurs in the coasting phase). For our choice of parameters, we will always be  in the second case, thus we can introduce the photospheric radius as  the distance such that  $\tau_{\rm{T}} =1$~\cite{Meszaros:1999gb}:
 	\begin{equation}
 		R_{\mathrm{\rm PH}} = \frac{\sigma_{\mathrm{T}} \tilde{L}_{\mathrm{iso}} \mathcal{R}}{4 \pi \eta^{3} m_{p} c^{3}}\ .
 		\label{eq:barionic_photosphere}
 	\end{equation}
 	The radiation coming from the photosphere is the first electromagnetic signal detectable from the fireball. It emerges  peaking at~\cite{Meszaros:1999gb}
	\begin{equation}
 		\label{eq: photospheric temperature}
 		k_{B}\tilde{T}_{\mathrm{\rm PH}} = k_{B} \left(\frac{\tilde{L}_{\mathrm{iso}}}{4 \pi R_{0}^{2} \sigma_{B}}\right)^{\frac{1}{4}} \left(\frac{R_{\mathrm{\rm PH}}}{R_{\mathrm{sat}}}\right)^{-\frac{2}{3}}\ ,
	\end{equation}
	where $\sigma_{B}$ is the Stefan-Boltzmann constant and $k_{B}$ the Boltzmann constant. The energy $\tilde{E}_{\rm PH}$ emerging from the photosphere is parametrized  through $\varepsilon_{\rm PH} =  \tilde{E}_{\rm{PH}}/\tilde{E}_{\rm iso}$. 
	
	Since the central engine responsible for the launch of the relativistic jet is expected to have an erratic activity, the produced outflow is unsteady and radially inhomogeneous. This causes internal collisions between shells of matter emitted with time lag $t_{v}$ to occur at a distance~\cite{Rees:1994nw}
	
	\begin{equation}
		\label{eq: IS radius}
 		R_{\mathrm{IS}} \simeq \frac{2 c t_{v} \Gamma^{2}}{1 + z}\ ;
 	\end{equation}
 	this is the  IS radius, where a fraction $\varepsilon_{\rm IS}$ of the total outflow energy ($\tilde{E}_{\rm iso}$) is dissipated, and particles are accelerated.

 	\item \textit{Fireball deceleration}:
 	 The fireball shell is eventually decelerated~\cite{Piran:1993jm,Piran:1999kx,vanEerten:2018amz}
 	 by the circumburst medium  that can either be  the interstellar medium  or the pre-ejected stellar wind from the progenitor before the collapse. Let us consider an external density profile~\cite{Panaitescu:2000bk}:
 	\begin{equation}
 		\label{eq: CMB density}
 		n_{b}(R) = A R^{-s}\ ,
 	\end{equation}
 	with $s = 0$ for a homogeneous medium and $s = 2$ for a wind ejected at  constant speed. For a thin shell~\cite{Sari:1995cb}, the deceleration radius is defined as the distance where the swept mass  from the circumburst medium is  $m_{\rm{CMB}} = M/\eta$ [or $\Gamma (R_{\rm{dec}}) = \eta/2$]~\cite{Blandford:1976uq}: 
 	\begin{equation}
 		\label{eq: deceleration radius}
 		R_{\rm{dec}} = \left(\frac{3 - s}{4 \pi}\frac{\tilde{E}_{\rm K, \,iso}}{m_{p} c^{2} A \eta^{2}}\right)^{1/(3 - s)}\ ;
 	\end{equation}
 	in alternative, $R_{\rm{dec}}$ can be obtained from the observed deceleration time $t_{\rm dec}$~\cite{Zhang:2018ond}:
 	\begin{equation}
 		\label{eq: deceleration time}
 		t_{\rm{dec}} \simeq 1.3 (1 + z) \frac{R_{\rm{dec}}}{\eta^{2} c}\ ,
 	\end{equation}
	where $\tilde{E}_{\rm K, \,iso} =  \tilde{E}_{\rm{iso}}-\tilde{E}_{\gamma, \rm{iso}}$ is the  isotropic equivalent kinetic energy of the ouflow after $\tilde{E}_{\gamma, \rm{iso}}$ has been radiated during the prompt phase. At $R_{\rm{dec}}$, an external  shock forms and propagates into the medium, hence the deceleration radius is essentially the initial external shock radius.
 \end{enumerate}

\subsubsection{Jet model with internal shocks}

\label{sec:ISmodel}

For long time, the IS model~\cite{Rees:1994nw,Kobayashi:1997jk,Daigne:1998xc} has been  considered  as the standard model for the prompt emission in the literature. Among the merits of this model there is its ability to naturally explain the variability of the lightcurves, to provide natural sites for the dissipation of the kinetic energy of the baryonic fireball, as well as sites for particle acceleration and non-thermal radiation.

The erratic activity of the central engine is responsible for the creation of an outflow that can be visualized as being composed of several shells. Collisions of such shells with different masses and/or Lorenz factors cause the dissipation of the kinetic energy of the jet at $ R_{\rm IS} $\footnote{If there is a large spread in the $\Gamma$ values of the shells, then $R_{\rm IS}$ can also spread a lot~\cite{Guetta:2000ye,Bustamante:2016wpu, Rudolph:2019ccl}.} (see Fig.~\ref{Fig:sketch fireball}).

Part of the dissipated energy, $ \varepsilon_{\rm IS} \tilde{E}_{\rm iso}$, is used for particle acceleration.  
Non-thermal electrons (protons) receive a fraction $\varepsilon_{e}$ 
($\varepsilon_{p}$), 
while a fraction $\varepsilon_{B}$ goes into the amplification of magnetic fields. In this scenario, electrons emit synchrotron radiation in the fast cooling regime. The radiated energy can thus be expressed as $E^{\prime}_{\gamma,\mathrm{iso}} = \varepsilon_{e} \varepsilon_{\rm IS}E^{\prime}_{\rm iso}$
and the magnetic field as
 \begin{equation}
 	B^\prime = \sqrt{8 \pi \frac{\varepsilon_{B}}{\varepsilon_{e}}\frac{E^\prime_{\gamma,\mathrm{iso}}}{V^\prime_{\mathrm{iso}}}}\ .
 \end{equation}
The protons co-accelerated with electrons interact with the prompt photons through photo-hadronic interactions and produce neutrinos, as discussed in Sec.~\ref{subsec:neutrino production}. Here $ V^{\prime}_{\rm iso} = 4\pi R^{2}_{\gamma} \Gamma c\tilde{t}_{\rm dur}$ represents the isotropic volume of the jet in the comoving frame. 

Within a more  realistic setup, various collisions between plasma shells occur along the jet.  Scenarios involving collisions of multiple shells have been  considered~\cite{Guetta:2000ye,Bustamante:2016wpu,Rudolph:2019ccl} and can lead to lower neutrino fluxes. 
Yet, in this work, since we aim to compare different jet models,  we adopt one representative shell with average parameters and spectral properties for simplicity.

\subsubsection{Jet model with a dissipative photosphere and internal shocks}

\label{sec:PH+IS}

In the class of photospheric models, it is assumed that the dominant radiation observed in the prompt phase is produced in the optically thick region below the photosphere~\cite{Beloborodov:2017use}. Depending on the presence of dissipative processes acting in the optically thick parts of the outflow, photospheric models can be classified in non-dissipative or dissipative ones.

In the presence of a non-dissipative photosphere, according to the standard fireball model, the thermal radiation advected with the flow and unaffected by the propagation is released at $R_{\mathrm{PH}}$, see Fig.~\ref{Fig:sketch fireball}. Depending on the dimensionless entropy of the outflow (see Eq.~\ref{eq: critical_dimensionless_entropy}), this component can be very bright or highly inefficient  and is characterized by the fraction $ \varepsilon_{\rm{PH}} = (\eta/ \eta_{\ast})^{8/3} $~\cite{Meszaros:1999gb}.

For a dissipative photosphere, strong subphotospheric dissipation is required in the optically thick inner parts of relativistic outflows in order to account for the detected non-thermal spectra~\cite{Thompson:1994zh,Giannios:2006mx,Gill:2014fwa,Rees:2004gt,Vurm:2012be,Vurm:2015yfa,Beloborodov:2012ys}. In this scenario, the spectral peak and the low-energy spectrum below the peak are formed by quasi-thermal Comptonization of seed photons by mildly relativistic electrons when the  Thomson optical depth of the flow is $ 1 \lesssim \tau_{T} \lesssim 100$~\cite{Giannios:2006jb,Thompson:2013yna}.
In the literature, several  sub-photopsheric dissipative mechanisms have been proposed, including ISs at small-radii~\cite{Rees:2004gt}, collisional nuclear 
processes~\cite{Beloborodov:2009be}, or dissipation of magnetic energy~\cite{Giannios:2006jb}. One of the most attractive features of these models is their ability to naturally explain the  observed small dispersion of the sub-MeV peak and the high prompt emission efficiency~\cite{Lazzati:2009xx,Lazzati:2013ym,Gottlieb:2019aae}, that the standard version of the IS model cannot easily explain. 

The scenario explored in this work considers the main prompt emission as being released at $ R_{\rm PH} $ with a non-thermal spectrum. These photons cross  the IS region and interact with energetic protons accelerated at the IS to produce neutrinos. We do not consider neutrino production below and at the photosphere, as this would result in neutrino energies well below the PeV range that we are interested in (i.e., GeV neutrinos produced in proton-neutron collisions in the ejecta~\cite{Bahcall:2000sa,Kashiyama:2013ata} or TeV neutrinos produced via $pp$ interactions of protons accelerated at sub-photospheric ISs~\cite{Wang:2008zm,Murase:2008sp,Xiao:2017blv,Murase:2013ffa,Murase:2013hh}). Indeed, photo-hadronic interactions in the opaque region do not lead to efficient production of high-energy neutrinos because of inefficient Fermi acceleration that limits the maximum proton energy to low values~\cite{Beloborodov:2016jmz}. We  stress that  the wording ``dissipative photosphere" in this work is meant to highlight the  non-thermal nature of the photospheric spectrum and it should not be associated with the neutrino production region below the photosphere, as usually done in the literature (see, e.g.~\cite{Wang:2008zm}).

\subsubsection{Jet model with three emission components}
\label{sec:3comp}

The three-component GRB model was introduced in Ref.~\cite{Guiriec:2015ppa}, where the authors found that  a thermal  component described by a black body  (BB) spectrum, a Band spectrum (sometimes statistically equivalent to a cut-off 
power-law, CPL) and a non-thermal  power-law (PL) spectrum  at high energies (with or without cut-off) represent a globally better description of the data than the Band  spectral fit for a number of bursts. We refer the reader to Appendix~\ref{sec:fitting_models} for details on the spectral energy distributions of photons.

As argued in Ref.~\cite{Guiriec:2015ppa}, the physical interpretation proposed for the three components is the following. The BB component, given its weakness, is interpreted as thermal photospheric emission of a magnetized jet not strongly affected by sub-photospheric dissipation. The non-thermal emission fitted by the Band (or CPL) component, given the observed variability, is assumed to be  produced in the optically thin region of the jet from relativistic electrons. The third PL (or CPL) component, which extends over at least $5$ decades in energy and sometimes emerges with a slight temporal delay with respect to the trigger of the burst, is the one with the least clear origin. Because of its initial temporal variability, it is assumed to be of internal origin; e.g., it might be due to inverse Compton processes, even if this scenario is not able to explain the extension of such a component to lower energies or the temporal delay. Finally, the fact that in some cases the PL component becomes dominant at the end of the bursts and lasts longer than the prompt emission led to identify it with the emergence of an early afterglow, which corresponds to the start of deceleration of the outflow.

\subsection{Poynting flux dominated jets}

When the central compact object is a rapidly rotating black hole threaded by open magnetic field lines, it is possible to tap   the black hole spin energy to produce Poynting-flux dominated jets~\cite{Blandford:1977ds}. The  electromagnetic luminosity of this jet is much larger than the kinetic luminosity associated to matter.

A characteristic parameter  is the magnetization  $ \sigma $, defined as the ratio of the Poynting luminosity and the kinetic luminosity: 
\begin{equation}
	\sigma(R) \equiv \frac{L_{B}}{L_{K}} = \frac{B^{2}(R)}{4 \pi \Gamma \rho(R) c^{2}} = \frac{B^{\prime 2}(R)}{4 \pi \rho^\prime(R) c^{2}}\ ,
	\label{eq:sigma}
\end{equation}
where $B^\prime(R)$ and $\rho^\prime(R)$ are the magnetic field strength and matter density in the comoving frame at a certain distance $R$ from the central engine. Hence, the total jet luminosity at any radius is $L(R) = [1 + \sigma(R)] L_{K}(R)$. 
In this work we consider two models for magnetized jets: the ICMART model and the gradual magnetic dissipation model, which we briefly introduce  below.

\subsubsection{ICMART model}
\label{ICMART}

\begin{figure}[]
	\centering
	\includegraphics[width=0.7\textwidth]{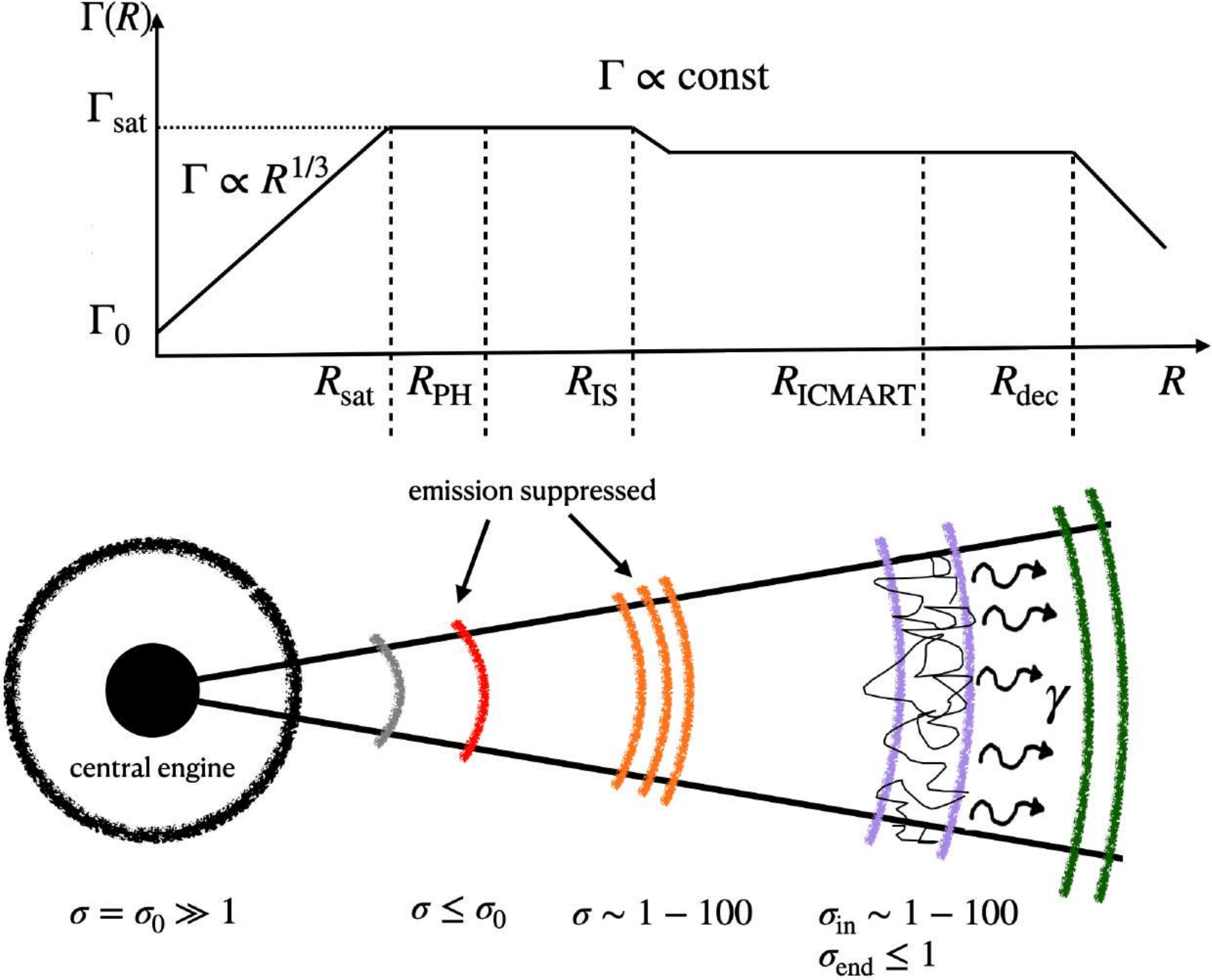}
	\caption{Schematic representation of a Poynting flux dominated jet (not in scale) in the ICMART model. The Lorentz factor $ \Gamma $ is shown  as a function of jet radius. The radiation from the photosphere ($ R_{\rm PH}$) and ISs ($ R_{\rm IS}$) is strongly suppressed and can be at most $1/(\sigma+1)$ of the total jet energy (see Eq.~\ref{eq:sigma});  typical values for the magnetization parameter $\sigma$ are shown. The emitting region is located at $R_{\rm ICMART}$, where magnetic reconnection causes a strong discharge of magnetic energy and the emission of gamma-rays. The magnetization at $R_{\rm ICMART}$ is $\sigma_{\rm in}$ and $\sigma_{\rm end}$ in the beginning and at the end of an ICMART event, respectively.  }
	\label{Fig:sketch ICMART}
\end{figure}
The ICMART model~\cite{Zhang:2010jt} considers Poynting flux dominated jets, whose energy is dissipated and radiated away at very large radii from the central engine, as shown in Fig.~\ref{Fig:sketch ICMART}.  The main motivation behind this model relies on the non-detection (or detection of a very weak) photospheric component in the spectra of some GRBs, hinting that the jet composition cannot  be largely Poynting flux dominated at the photosphere.

The  GRB central engine intermittently ejects  an unsteady jet with variable Lorentz factor and with a nearly constant degree of magnetization $\sigma_{0}\equiv \sigma(R_{0})$. Such a jet is composed by many discrete magnetized shells which collide at  $R_{\mathrm{IS}}$ (see Eq.~\ref{eq: IS radius} and Fig.~\ref{Fig:sketch ICMART}). Yet,  the kinetic energy dissipated at the ISs is smaller by a factor [$1+ \sigma(R_{\mathrm{IS}})$] with respect to the energy available in the traditional IS model. Hence,  the total energy emitted in radiation could be completely negligible at this stage.

In the optically thin region, the  early internal collisions  have the role of altering, and eventually destroying, the ordered magnetic field configuration, triggering the first reconnection event. The ejection of plasma from the reconnection layer would disturb the nearby ambient plasma and produce turbulence, facilitating more reconnection events which would lead to a runaway catastrophic release of the stored magnetic field energy at the radius defined as $R_{\mathrm{ICMART}}$. This would correspond to one ICMART event, which would compose one GRB pulse. Other collisions that trigger other reconnection-turbulence avalanches would give rise to other pulses. 

This model successfully reproduces  the observed GRB lightcurves with both fast and slow components~\cite{Zhang:2013ycn}. The slow component, related to the central engine activity, would be caused by the superposition of emission from all the mini-jets due to multiple reconnection sites, while the erratic fast component would be related to the mini-jets  pointing towards the observer.

\subsubsection{Magnetized jet model with gradual dissipation}
\label{sec:MAG-DISS}

\begin{figure}[]
	\centering
	\includegraphics[width=0.7\textwidth]{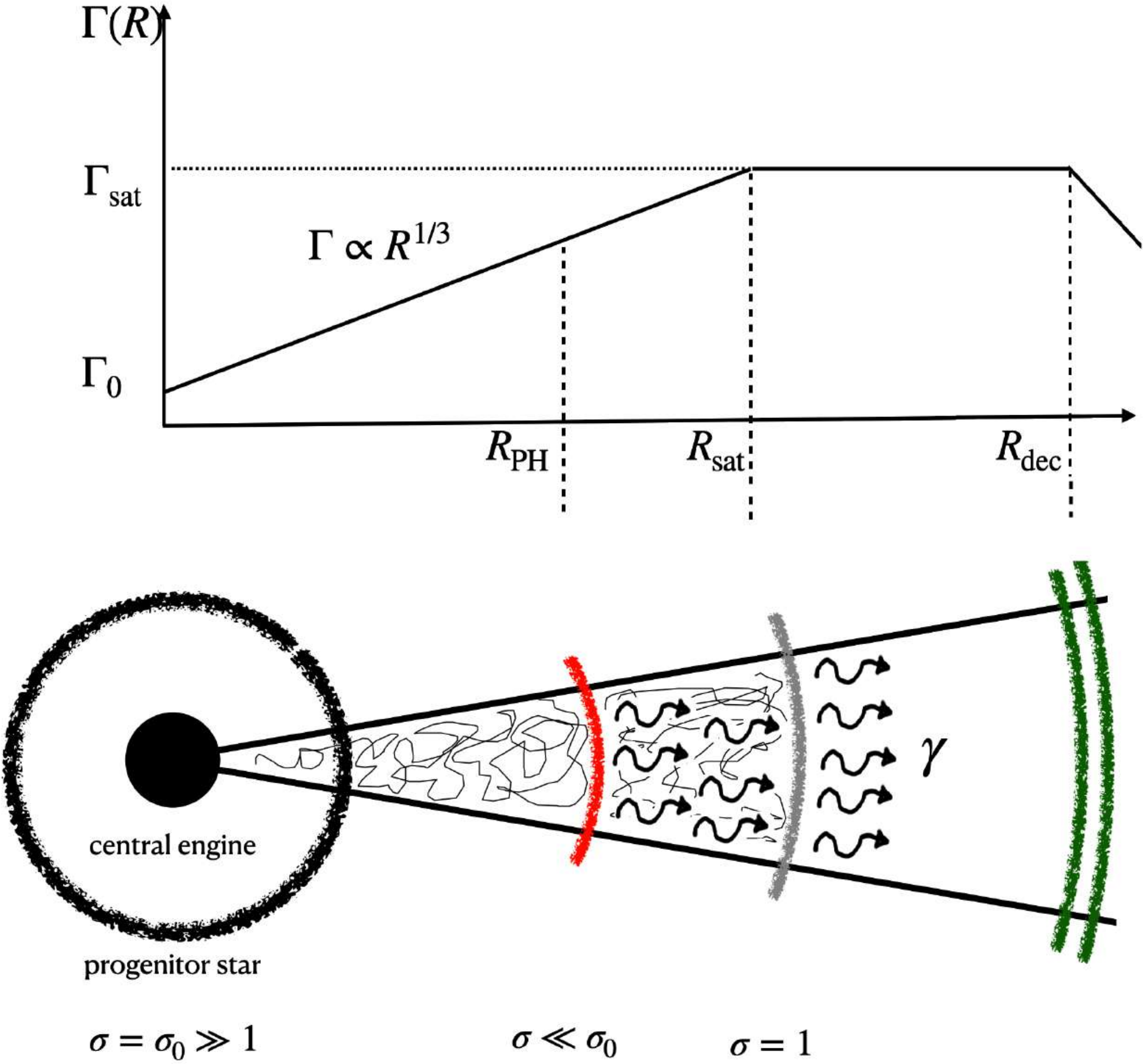}
	\caption{Schematic representation of a Poynting flux dominated jet (not in scale) in the gradual energy dissipation model. The Lorentz factor $ \Gamma $ is shown  as a function of jet radius. The radiation from the photosphere ($ R_{\rm PH}$) can be very bright, depending on the initial magnetization $\sigma_{0}$ of the outflow. Typical values for the magnetization parameter $\sigma$ are shown. The emitting region is located between $R_{\rm PH}$ and $R_{\rm sat}$, where magnetic reconnection causes the dissipation of magnetic field energy, the emission of thermal gamma-rays at $R_{\rm PH}$ and synchrotron radiation from accelerated electrons in the optically thin region up to $R_{\rm sat}$. 
	}
	\label{Fig:sketch mag}
\end{figure}

In this  scenario, the energy dissipation through reconnection starts below the jet photosphere and occurs gradually over a wide range of radii~\cite{Drenkhahn:2001ue,Drenkhahn:2002ug}, as schematically shown in Fig.~\ref{Fig:sketch mag}. Following  Refs.~\cite{Beniamini:2017fqh,Gill:2020oon}, 
we consider  magnetized outflows with a striped-wind magnetic field structure, where energy is gradually  dissipated through magnetic reconnection until the saturation radius. This model can naturally explain  the  double-hump electromagnetic spectra sometimes observed~\cite{Gill:2020oon}. 

The jet is injected at $R_{0} \sim 10^{7}$~cm  with magnetization $\sigma_{0}\gg1$ and Lorentz factor $\Gamma_{0} = \sqrt{\sigma_{0} + 1}\approx \sigma_0$.  As it propagates, the magnetic field lines of opposite polarity reconnect, causing the magnetic energy to be dissipated at a rate~\cite{Drenkhahn:2001ue}:
\begin{equation}
	\label{eq:E_B_diss_rate}
	\dot{E}_{\mathrm{diss}} = -\frac{dL_{B}}{dR} = - \frac{d}{dR}\left(\frac{\sigma}{\sigma + 1} L\right)\propto R^{1/3}\ ,
\end{equation}
where $\sigma(R)$ is obtained from the conservation of the total specific energy ${\Gamma(R) \sigma(R) = \Gamma_{0} \sigma_{0}}$. 

The Lorentz factor of the flow evolves as~\cite{Drenkhahn:2001ue}
\begin{equation}
	\Gamma(R) = \Gamma_{\mathrm{sat} }\left(\frac{R}{R_{\mathrm{sat}}}\right)^{1/3}\ ,
\end{equation}
until the saturation radius $R_{\mathrm{sat}} = \lambda \Gamma^{2}_{\rm sat}$ (see Fig.~\ref{Fig:sketch mag}), where $\lambda$ is connected to the characteristic length scale over which the magnetic field lines reverse polarity.  This length scale can be related to the angular frequency of the central engine (e.g., of millisecond magnetars) or with the size of the magnetic loops threading the accretion disk~\cite{Parfrey:2014wga}.

Motivated by results of particle-in-cell (PIC) simulations of magnetic reconnection in magnetically dominated electron-proton plasmas~\cite{ 2015MNRAS.450..183S, 2016ApJ...818L...9G,Werner:2016fxe}, we assume that half of the dissipated energy in Eq.~\ref{eq:E_B_diss_rate} is converted in kinetic energy of the jet, while the other half goes into particle acceleration and is redistributed among electrons and protons\footnote{Rough energy equipartition between magnetic field, protons and electron-positron pairs is also found in kinetic simulations of reconnection in pair-proton plasmas~\cite{2019ApJ...880...37P}.}.
 In particular, the fraction of energy which goes into electrons is~\cite{Werner:2016fxe}
\begin{equation}
	\label{eq:varepsilon_e_magnetic_model}
	\varepsilon_{e} \approx \frac{1}{4} \left(1 + \sqrt{\frac{\sigma}{10 + \sigma}}\right)\ ,
\end{equation}
while the one that goes into protons 
has been extracted from Fig.~20 of Ref.~\cite{Werner:2016fxe} and is $\varepsilon_{p}\sim 1-\varepsilon_e$.
A fraction $\xi$ of electrons injected into the dissipation region are accelerated into a power-law distribution $n^{\prime}_{e}(\gamma^{\prime}_{e}) \propto \gamma^{\prime- k_{e}}_{e}$ in the interval $[\gamma^{\prime}_{e, \mathrm{min}}, \gamma^{\prime}_{e, \mathrm{max}}]$ with the minimum electron Lorentz factor being
\begin{equation}
	\gamma^{\prime}_{e, \mathrm{min}}(R) = \frac{k_{e} - 2}{k_{e} - 1} \frac{\varepsilon_{e}}{2 \xi} \sigma(R) \frac{m_{p}}{m_{e}}\ ,
\end{equation}
 and $\gamma^{\prime}_{e, \mathrm{max}}$ is the maximum electron energy obtained by equating the acceleration time and the total cooling time. 
 The power-law slope of the accelerated particles in relativistic reconnection depends on the plasma magnetization in a way that harder spectra ($k_e <2$) are obtained for $\sigma \gg1$ \cite{Sironi:2014jfa, Guo2014, 2016ApJ...816L...8W}. Here, we adopt the following parameterization for the electron power-law slope~\cite{Werner:2016fxe}: 
\begin{equation}
	\label{eq: electron_slope_magnetic_model}
	k_{e}(\sigma) \approx 1.9 + 0.7/ \sqrt{\sigma}\ .
\end{equation}
The  proton spectrum will be discussed in detail in Sec.~\ref{subsubsec: Magnetic jet model with gradual dissipation}.

\subsection{Proton synchrotron model}
\label{sec:psync}

Recently, Refs.~\cite{Oganesyan:2017ork,Oganesyan:2019fpa,Ravasio:2019kiw} have analyzed the spectra of a sample of GRBs for which data down to the soft X-ray band and, in some cases, in the optical are available. This extensive work has established the common presence of a spectral break in the low energy tail of the prompt spectra and led to realize that the spectra could be fitted by three 
power-laws. The spectral indices below and above the break are found to be
$\alpha_{1} \simeq -2/3$ and $\alpha_{2} \simeq -3/2$ respectively, while the photon index of the third PL is $ \beta <2 $. The values of all photon indexes are consistent with the predicted values for the synchrotron emission in a marginally fast cooling regime~\cite{Daigne:2010fb}. However,  if electrons are responsible for the prompt emission, then the parameters of the jet have to change drastically with respect to the standard scenario, in which the emission takes place at relatively small radii and with strong magnetic fields in situ. One possible way out to this has been discussed in Ref.~\cite{Ghisellini:2019lgz}, where  protons are considered to be the  particles which radiate synchrotron emission in the marginally fast cooling regime; in this way, it is possible to  recover the typical emitting region size at $R_{\gamma} \simeq 10^{13}$~cm.

\section{Main model ingredients}
\label{sec:ingredients}
 In this section, we outline  some of the quantities  characterizing the energetics and  geometry of the jet for all  models. We also introduce the target particle distributions. 
\subsection{Reference model parameters}
\label{subsec: Reference jet parameters}
The gamma-ray emission is assumed to originate from an isotropic volume $V^{\prime}_{\rm{iso, s}} = 4 \pi R_{\gamma}^2 \Delta ^{\prime}_{s}$, where ${\Delta^{\prime}_{s}= R_{\gamma}/2\Gamma}$ is the comoving thickness of the emitting shell and $R_{\gamma}$ is the distance from the central engine where  the electromagnetic radiation is produced. 
Dissipation--whether it occurs in the photosphere, in the optically thin region (e.g., ISs) or external shocks--causes  the conversion of a fraction $\varepsilon_{d}$ of the total jet energy $\tilde{E}_{\rm{iso}}$ into thermal energy, bulk kinetic energy, non-thermal particle energy, and magnetic energy. The energy stored in relativistic electrons, protons, and magnetic fields in the emitting region can be parameterized through the fractions $\varepsilon_{p}$, $\varepsilon_{e}$ and $\varepsilon_{B}$,  respectively.

These parameters ignore the detailed microphysics at the plasma level, but allow to establish a direct connection with the observables. The dissipation efficiency $\varepsilon_{\rm IS}=0.2$ has been chosen by following Refs.~\cite{Kobayashi:1997jk,Guetta:2000ye}. In addition, for the IS models, we rely on  PIC simulations of mildly relativistic shocks in electron-ion plasma. Recently, a relatively long 2D PIC simulation has been performed~\cite{Crumley:2018kvf} and it has been shown that quasi-parallel shocks can be efficient particle accelerators. The energy fractions  going into non-thermal protons, electrons and the turbulent magnetic field are found to be $\varepsilon_{p}\simeq 0.1$, $\varepsilon_{e}\simeq 5\times 10^{-4}$ and $\varepsilon_{B}\gtrsim 0.1$, respectively. These results are valid for a shock with Lorentz factor $\Gamma_{\rm sh}=1.5$. We know, on the other hand, that efficient energy dissipation through ISs can take place only if a large spread in  Lorentz gamma factors is present within a kinetic dominated jet~\cite{Guetta:2000ye}, which would lead to   $1 \lesssim\Gamma_{\rm sh}\lesssim 5$. Unfortunately, as of today, the midly relativistic regime is poorly explored, and a study of the transition regime from non relativistic to ultrarelativistic is still necessary. Nevertheless, the results from~\cite{Crumley:2018kvf} can be considered as indicative for our case. Given that for the relativistic regime it has been found $\varepsilon_{e}\lesssim 0.1$~\cite{Sironi:2010rb}, we conservatively adopt $\varepsilon_{e}=0.01$, while $\varepsilon_{p}=0.1$ and $\varepsilon_{B}=0.1$.
For the magnetized jet models, instead, these parameters are found to depend on the magnetization of the jet, as we will see later.
Finally, at the relativistic external shock, in the deceleration phase, we adopt $\varepsilon_{e} = 4\times 10^{-2}$, $\varepsilon_{B} = 10^{-4}$ and $\varepsilon_{p} = 1 - \varepsilon_{e} - \varepsilon_{B}$, which turn out to be constrained by observations for a number of cases~\cite{Kumar:2009ps,Santana:2013saa,Beniamini:2016hzc}. In addition, we use $k_{e}=k_{p}=2.5$ for the  power slope, motivated by PIC simulations for ultra-relativistic shocks~\cite{Sironi:2013ri}.

For what concerns the energetics of our reference jet, motivated by recent observations of  GRB afterglows~\cite{Wang:2015vpa}, we choose  ${\tilde{E}_{\rm{iso}} = 3.4 \times 10^{54}}$~ergs, where a typical opening angle of $\theta_{j}= 3$ degrees is  adopted. Our benchmark Lorenz factor  is $\Gamma=\Gamma_{\rm sat} = 300$~\cite{Racusin:2011jf,Ghirlanda:2017opl}.
The duration of the burst is taken to be $\tilde{t}_{\rm dur} = 100 \,{\rm s}/(1 + z)$, where $z = 2$ is the redshift we adopt for our reference GRB. Finally we use $t_{v} = 0.5$~s as the variability timescale of the GRB lightcurve~\cite{Albert:2020lvs}. 
The parameters adopted for each model for our benchmark GRB are summarized in Table~\ref{Table: all model parameters}. 
\begin{table}[ht]
	\caption{\label{Table: all model parameters}Characteristic parameters assumed for  our benchmark GRB jet for the scenarios considered in this paper: internal shock (IS) model, dissipative photosphere model with internal shocks (PH-IS), three components model (3-COMP), ICMART model, magnetized jet model with gradual dissipation (MAG-DISS), and proton synchrotron model (p-SYNCH). In the case of quantities varying along the jet, the variability range is reported.  For the magnetic model with gradual dissipation,  the  electron fraction, the electron power-law index,  and the  proton power-law index are defined in   Eqs.~\ref{eq:varepsilon_e_magnetic_model}, and \ref{eq: electron_slope_magnetic_model}, respectively.}
	\begin{center}
	\begin{adjustbox}{width=1.0\textwidth}
	\resizebox{\textwidth}{!}{
	\begin{tabular}{c|c|cccccc}
			\toprule
			\toprule
			{\it Parameter} & {\it Symbol} &\multicolumn{6}{c}{{\it Model}} \\
			&& IS & PH-IS & 3-COMP & ICMART & MAG-DISS & p-SYNCH\\
            \toprule
			Total jet energy & $ \tilde{E}_{\rm iso}$ & \multicolumn{5}{c}{ $ 3.4\times 10^{54} $~erg} &\multicolumn{1}{|c}{ n/a }  \\
			\cline{3-8}
			Jet opening angle  & $ \theta_{j} $& \multicolumn{6}{c}{ $ 3^{\circ} $}\\  
			\cline{3-8}
			Lorentz boost factor & $ \Gamma $ & \multicolumn{6}{c}{300} \\
			\cline{3-8}
			Redshift  & $ z $ &\multicolumn{6}{c}{2} \\
			\cline{3-8}
			Duration of the burst  & $ t_{\rm dur} $ &\multicolumn{6}{c}{$ 100 $~s} \\
			\cline{3-8}
			Variability time scale  & $ t_{v} $ &\multicolumn{6}{c}{$ 0.5 $~s} \\
			\cline{3-8}
			Dissipation efficiency & $ \varepsilon_{d} $ & \multicolumn{2}{c|}{$\varepsilon_{\rm IS}= 0.2$} &\multicolumn{1}{c|}{n/a} & \multicolumn{1}{c|}{$ \varepsilon_{d}=0.35$}& \multicolumn{1}{c|}{$0.24$} & n/a\\
			\cline{3-8}
			Electron energy fraction & $ \varepsilon_{e} $&  \multicolumn{3}{c|}{$0.01$}&\multicolumn{1}{c|}{$ 0.5 $} & \multicolumn{1}{c|}{$0.35-0.36$} & n/a\\
			\cline{3-8}
			Proton energy fraction & $ \varepsilon_{p} $& \multicolumn{3}{c|}{$0.1$}&\multicolumn{1}{c|}{$ 0.5 $} & \multicolumn{1}{c|}{$0.64-0.65$} & n/a\\
			\cline{3-8}
			Electron power-law index & $ k_{e} $&\multicolumn{2}{c|}{$2.2$} & \multicolumn{2}{c|}{n/a} & \multicolumn{1}{c|}{$2.4-2.6$} & n/a\\
			\cline{3-8}
			Proton power-law index & $ k_{p} $ & \multicolumn{3}{c|}{$2.2$}&\multicolumn{1}{c|}{$2$}  &\multicolumn{1}{c|}{$2.4-2.6$} & $2.6$\\
			\cline{3-8}
			Magnetization at $R_{\gamma}$ & $\sigma$&  \multicolumn{3}{c|}{n/a} & \multicolumn{1}{c|}{$45$}& \multicolumn{1}{c|}{$1.35 - 1.81$} & {n/a} \\
			\bottomrule
		\end{tabular}}
		\end{adjustbox}
	\end{center}
\end{table}

A  useful quantity that allows a comparison among different models  is the radiative efficiency of the jet, defined as the fraction of the total jet energy which is radiated in photons~\cite{LloydRonning:2004gv}:
\begin{equation}
	\label{eq: radiation_efficiency}
	\eta_{\gamma} = \frac{\tilde{E}_{\gamma, \rm{iso}}}{\tilde{E}_{\rm iso}}\ .
\end{equation}
For example $\eta_{\gamma} = \varepsilon_{\rm{PH}}$ when the dominant radiation is of photospheric origin or $\eta_{\gamma} = \varepsilon_{\rm IS} \varepsilon_{e}$ when the radiation is produced at the IS, assuming a fast cooling regime for  electrons.

\subsection{Spectral energy distribution of  protons}
\label{subsec:proton_cool_times}

For the purposes of this work, it is sufficient to  assume that protons and electrons in the dissipation site are accelerated via Fermi-like mechanisms\footnote{In the reconnection region there are various particle acceleration sites, see e.g.~Ref.~\cite{2015ApJ...815..101N}. It remains a matter of active research what is the dominant process responsible for the formation of the power-law, see e.g.~Refs.~\cite{Sironi:2014jfa, Guo2014, 2015ApJ...815..101N, 2018MNRAS.481.5687P, 2020ApJ...899..151K}}. 
The accelerated particles acquire
a non-thermal energy distribution that can be phenomenologically described as~\cite{Lipari:2007su}:
\begin{equation}
	n^\prime_{p}(E^\prime_{p}) = A  E^{\prime -k}_{p}\ \exp{\left[-\left(\frac{E^\prime_{p}}{E^\prime_{p,\rm{max}}}\right)^{\alpha_{p}}\right]}\Theta(E^\prime_{p} - E^\prime_{p,\rm{min}})\ ,
	\label{proton_spectrum}
\end{equation}
where 
$A = U^\prime_{p} \left[\int_{E^\prime_{p,\rm{min}}}^{E^\prime_{p,\rm{max}}} n^\prime_{p}(E^\prime_{p}) E^\prime_{p} dE^\prime_{p}\right]^{-1}$
is the normalization of the spectrum (in units of $\rm{GeV}^{-1} \rm{cm}^{-3}$) and $\Theta$ is the Heaviside function, with $U^\prime_{p} = \varepsilon_{p} \varepsilon_{d} E^\prime_{\rm{iso}}$ being the fraction of the  dissipated jet energy that goes into acceleration of protons. The power-law index is found to be $k \approx 2.3$ in the ultra-relativistic shock limit in semi-analytical and Monte Carlo simulations, although it is predicted to be steeper from particle-in-cell simulations~\cite{Sironi:2013ri}. 
The power-law index is instead $k = 2$ for a non-relativistic shock~\cite{Matthews:2020lig}, while it depends on the jet magnetization for magnetically dominated jets, as we will see later. The exponential cut-off with $\alpha_{p} $ is due to energy losses of protons and we adopt $\alpha_{p} = 2$ following Ref.~\cite{Hummer:2010vx}, $E^\prime_{p, \mathrm{min}}$ is the minimum energy of the protons that are injected within the acceleration region, and $ E^\prime_{p, \mathrm{max}} $ is the maximum  proton energy. The latter is constrained by the Larmor radius being smaller than the size of the acceleration region, or imposing that the acceleration timescale, 
\begin{equation}
	t^{\prime -1}_{p,\rm{acc}} = \frac{\zeta c e B^\prime}{E^\prime_{p}}\ ,
\end{equation}
is shorter than the total cooling timescale for protons. Here $\zeta =1$ is the acceleration efficiency adopted throughout this work. The total cooling timescale is given by
\begin{equation}
	t^{\prime-1}_{p,\mathrm{cool}} = t^{\prime -1}_{\mathrm{sync}} + t^{\prime -1}_{p,\mathrm{IC}} + t^{\prime -1}_{p,\mathrm{BH}} + t^{\prime -1}_{p\gamma} + t^{\prime -1}_{p,\mathrm{hc}} + t^{\prime -1}_{p,\mathrm{ad}}\ ;
\end{equation}
where  $t^\prime_{\rm{sync}}$, $t^\prime_{p,\rm{IC}}$, $t^\prime_{p,\rm{BH}}$, $t^\prime_{p\gamma}$, $t^\prime_{p,\rm{hc}}$, $t^\prime_{p,\rm{ad}}$ are the proton synchrotron (sync), inverse Compton (IC), Bethe-Heitler ($p\gamma \rightarrow p e^{+} e^{-}$, BH), hadronic (hc) and adiabatic (ad) cooling times, respectively. They are defined as follows~\cite{Dermer:2009zz,Razzaque:2005bh,Gao:2012ay}:

\begin{eqnarray}
	t^{\prime -1}_{p,\mathrm{sync}} &=& \frac{4\sigma_{T} m^{2}_{e} E^\prime_{p} B^{\prime 2}}{3 m^{4}_{p} c^{3} 8 \pi}\ ,\label{t_sync}\\
	t^{\prime -1}_{p,\mathrm{IC}} &=& \frac{3 (m_{e} c^{2})^{2} \sigma_{T} c}{16 \gamma^{\prime 2}_{p} (\gamma^{\prime}_{p} - 1) \beta^{\prime}_{p}}\int_{E^{\prime}_{\gamma,\mathrm{min}}}^{E^{\prime}_{\gamma,\mathrm{max}}} \frac{d E^{\prime}_{\gamma}}{E^{\prime 2}_{\gamma}} F(E^{\prime}_{\gamma}, \gamma^{\prime}_{p}) n^{\prime}_{\gamma}(E^{\prime}_{\gamma})\ , \label{t_IC}\\
	t^{\prime -1}_{p,\mathrm{BH}} &=& \frac{7 m_{e} \alpha \sigma_{T} c}{9 \sqrt{2} \pi m_{p} \gamma^{\prime 2}_{p}} \int_{\gamma^{\prime -1}_{p}}^{\frac{E^\prime_{\gamma,\mathrm{max}}}{m_{e}c^{2}}} d\epsilon^\prime \frac{n^\prime_{\gamma}(\epsilon^\prime)}{\epsilon^{\prime 2}} \left\{(2 \gamma^\prime_{p} \epsilon^\prime)^{3/2} \left[\ln(\gamma^\prime_{p} \epsilon^\prime) - \frac{2}{3}\right] + \frac{2^{5/2}}{3}\right\}\ ,\label{t_BH}\\
	t^{\prime -1}_{p\gamma} &=& \frac{c}{2\gamma^{\prime 2}_{p}}\int_{\frac{E_{\mathrm{th}}}{2 \gamma^\prime_{p}}}^{\infty} d E^\prime_{\gamma} \frac{n^\prime_{\gamma}(E^\prime_{\gamma})}{E^{\prime 2}_{\gamma}} \int_{E_{\mathrm{th}}}^{2 \gamma^\prime_{p} E^\prime_{\gamma}} dE_{r} E_{r}\sigma_{p\gamma}(E_{r}) K_{p\gamma}(E_{r})\ , \label{t_pgamma}\\
	t^{\prime -1}_{\mathrm{hc}} &=& c n^\prime_{p} \sigma_{pp} K_{pp}\ , \label{t_hc}\\
	t^{\prime -1}_{p,\mathrm{ad}} &=& \frac{ c \Gamma}{R}\ .\label{t_ad}
\end{eqnarray}
In the definitions above,  $  \epsilon^\prime = E^{\prime}_{\gamma}/m_{e} c^{2}$, $ \gamma^{\prime}_{p} = E^{\prime}_{p}/ m_{p} c^{2} $, and ${\alpha = 1/137}$ is the fine structure constant. The cross sections $\sigma_{p\gamma}$ and $\sigma_{pp}$, for  $p\gamma$ and $pp$ interactions respectively, are taken from Ref.~\cite{Zyla:2020zbs}. The function $F(E^{\prime}_{\gamma},\gamma^{\prime}_{p})$ is provided in Ref.~\cite{PhysRev.137.B1306}, while $K_{p\gamma}$ is the inelasticity of $p\gamma$ collisions~\cite{Dermer:2009zz}:
\begin{equation}
	K_{p\gamma}(E_{r}) = \begin{cases}
		0.2\,\,\,\,E_{\mathrm{th}} < E_{r} < 1\ \mathrm{GeV}\\
		0.6\,\,\,\, E_{r} > 1\ \mathrm{GeV}\ ,
	\end{cases}
\end{equation}
where $E_{r} = \gamma^\prime_{p} E^\prime_{\gamma}(1 - \beta^\prime_{p}\cos\theta^\prime)$ is the relative energy between a proton with gamma factor $\gamma^\prime_{p}$ and a photon of energy $E^\prime_{\gamma}$, whose directions form an angle $\theta^\prime$ in the comoving system, $E_{\mathrm{th}} = 0.15\,\rm{GeV}$ is the threshold for the photo-hadronic interaction, $n^\prime_{\gamma}(E^\prime_{\gamma})$ is the target photon density field (in units of $\mathrm{GeV}^{-1} \mathrm{cm}^{-3}$), $K_{pp} = 0.8$,  and $n^\prime_{p}$ is the comoving proton density defined as $n^{\prime}_{p}= n^{\prime}_{b}/2$, where $n^{\prime}_{b}$ is the baryonic density defined in Eq.~\ref{eq: baryon_density}.
As we will see in Sec.~\ref{susubsec: P_SYNCH_result}, the proton synchrotron scenario is such that  the properties of the proton distribution (e.g., minimum energy, power-law slope), as well as the shape of energy distribution itself, can be directly inferred  from the observed GRB prompt spectra.

\section{Neutrino production in the gamma-ray burst jet}
\label{subsec:neutrino production}

The simultaneous presence of a high density target photon field in the site of proton acceleration--that can be radiated  by  co-accelerated electrons, by protons themselves or have an external origin--leads to an efficient production of high-energy neutrinos through photo-hadronic interactions. Since the number of target photons is always much larger than the number density of non-relativistic (cold) protons in all cases of study, we neglect the $pp$ contribution.

Photo-hadronic interactions lead to charged pion and kaon (as well as neutron) production, which subsequently cool and  decay  in muons and neutrinos.
According to the standard picture,  pion production occurs through the $\Delta(1232)$ resonance channel: 
\begin{equation}
	p\,\,+\,\,\gamma\,\,\longrightarrow \,\,\Delta^{+}\,\,\longrightarrow \,\, \begin{cases}
		n+\pi^{+}\quad 1/3 \,\, \text{of all cases}\\
		p+\pi^{0}\quad \,2/3 \, \,\,\text{of all cases}
	\end{cases}
	\label{delta_resonance_channel}
\end{equation}
followed by the decay chain
\begin{align}
	\pi^{+}  \rightarrow &\mu^{+} + \nu_{\mu} \label{pi_plus_decay}\\
	& \mu^{+} \rightarrow \bar{\nu}_{\mu} + \nu_{e}\,+e^{+}\ .
\end{align}
In order to accurately estimate the neutrino spectral energy distribution and the related neutrino flavor ratio, we rely on the photo-hadronic interaction model of Ref.~\cite{Hummer:2010vx} (model Sim-B and Sim-C) based on  SOPHIA~\cite{Mucke:1999yb}. The latter 
includes higher resonances, direct and multi-pion production contributions. 
Note that, although we compute the neutrino and antineutrino spectral distributions separately, in the following we do not distinguish between them unless otherwise specified.

Given the photon  and proton energy distributions in the comoving frame, $n^\prime_{\gamma}(E^\prime_{\gamma})$ and $n^\prime_{p}(E^\prime_{p})$, the production rate of secondary particles  is given by~\cite{Hummer:2010vx} (in units of $\rm{GeV}^{-1}\rm{cm}^{-3}\rm{s}^{-1}$)
\begin{equation}
	Q^\prime_{l}(E^\prime_{l}) = \int_{E^\prime_{l}}^{\infty} \frac{dE^\prime_{p}}{E^\prime_{p}} n^\prime_{p}(E^\prime_{p}) \int_{E_{\mathrm{th}}/2 \gamma^\prime_{p}} ^{\infty} dE^\prime_{\gamma} n^\prime_{\gamma}(E^\prime_{\gamma}) c R_{\alpha}(x,y)\ ,
\end{equation}
where $x = E^\prime_{l}/E^\prime_{p}$ is the fraction of proton energy going into daughter particles, $y = \gamma^\prime_{p} E^\prime_{\gamma}$, and $l$ stands for $\pi^{+}$, 
$\pi^{-}$, $\pi^{0}$, and $K^{+}$.
Since kaons suffer less from radiative cooling than charged pions due to their larger mass and shorter lifetime, their contribution to the neutrino flux becomes important at high energies~\cite{Asano:2006zzb,Tamborra:2015qza,Petropoulou:2014lja}, whilst it is sub-leading at lower energies, given the low branching ratio for their production. The ``response function'' $R_{l}(x,y)$  contains all the information about the interaction type (cross section and multiplicity of the products); we refer the interested reader to Ref.~\cite{Hummer:2010vx} for more details. 

Once produced, the charged mesons undergo different energy losses before decaying into neutrinos. Their energy distribution at decay is 
\begin{equation}
	Q^{\prime \mathrm{dec}}_{l}(E^\prime_{l}) = Q^\prime_{l}(E^\prime_{l}) \left[1 - \exp\left(-\frac{t^\prime_{l, \mathrm{cool}}m_{l}}{E^\prime_{l} \tau^\prime_{l}}\right)\right]\ ,
	\label{cooled_Q}
\end{equation}
with $t^\prime_{l, {\rm{cool}}}$ being the cooling time scale and $\tau^\prime_{l}$ the lifetime of the meson $l$. The neutrino energy distribution originating from the decay processes like the one in Eq.~\ref{pi_plus_decay} is 
\begin{equation}
	Q^\prime_{\nu_{\alpha}}(E^\prime_{\nu_{\alpha}}) = \int_{E^\prime_{\nu_{\alpha}}}^{\infty} Q^{\prime \mathrm{dec}}_{l}(E^\prime_{l}) \frac{1}{E^\prime_{l}} F_{l\rightarrow \nu_{\alpha}}\left(\frac{E^\prime_{\nu_{\alpha}}}{E^\prime_{l}}\right)\ ,
\end{equation}
where $F_{l\rightarrow \nu_{\alpha}}$ is defined in Ref.~\cite{Lipari:2007su} for ultra-relativistic parent particles. The same procedure is followed for  antineutrinos.

The steps above also allow to compute  the spectra of charged muons. Again, the cooled muon spectra are derived as in Eq.~\ref{cooled_Q} and the neutrinos generated by the muon decay are computed following Ref.~\cite{Lipari:2007su}. 

The total neutrino injection rate $Q^\prime_{\nu_{\alpha}}(E^\prime_{\nu_{\alpha}})$ at the source is obtained by summing over the contributions from all channels. 
Finally, the   fluence for the flavor $\nu_\alpha$ at Earth from a  source at redshift $z$ is (in units of $\mathrm{GeV}^{-1} \mathrm{cm}^{-2}$) 
\begin{equation}
	\label{neutrino_fluence_earth}
	\Phi_{\nu_{\alpha}}(E_{\nu_{\alpha}},z) = \hat{N} \frac{(1 + z)^2}{4 \pi d^{2}_{L}(z)}\underset{\beta}{\sum} P_{\nu_{\beta} \rightarrow \nu_{\alpha}}(E_{\nu_{\alpha}}) Q^\prime_{\nu_{\beta}}\left[\frac{E_{\nu_{\alpha}}(1 + z)}{ \Gamma}\right]\ ,
\end{equation}
where~\cite{Anchordoqui:2013dnh}
\begin{align}
	P_{\nu_{e}\rightarrow\nu_{\mu}} &= P_{\nu_{\mu}\rightarrow\nu_{e}} = P_{\nu_{e}\rightarrow\nu_{\tau}} = \frac{1}{4}\sin^{2}2\theta_{12}\ ,\\
	P_{\nu_{\mu}\rightarrow\nu_{\mu}}&= P_{\nu_{\mu}\rightarrow\nu_{\tau}} = \frac{1}{8}(4-\sin^{2}2\theta_{12})\ ,\\
	P_{\nu_{e}\rightarrow\nu_{e}}&= 1-\frac{1}{2}\sin^{2}2\theta_{12}\ ,
\end{align}
with $\theta_{12} \simeq 33.5$ degrees~\cite{Esteban:2020cvm}, $P_{\nu_{\beta}\rightarrow \nu_{\alpha}} = P_{\bar{\nu}_{\beta}\rightarrow \bar{\nu}_{\alpha}}$, and $ \hat{N} = V^{\prime}_{\rm iso, s} t_{\rm dur}$~\cite{Baerwald:2011ee} being the normalization factor depending on the volume of the interaction region. The  luminosity distance   $ d_{L}(z) $ is  defined in a flat $\Lambda$CDM  cosmology as
\begin{equation}
	\label{luminosity_distance}
	d_{L}(z) = (1+z) \frac{c}{H_{0}} \int_{0}^{z} \frac{dz^\prime}{\sqrt{\Omega_{\Lambda}+\Omega_{M}(1+z^\prime)^3}}
\end{equation}
with $\Omega_{M} = 0.315$, $\Omega_{\Lambda} = 0.685$ and the Hubble constant $H_{0} = 67.4$~km s$^{-1}$ Mpc$^{-1}$~\cite{Aghanim:2018eyx}.

\section{Results: Gamma-ray burst neutrino emission}
\label{sec:emission_models}

Each of the dissipation mechanisms introduced in Sec.~\ref{sec: Properties of Astro jets}, according to the radius at which it takes place, leads to different photon energy distributions. In Appendix~\ref{sec:fitting_models} we report the empirical functions usually adopted to fit the observed photon spectra. For each of the GRB models considered in this section, we assume that the spectral energy distribution of photons is either given by one of the fitting functions or a combination of them. 
In this section, we 
investigate the neutrino production in the prompt phase for each scenario.

\subsection{Kinetic dominated jets}

\subsubsection{Jet model with internal shocks}
\label{subsubsec: simple IS model}
\begin{figure}[t]
	\centering
	\includegraphics[width=0.49\textwidth]{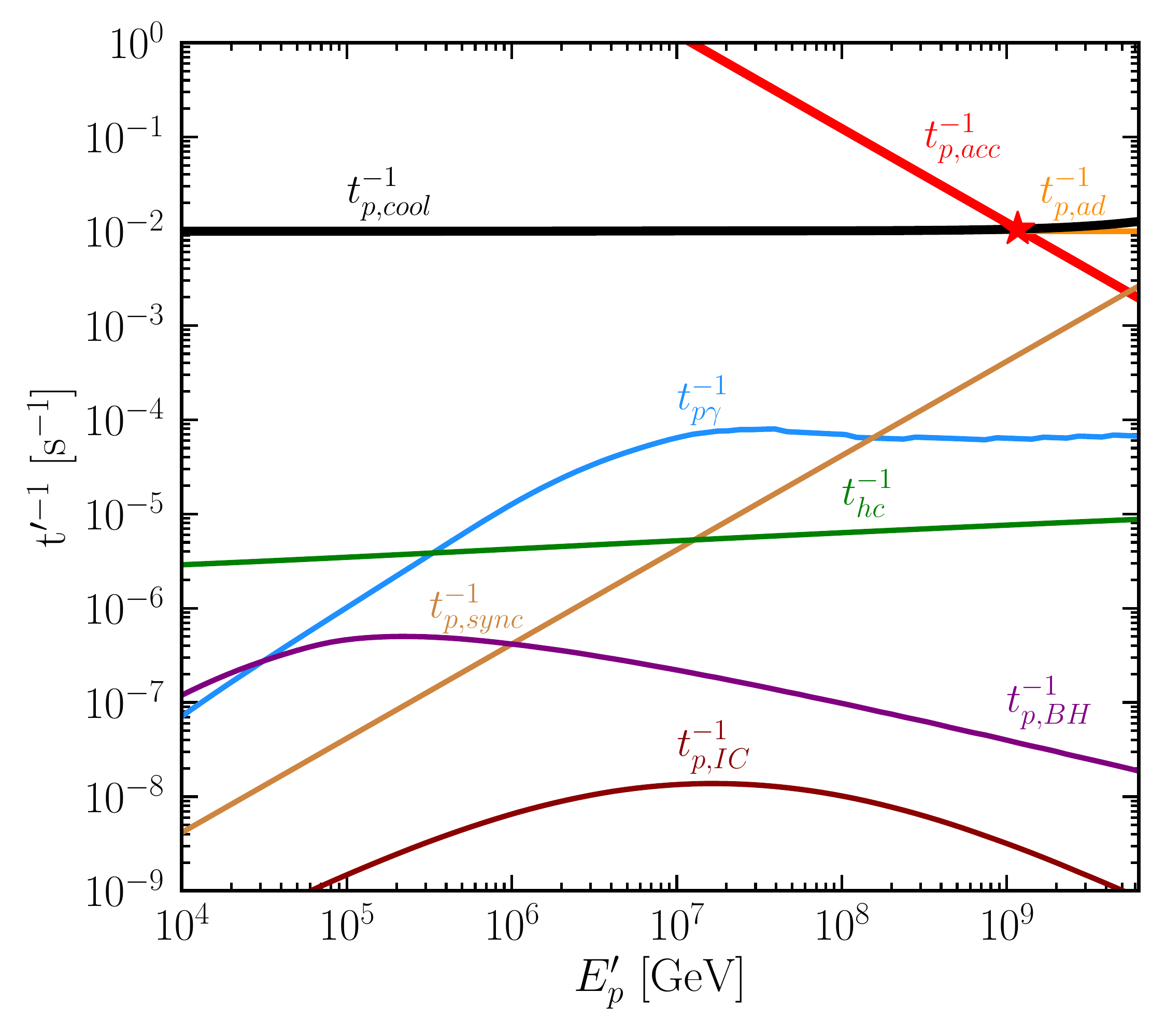}
	\includegraphics[width=0.49\textwidth]{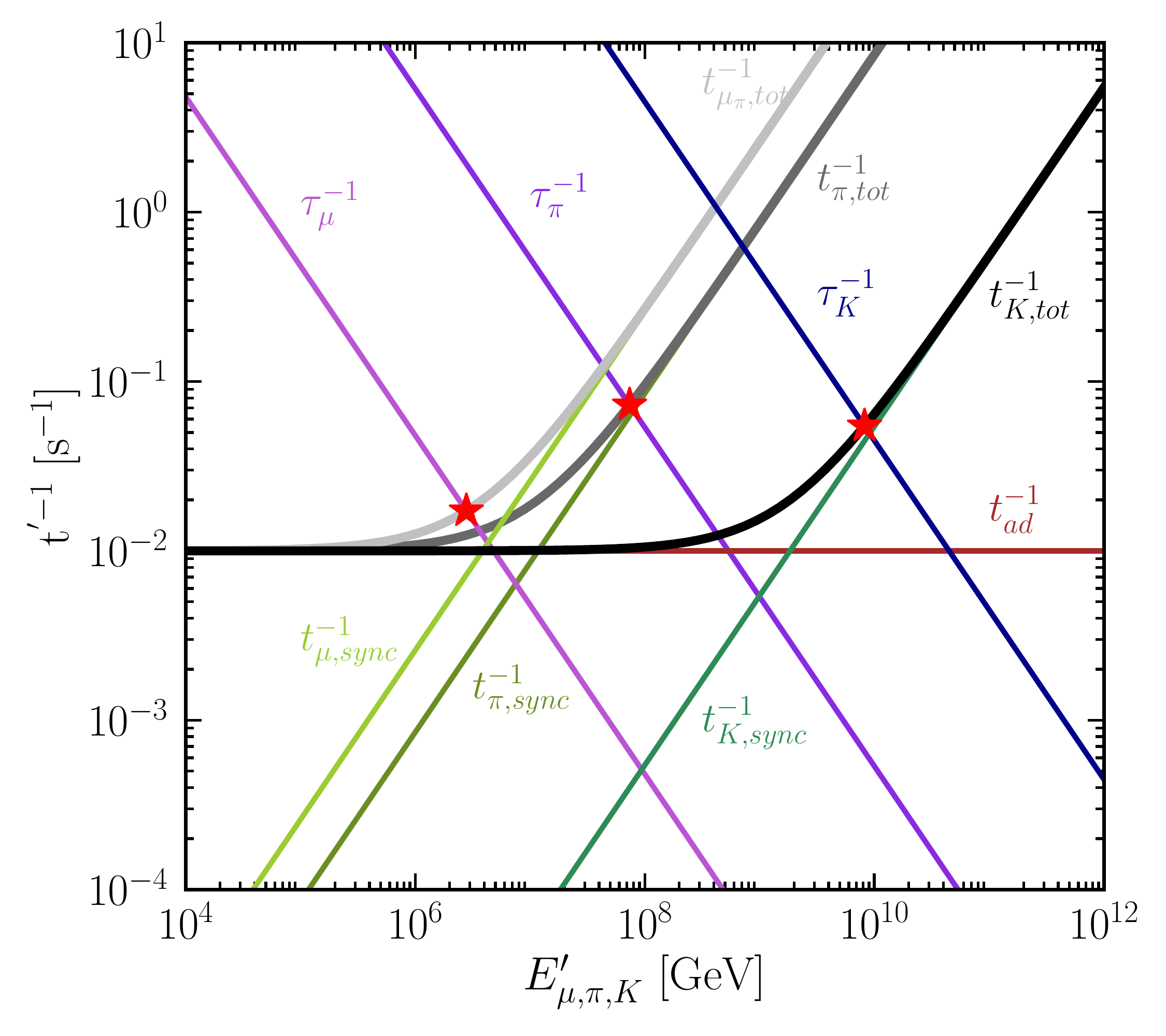}
	\caption{{\it Left:} Inverse cooling timescales for protons at the IS radius 
		as  functions of the proton energy in the comoving frame for our benchmark  GRB, see Table~\ref{Table: all model parameters}.
		The thin solid lines mark the individual cooling processes introduced in Sec.~\ref{subsec:proton_cool_times}; the thick black and red solid lines represent the total cooling timescale
		and the acceleration timescale,  respectively. The red star marks the maximum comoving  proton energy  such that $t^{\prime -1}_{p,{\rm cool}} = t^{\prime -1}_{p, {\rm acc}}$. Protons are mainly cooled by adiabatic expansion and $p\gamma$ interactions. 
		{\it Right:}  Analogous to the left panel, but for the inverse cooling timescales for pions, muons, and kaons. The dominant energy losses in this case are adiabatic cooling at low energies and synchrotron cooling at higher energies.}
			\label{Fig:t_p_cool_BAND}
\end{figure}

We focus on the IS model introduced in Sec.~\ref{sec:ISmodel} with the photon spectrum produced at the IS radius and described by the Band function in Eq.~\ref{eq:BAND function}. The radiative efficiency  is $\eta_{\gamma} = \varepsilon_{\rm IS} \varepsilon_{e} \simeq0.002$ (see Table~\ref{Table: all model parameters}). 

In order to establish the relative importance of the various energy loss processes in this scenario, we compute the proton and the secondary particle ($K^{\pm}$, $\pi^{\pm}$ and $\mu^{\pm}$) cooling times as illustrated in Sec.~\ref{subsec:proton_cool_times}. The cooling times are shown in Fig.~\ref{Fig:t_p_cool_BAND} as functions of the particle energy in the comoving frame.  With the parameters adopted for our benchmark GRB, protons are mainly cooled by adiabatic expansion up to $E^{\prime}_{p,\rm max}$ (left panel of Fig.~\ref{Fig:t_p_cool_BAND}), with the second dominant energy loss mechanisms being photo-hadronic interaction at intermediate energies, and synchrotron loss at higher energies. For mesons and muons (right panel of Fig.~\ref{Fig:t_p_cool_BAND}) adiabatic and synchrotron cooling at low and high energies, respectively, are the two dominant cooling processes.

\begin{figure}[tpb]
	\centering
	\includegraphics[width=0.49\textwidth]{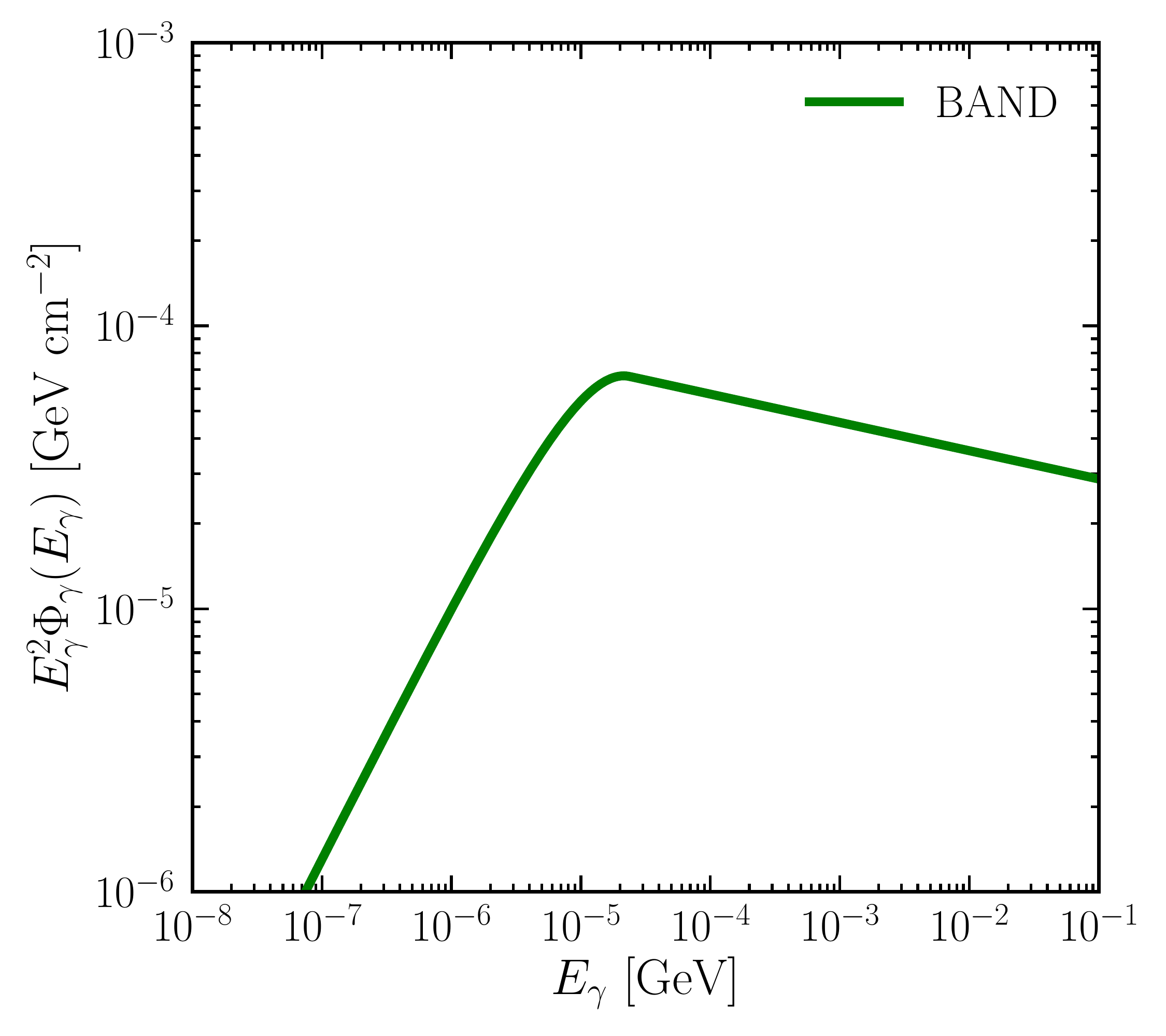}
	\includegraphics[width=0.49\textwidth]{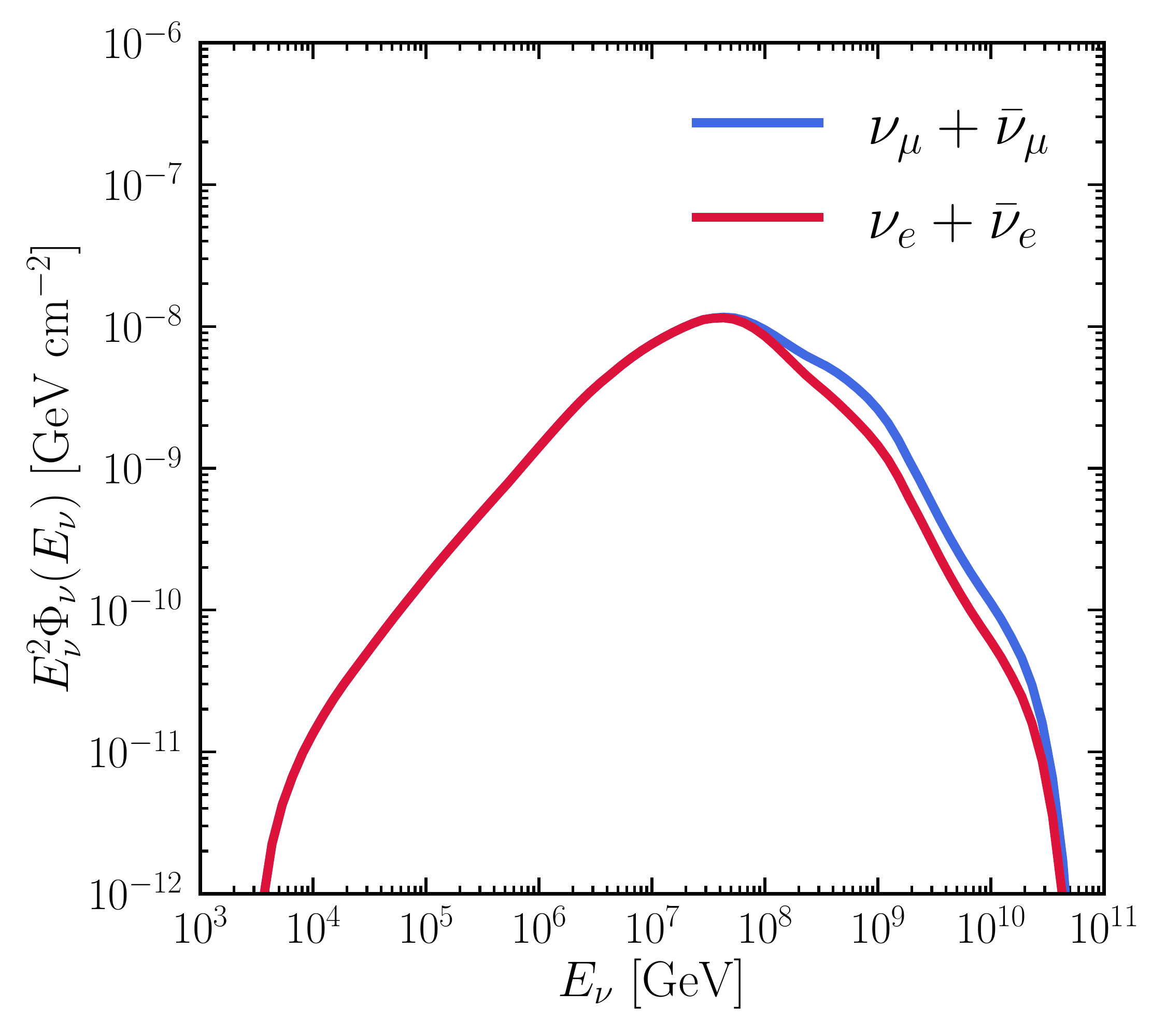}
	\caption{{\it Left:} Band photon fluence observed at Earth for our benchmark GRB in the IS model, see Table~\ref{Table: all model parameters}. {\it Right:}  Correspondent $\nu_\alpha+\bar{\nu}_\alpha$ fluence in the observer frame (in red for the electron flavor and in blue for the muon flavor) in the presence of  flavor conversions. The  fluence for the muon flavor peaks at $E^{\rm peak}_{\nu}= 4\times 10^{7}$~GeV. 
	}
	\label{Fig:neutrino_fluence_earth&ngamma_BAND}
\end{figure}

Following Sec.~\ref{subsec:neutrino production}, we compute the neutrino production rate in the comoving frame at $ R_{\rm IS}$ and the correspondent fluence at Earth including flavor conversions. The results are shown in the right panel of Fig.~\ref{Fig:neutrino_fluence_earth&ngamma_BAND}, while the photon spectrum described by the Band function is shown in the left panel.
Our results are in good agreement with analogous estimations reported in the literature for comparable input parameters, see e.g.~Ref.~\cite{Baerwald:2013pu}. In this scenario, the  fluence for the muon flavor peaks at $E^{\rm peak}_{\nu}= 4\times 10^{7}$~GeV and rapidly declines at higher energies. The effects due to the cooling of kaons are not visible because, as  shown in Fig.~\ref{Fig:t_p_cool_BAND}, the maximum proton energy is more than one order of magnitude lower than the one at which kaons  cool by synchrotron radiation, and the pion cooling starts around $\lesssim E^{\prime}_{p,\rm max}/4$.

\subsubsection{Jet model with a dissipative photosphere and internal shocks}
\label{subsubsec: Dissipative photosphere + IS}

We now explore  the model introduced in Sec.~\ref{sec:PH+IS} and consider a jet with an efficient photospheric emission, described by a Band spectrum peaking at the energy given by Eq.~\ref{eq: Amati_peak}, and  undergoing further IS dissipation. At  $R_{\rm IS}$,  protons and electrons are efficiently accelerated and turbulent magnetic fields may build up. In this scenario,  electrons cool, other than  by emitting synchrotron radiation, also by Compton up-scattering of the non-thermal photospheric photons. As we are interested in investigating the case where the photospheric emission is dominant in the MeV energy range, we consider  Case (I) of  Table 1 of Ref.~\cite{Toma:2010xw}, corresponding to  the luminosity hierarchy $L_{\mathrm{PH}} \gg L_{\mathrm{UP}} \gg L_{\mathrm{SYNC}}$, where $L_{\mathrm{PH}}$, $L_{\mathrm{UP}}$, and $L_{\mathrm{SYNC}}$  stand for the photospheric luminosity, up-scattered photospheric luminosity of the accelerated electrons at $R_{\mathrm{IS}}$, and synchrotron luminosity radiated by the electrons at $R_{\rm IS}$, respectively. 

Following Ref.~\cite{Toma:2010xw}, we define 
\begin{equation}
	x = \frac{\varepsilon_{\rm IS} \varepsilon_{B}}{\varepsilon_{\mathrm{PH}}} \quad \mathrm{and} \quad Y = \frac{U^\prime_{\mathrm{SYNC}}}{U^\prime_{\mathrm{B}}} = \frac{4}{3} \frac{(k_{e}-1)}{(k_{e}-2)} \tau_{T} \gamma^\prime_{e, \mathrm{min}} \gamma^\prime_{e, \mathrm{cool}} h\ ,
\end{equation}
where $Y$ is the Compton parameter, $k_{e}$ is the slope of the electron energy distribution, $h$ is a function of $\gamma^\prime_{e, \mathrm{min}}$ and $\gamma^\prime_{e, \mathrm{cool}}$ and depends on the cooling regime, $\gamma^\prime_{e, \mathrm{min}}$ is the minimum Lorentz factor of the electrons injected in the acceleration region
\begin{equation}
	\label{eq: gamma_e_min in shocks}
	\gamma^\prime_{e, \mathrm{min}} = \frac{m_{p}}{m_{e}} \frac{k_{e} - 2}{k_{e} - 1} \mathcal{R}^{-1} \xi^{-1} \varepsilon_{\rm IS} \varepsilon_{e}\ , 
\end{equation}
with $\xi$ being the fraction of electrons accelerated  at the shock and $\mathcal{R}$ being the number of leptons per baryon. Finally, $\gamma^{\prime}_{e, \mathrm{cool}}$ is the electron cooling Lorentz factor obtained from $\gamma^\prime_{e, \mathrm{cool}} m_{e} c^{2} = P(\gamma^\prime_{e, \mathrm{cool}}) t^\prime_{\mathrm{ad}}$ and given by
\begin{equation}
	\gamma^\prime_{e, \mathrm{cool}}(R) \simeq \frac{3m_{e} \mathcal{R}}{4m_{p} \tau_{T} \varepsilon_{\rm PH}} \frac{1}{x (1 + Y) + 1}\ ;
\end{equation}
$t^\prime_{\mathrm{ad}}$ being  the adiabatic cooling timescale and
${P(\gamma^\prime_{e, \mathrm{cool}}) = {4}/{3} \sigma_{\mathrm{T}} c \gamma^{\prime 2}_{e, \mathrm{cool}}(U^\prime_{\mathrm{B}} + U^\prime_{\mathrm{SYNC}} + U^\prime_{\mathrm{PH}})}$  the cooling rate for  electrons. The conditions we need to fulfill in order to satisfy $L_{\mathrm{PH}} \gg L_{\mathrm{UP}} \gg L_{\mathrm{SYNC}}$ are
\begin{equation}
	\eta < \eta_{\ast}\ , \quad x \ll 1\ ,\quad x Y \ll 1\ ,\quad Y = \frac{\varepsilon_{\rm IS} \varepsilon_{e} h}{\varepsilon_{\rm{PH}}} \ll 1\ . \label{eq:1_up_condition}
\end{equation}
In this way, it is possible to estimate  $L_{\mathrm{UP}} = Y L_{\mathrm{PH}}$ and $L_{\mathrm{SYNC}} = x Y L_{\mathrm{PH}}$. 

We adopt the electron slope $k_{e} = 2.2$ and fix $\varepsilon_{\rm IS}$  by relying on the observations in the optical band; 
by assuming that  the synchrotron extended emission in this range should not be brighter than what is typically observed, the following constraint on the  flux should hold:  $F^{\mathrm{sync}}_{\nu}(E_{\gamma,\mathrm{opt}}) < 100$~mJy with $E_{\gamma,\rm opt} = 2$~eV~\cite{Samuelsson:2018fan}. In our case, $\varepsilon_{\rm IS} = 0.2$  satisfies such a condition. The  radiative efficiency of this GRB is  $ \eta_{\gamma} = (\tilde{E}_{\rm PH} + \tilde{E}_{\rm SYNC} + \tilde{E}_{\rm UP})/\tilde{E}_{\rm iso} = \tilde{E}_{\rm PH}(1+Y+xY)/\tilde{E}_{\rm iso}\simeq 0.2$. 
Since the high-energy photopsheric photons are absorbed by the $e^{\pm}$ pair creation at $R_{\rm PH}$, we use a cut-off for the Band spectrum at $R_{\rm PH}$, defined in Eq.~\ref{eq: E_cutoff}.

\begin{figure}[tpb]
	\begin{center}
		\includegraphics[width=0.49\textwidth]{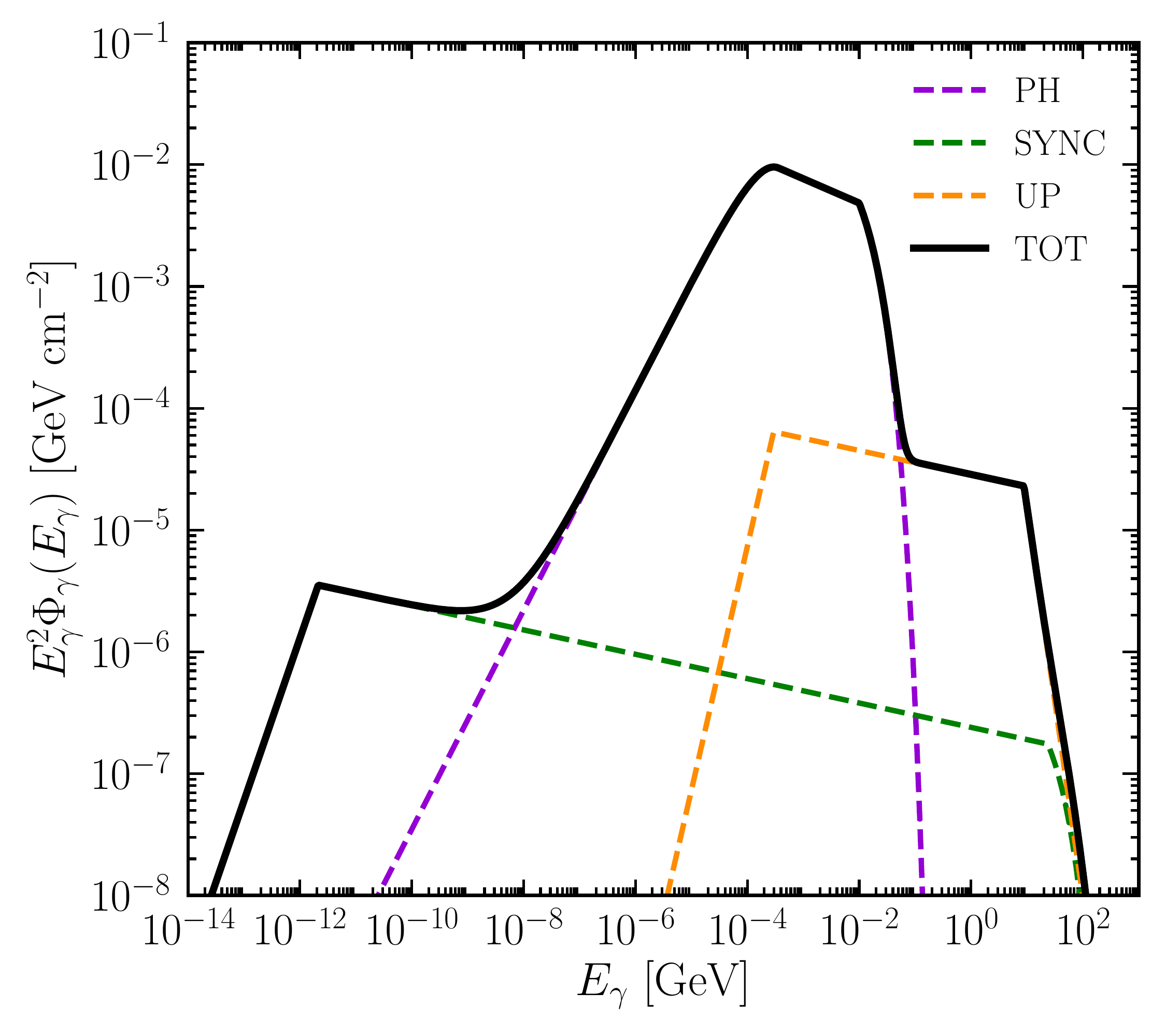}
		\includegraphics[width=0.49\textwidth]{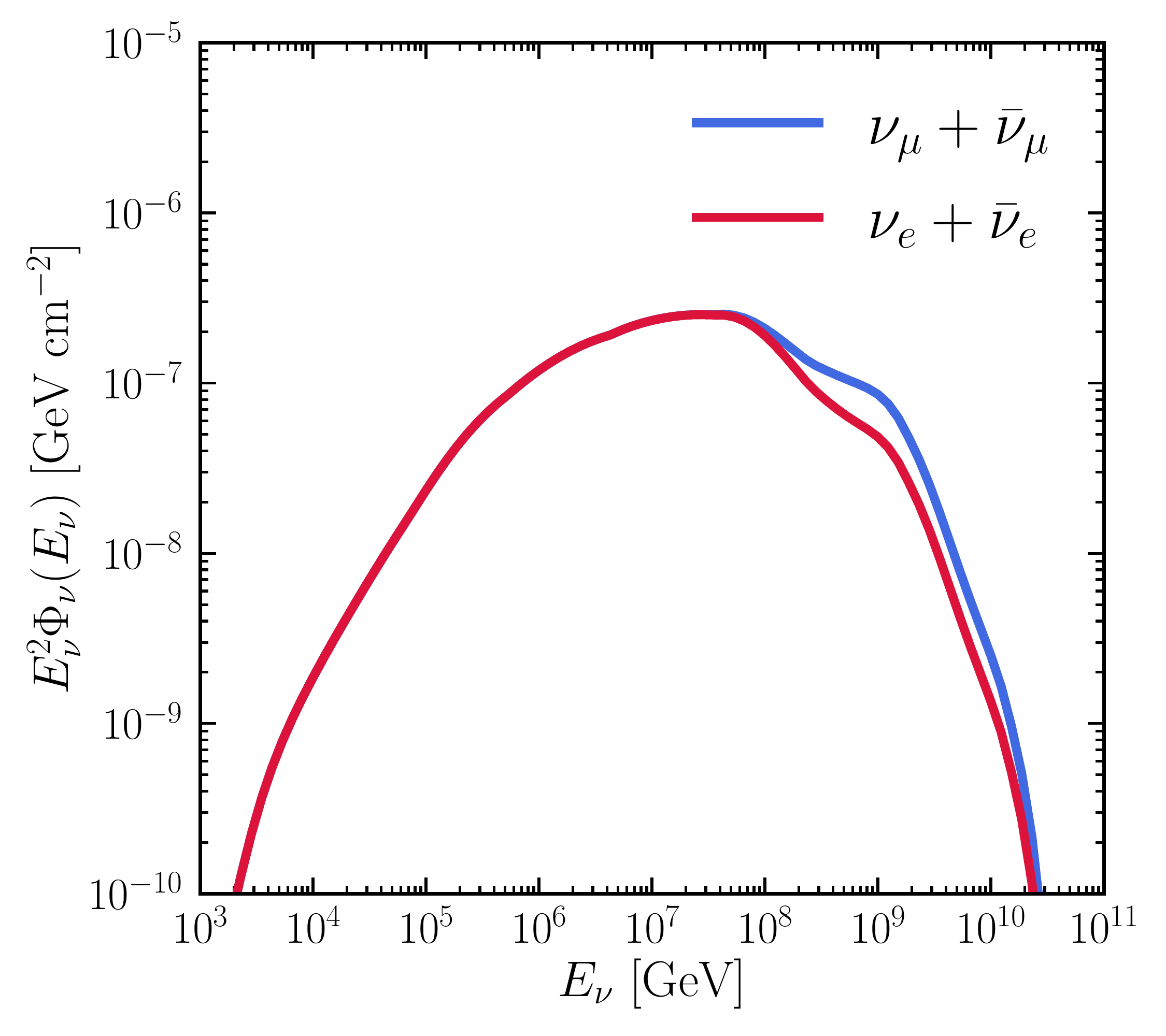}
	\end{center}
	\caption{{\it Left:} Photon fluence  observed at Earth for the IS model with dissipative photosphere. The photospheric emission (PH, violet line), the photospheric up-scattered emission (UP, orange line), and the synchrotron emission  of electrons accelerated at $R_{\mathrm{IS}}$ (SYNC, green  line) are plotted together with  the total photon fluence (in black).
	{\it Right:} $\nu_\alpha+\bar{\nu}_\alpha$ fluence in the  observer frame produced at $R_{\mathrm{IS}}$ with flavor oscillations included (in red for electron and blue for muon flavors).	The astrophysical parameters for this GRB are reported in Table~\ref{Table: all model parameters}, and  $\tilde{E}_{\mathrm{PH}} = 6.8 \times 10^{53}$~erg, $\tilde{E}_{\mathrm{SYNC}} \simeq 5.4\times 10^{50}$~erg, $\tilde{E}_{\mathrm{UP}} \simeq 6.8 \times 10^{51}$~erg,  $\mathcal{R} = 1$, $\xi$=1. The  fluence for the muon flavor peaks at $E^{\rm peak}_{\nu}=3.2 \times 10^{7}$~GeV and its normalization is larger than the one in Fig.~\ref{Fig:neutrino_fluence_earth&ngamma_BAND}, given a higher photon number density in the acceleration region.}
	\label{Fig:neutrino_fluence_earth&ngamma_BAND_SYNCH_EIC}
\end{figure}

We define the total photon energy distribution in the comoving frame at $R_{\rm IS}$ as
\begin{equation}
    n^{\prime}_{\gamma,\mathrm{tot}}(E^{\prime}_{\gamma})=\bigg(\frac{R_{\rm{PH}}}{R_{\rm{IS}}}\bigg)^{2} n^{\prime}_{\gamma,\rm{PH}}(E^{\prime}_{\gamma})+n^{\prime}_{\gamma,\rm{SYNC}}(E^{\prime}_{\gamma})+n^{\prime}_{\gamma,\rm{UP}}(E^{\prime}_{\gamma})
\end{equation}
and compute the cooling processes of protons at the IS. Because of the  more intense photon field at $R_{\rm IS}$ with respect to the simple IS scenario, $t_{p\gamma}$ and $t_{\rm BH}$ are shorter, while the synchrotron losses are  negligible.

Figure~\ref{Fig:neutrino_fluence_earth&ngamma_BAND_SYNCH_EIC} shows the resultant  photon (on the left panel) and neutrino (on the right panel) fluences for the IS shock scenario with a dissipative photosphere. The black curve in the left panel represents the overall photon fluence. 
The total spectrum is consistent with Fermi observations~\cite{FermiLAT:2018ksx}, being the high energy component subdominant with respect to the Band one.
From the right panel of Fig.~\ref{Fig:neutrino_fluence_earth&ngamma_BAND_SYNCH_EIC}, one can see that the fluence peaks at $E^{\rm peak}_{\nu}=3.2\times 10^{7}$~GeV and its normalization is only a factor $\mathcal{O}(10)$ larger than the one of the IS model  (see  Fig.~\ref{Fig:neutrino_fluence_earth&ngamma_BAND}), despite the larger  available photon energy [a factor $\mathcal{O}(100)$]. The reason for this is the higher $E^{\prime}_{\gamma, \rm peak}$ in the IS  scenario with dissipative photosphere, which affects  the ratio of the photon number densities at $E^{\prime}_{\gamma, \rm peak}$. It should also be noted that here we do not take into account the anisotropy of the incoming photospheric photon field at $R_{\rm IS}$, an effect that would cause a further reduction in the neutrino production efficiency, as pointed out also in Ref~\cite{Zhang:2012qy}.

\subsubsection{Jet model with three emission components}
\label{subsubsec: IS model with three components}

We are interested in a representative GRB of the class of bursts introduced in Sec.~\ref{sec:3comp}, hence we adopt average values for the spectral index and intensity of each component. To this purpose, we rely on  Refs.~\cite{Guiriec:2015ppa,Guiriec:2016bae,Arimoto:2016vhy}.

Once the outflow becomes transparent to radiation, a BB component is emitted at $R_{\rm PH}$, with  spectral index $\alpha_{\rm BB} = 0.4 $~\cite{Acuner:2019rif} and peaks at the temperature defined in Eq.~\ref{eq: photospheric temperature}. Subsequently, the kinetic energy of the outflow is dissipated at the ISs, and the main spectral component (CPL1) is produced. The latter is described by a CPL with spectral index $\alpha_{\rm CPL1} = -1$. An additional cut-off power-law (CPL2) begins to appear after a slight delay with respect to CPL1, with the cut-off shifting to higher energies until its disappearance. At later times, this additional component is well described by a simple PL, and we associate it to the beginning of the afterglow. With this choice, we take into account both  interpretations of the additional energetic component, namely the internal or external origin of CPL2.

At the deceleration radius $R_{\rm dec}$ (see Eq.~\ref{eq: deceleration radius}), the external shock starts
accelerating protons and electrons of the wind and the magnetic field builds up. Motivated by the afterglow modeling~\cite{Beniamini:2015eaa}, we use the following values for the energy fractions: 
$\varepsilon_{e} = 4\times 10^{-2}$, $\varepsilon_{B} = 10^{-4}$ and $\varepsilon_{p} = 1 - \varepsilon_{e} - \varepsilon_{B}$~\cite{Beniamini:2016hzc}, compatible with our choice for the prompt efficiency. 

We consider a wind type circumburst medium with $A= 3\times 10^{34}\,\rm{cm}^{-1}$~\cite{Razzaque:2013dsa} and an adiabatic blastwave, with $\Gamma(t)=\Gamma (t_{\rm dec}/4 t)^{1/4}$~\cite{Razzaque:2013dsa} and $R(t)=2 \Gamma^{2}(t) c t/(1+z)$ describing the temporal  evolution of the Lorenz factor and the radius of the forward shock after $t_{\rm dec}$, respectively. The energy of the accelerated particles in the blastwave, at a time $t$ after the deceleration, is $\tilde{U}_{p} = 4 \pi \varepsilon_{p} A R(t)m_{p} c^{2} [\Gamma^{2}(t) - 1]$. By relying on the temporal evolution of the  bright GRB investigated in Ref.~\cite{Ajello:2019avs}, we consider a simple PL produced at the forward shock, with power slope $\alpha_{\rm PL} = -1.8$ (Eq.~\ref{simple_PL}), and normalize it to $\tilde{E}_{\rm PL} = \varepsilon_{e}/\varepsilon_{p} \tilde{U}_{p}$. The photon field  target for $p\gamma$ interactions at the forward shock  is the sum of the PL, BB and CPL1 components; the latter two being Lorentz transformed in the comoving frame of the blastwave. 

Since we are interested in computing the neutrino fluence  emitted at the forward shock during the prompt phase, we take a representative average radius $R_{\ast}$ in logarithmic scale between $R_{\rm dec}$ and $R(t_{\rm dur}-t_{\rm dec})$. The  photon energy distribution is 
\begin{equation}
\label{eq: n_gamma_tot_FS}
    n^{\prime}_{\rm tot}(E^{\prime}_{\gamma},R_{\ast})= n^{\prime}_{\rm PL}(E^{\prime}_{\gamma})+\left(\frac{R_{\rm PH}}{R_{\ast}}\right)^{2} n^{\prime}_{\rm BB}\left(\frac{E^{\prime}_{\gamma}}{\Gamma_{\rm r}}\right)+\left(\frac{R_{\rm IS}}{R_{\ast}}\right)^{2} n^{\prime}_{\rm CPL1}\left(\frac{E^{\prime}_{\gamma}}{\Gamma_{\rm r}}\right)\ ,
\end{equation}
where $\Gamma_{\rm r}$ is the relative Lorenz factor between $\Gamma$ and $\Gamma_{\ast}\equiv \Gamma(R_{\ast})$.

\begin{figure}[]
	\centering
	\includegraphics[width=0.49\textwidth]{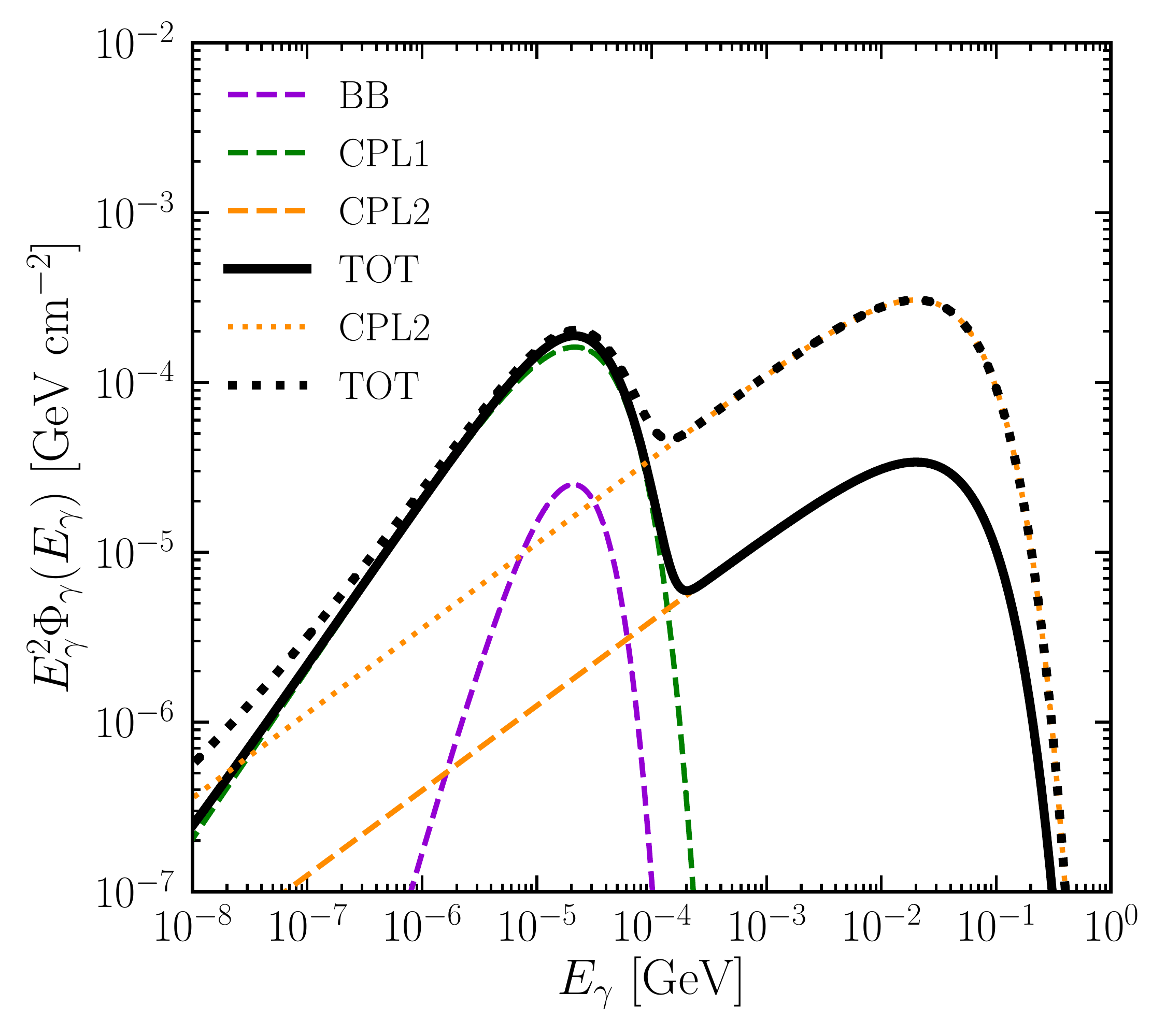}
	\includegraphics[width=0.49\textwidth]{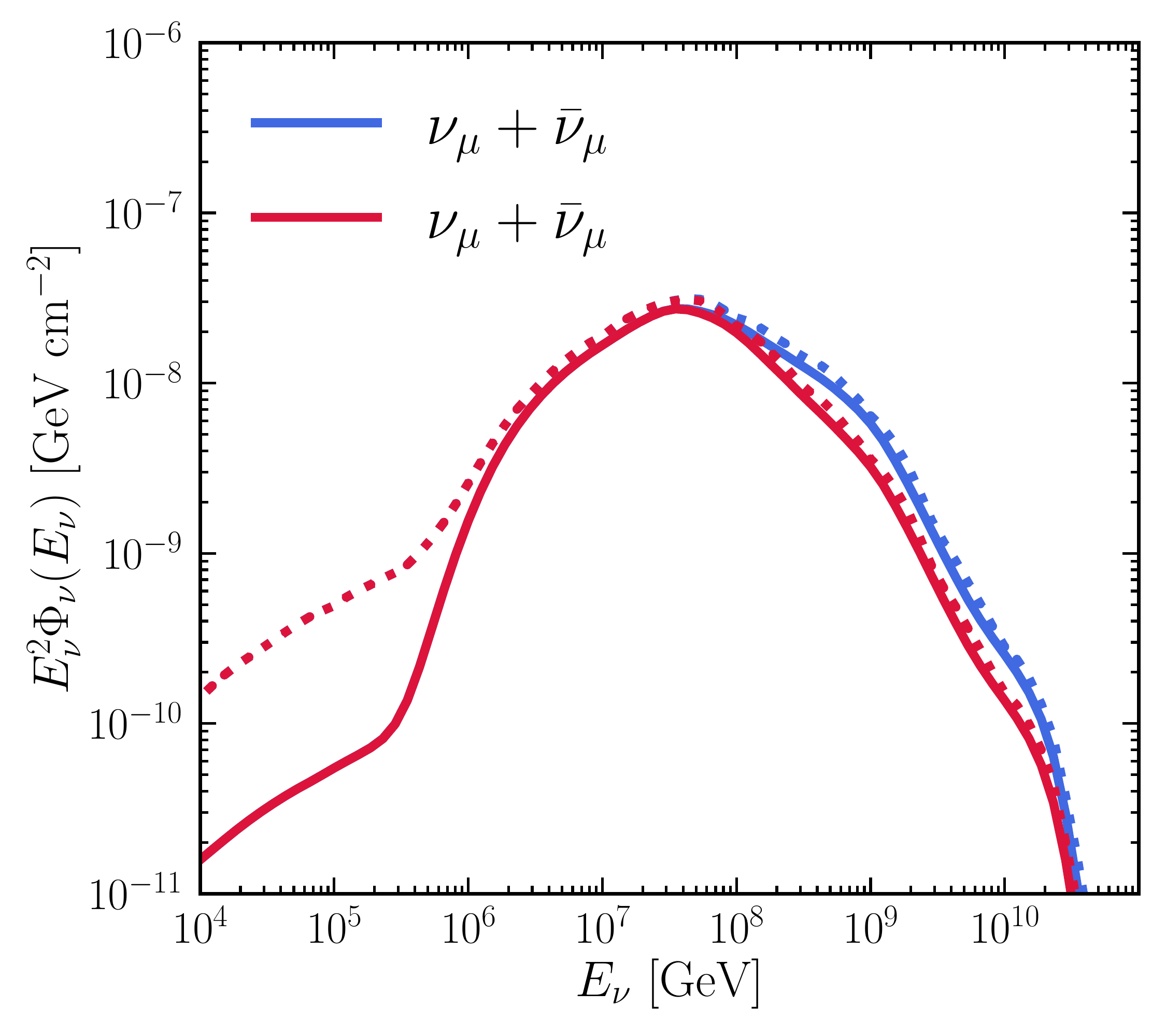}
	\includegraphics[width=0.49\textwidth]{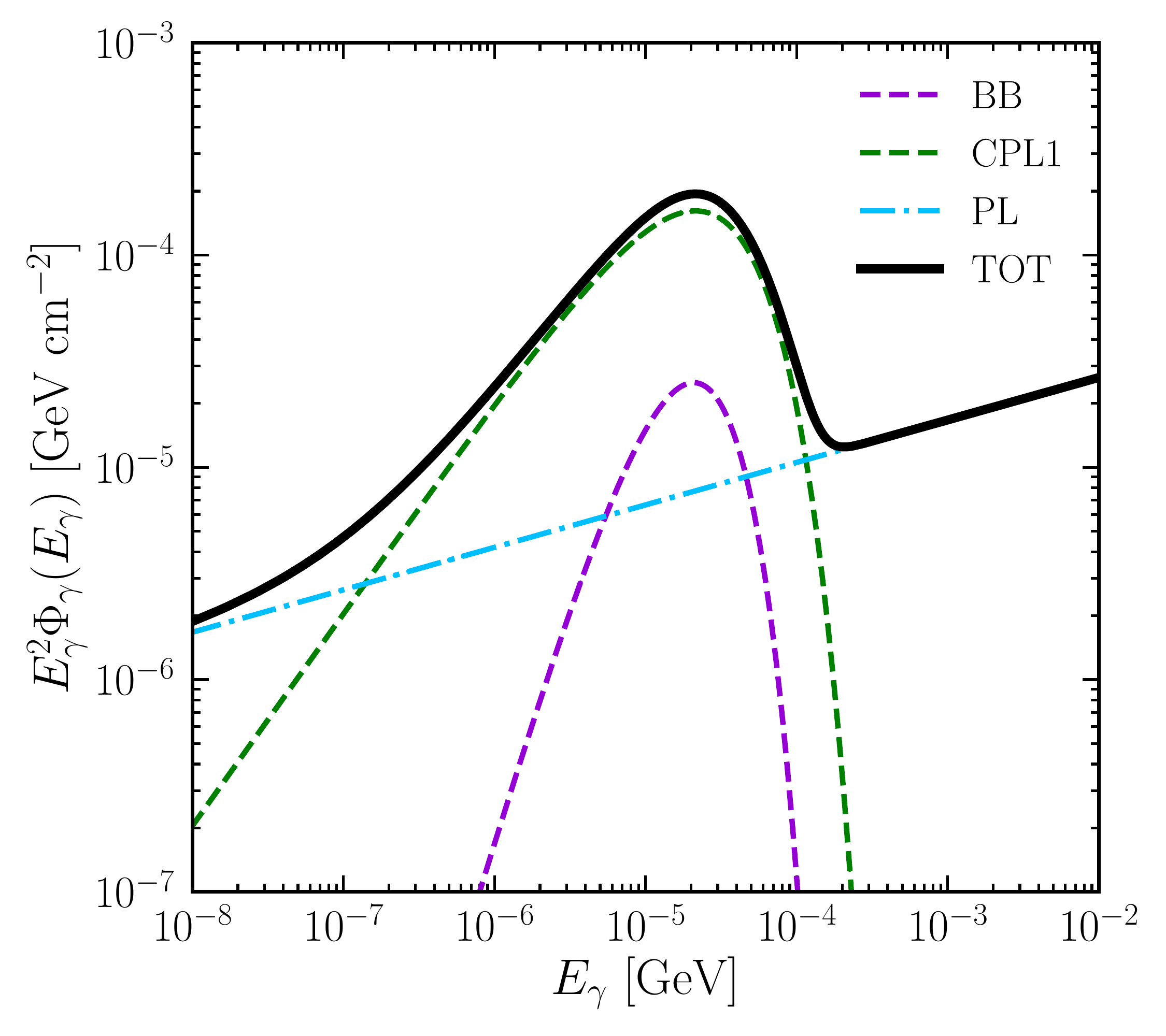}
	\includegraphics[width=0.49\textwidth]{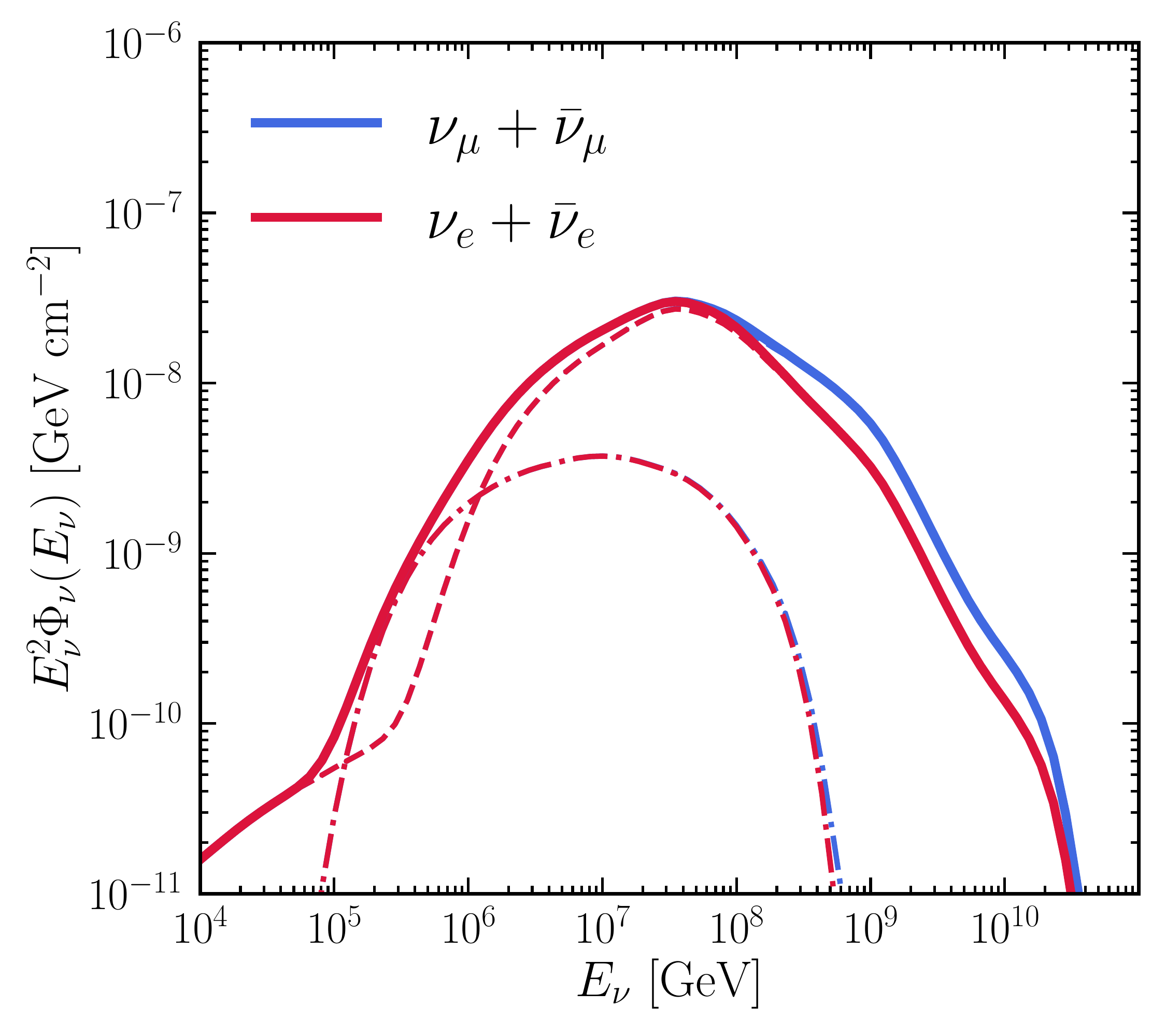}
	\caption{{\it Top left:} Photon fluence in the prompt phase for the  model with three components. It is composed by a thermal BB component (violet dashed curve), a cut-off power law CPL1 (green dashed curve), and a second cut-off power law CPL2 (in orange dashed for  case (I) such $\tilde{E}_{\mathrm{CPL1}} = 3 \tilde{E}_{\mathrm{CPL2}}$; in orange   dotted for  case (II) with $\tilde{E}_{\mathrm{CPL1}} = 1/3 \tilde{E}_{\mathrm{CPL2}}$). Furthermore  $\tilde{E}_{\mathrm{BB}} = 0.1 \tilde{E}_{\mathrm{CPL1}}$. The total fluence is plotted in black (solid and dotted lines). {\it Top right:} Correspondent  $\nu_\alpha+\bar{\nu}_\alpha$ fluence in the  observer frame  with flavor oscillations included (in red for electron and blue for muon flavors). The solid line represents the total contribution during the prompt for case (I), while the dotted one is for  case (II).
	The  fluence for the muon flavor peaks at $E^{\rm peak}_{\nu}= 4.2\times 10^{7}$~GeV for the case (I). The  low energy tail is affected by the interaction of protons with CPL2.
	{\it Bottom left:} Photon fluence for the scenario such that the emission from the forward shock (PL, blue dash-dotted line) starts during the prompt phase (dashed, BB+CPL1). {\it Bottom right:} Corresponding $\nu_{\alpha}+\bar{\nu}_{\alpha}$ fluence (dashed curve for the IS emission, dash-dotted for the forward shock, and solid line for the total).
	\label{Fig:neutrino_fluence_earth&ngamma_from_3_component_model_IS}}
\end{figure}

The BB component is always subdominant, while CPL1 and  CPL2 are expected to vary in absolute and relative intensity  from burst to burst; this is true also in the same GRB, once the temporal evolution is considered. In order to investigate to what extent the neutrino spectrum may be affected by these factors, we considered two scenarios of study for the prompt phase: case (I)  such that the energetics of the three components is $\tilde{E}_{\mathrm{BB}} \simeq 0.1 \tilde{E}_{\mathrm{CPL1}}$ and  $\tilde{E}_{\mathrm{CPL1}}= 3 \tilde{E}_{\mathrm{CPL2}}$ (solid black line in the top left panel of Fig.~\ref{Fig:neutrino_fluence_earth&ngamma_from_3_component_model_IS}) and case (II)  with $\tilde{E}_{\mathrm{CPL1}}= 1/3 \tilde{E}_{\mathrm{CPL2}}$ (dotted black line in the top left panel of Fig.~\ref{Fig:neutrino_fluence_earth&ngamma_from_3_component_model_IS}). The three cut-off power-laws (BB, CPL1, and CPL2) follow  Eq.~\ref{cutoff_PL} with  peak energies  $E_{\rm BB,peak}\simeq 2\times 10^{-5}$~GeV, $E_{\rm CPL1,peak}\simeq 7 \times 10^{-5}$~GeV, and $E_{\rm CPL2,peak}\simeq 2 \times 10^{-2}$~GeV, respectively. These values are consistent with the ones in Refs.~\cite{Guiriec:2015ppa,Ajello:2019avs}. With this set of parameters, the hierarchy and intensity of the various cooling processes is analogous to the  IS case  for the prompt phase (Fig.~\ref{Fig:t_p_cool_BAND}), while  adiabatic cooling is  the dominant cooling process by many orders of magnitude at $R_{\ast}$. 
For what concerns the forward shock, we assume a differential number density of protons $n^\prime_{p}(E^{\prime}_{p}) \propto E^{\prime\ -2.5}_{p}$~\cite{Sironi:2013ri} injected between the minimum energy $E^{\prime}_{p,\rm min} = m_{p} c^{2} \Gamma_{\ast}$ and the maximum $E^{\prime}_{p,\mathrm{max}}$, derived from the condition that the proton acceleration time $t^\prime_{\mathrm{acc}}$ is limited by the adiabatic time $t^\prime_{\mathrm{ad}}$ (see Sec.~\ref{subsec:proton_cool_times}). 

The neutrino fluence at the IS and forward shock is displayed in Fig.~\ref{Fig:neutrino_fluence_earth&ngamma_from_3_component_model_IS}. In the top left panel, the solid line represents the neutrino fluence for  case (I), while the dotted line stands for  case (II). The enhancement of the energetic component CPL2 of almost one order of magnitude leads to a negligible impact on the neutrino energy distribution, producing only a slight increase of the fluence at low energies.

The bottom  panels of Fig.~\ref{Fig:neutrino_fluence_earth&ngamma_from_3_component_model_IS}  display  the photon fluence and the correspondent neutrino fluence when the emission from the forward shock  starts during the prompt phase. 
The dashed lines represent the neutrino fluence produced at the IS from the interaction of accelerated protons and the photon field (BB+CPL1), while the dash-dotted line represents the neutrino outcome from the forward shock  at $R_{\ast}$, where we rely on Eq.~\ref{eq: n_gamma_tot_FS} for the photon field. The solid line describes the total neutrino fluence expected during the prompt phase. The forward shock contribution is significantly higher than what  expected for the afterglow phase~\cite{Razzaque:2013dsa}. This is mainly due to a much larger photon number density in the acceleration region. Furthermore, given the very low magnetic field and its inefficiency to accelerate particles to very high energies, the cutoff in the neutrino spectrum occurs at a lower energy  compared to the prompt case (see dot-dashed line). The overall intensity at peak energy of the neutrino emission in this scenario is slightly larger than in the simple IS case.

\subsection{Poynting flux dominated jets}

\subsubsection{ICMART model}

\begin{figure}[b]
	\begin{center}
		\includegraphics[width=0.49\textwidth]{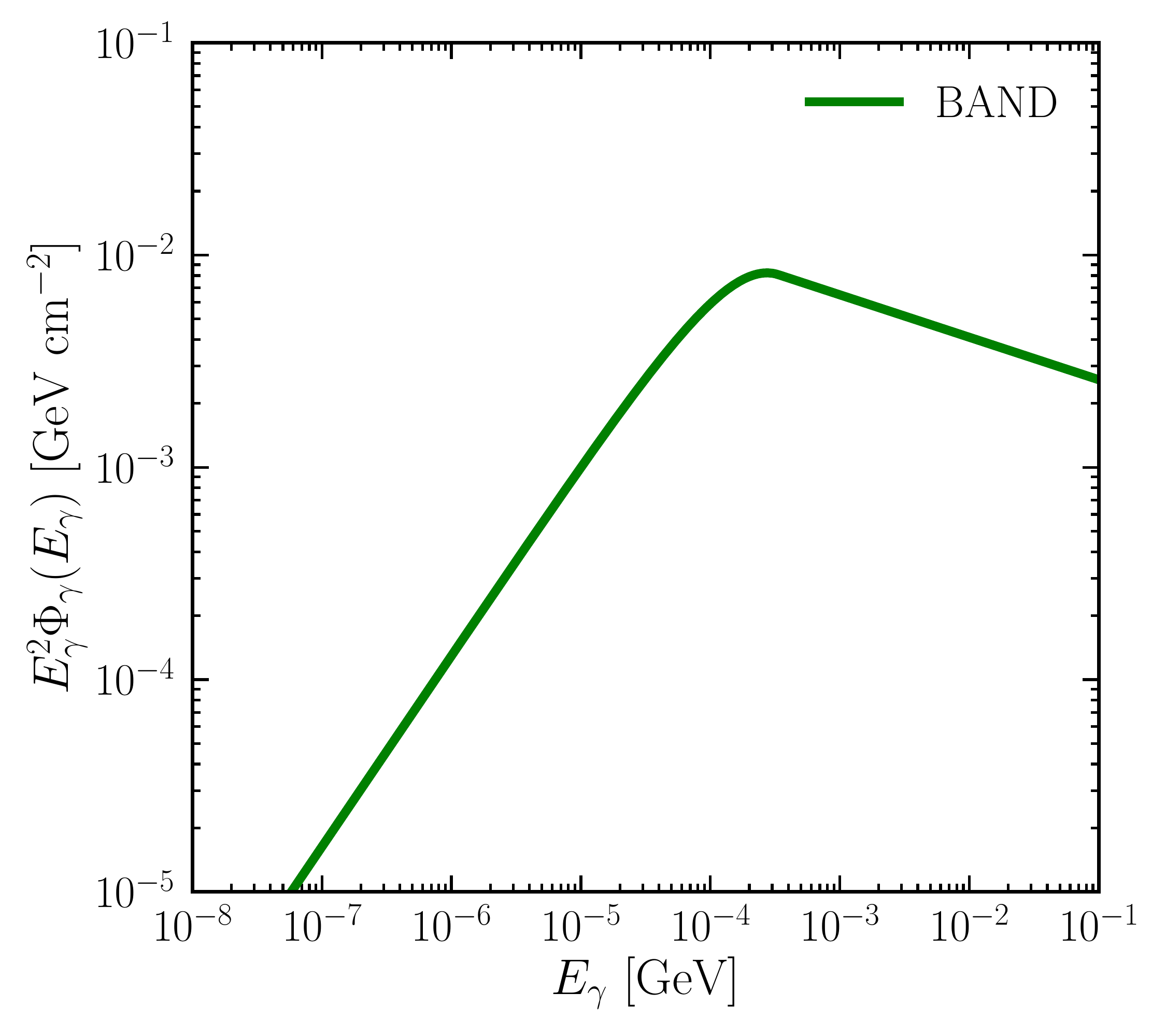}
		\includegraphics[width=0.49\textwidth]{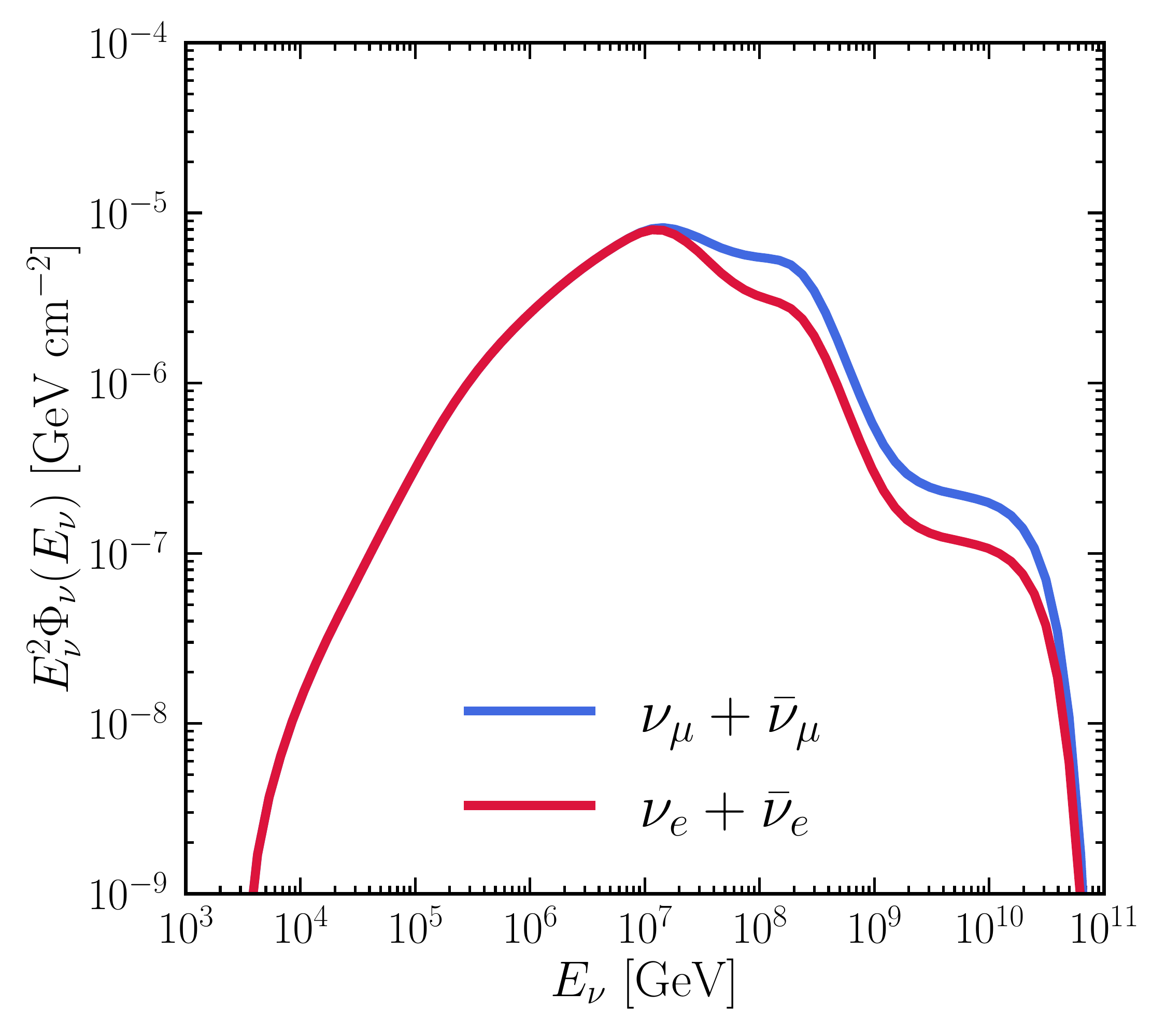}
	\end{center}
	\caption{{\it Left:} Band photon fluence observed at Earth and emitted at $R_{\rm ICMART} = 10^{15}$~cm for the ICMART model. The parameters of this GRB are reported in   Table~\ref{Table: all model parameters} and $\eta_{\gamma}= 0.17$. {\it Right:} Correspondent $\nu_\alpha+\bar{\nu}_\alpha$ fluence in the observer frame in the presence of  flavor conversions for the  ICMART model (in blue and red for the muon and the electron flavors, respectively).  The muon neutrino fluence peaks at $ E^{\rm peak}_{\nu} = 1.3\times 10^{7}$~GeV; the high energy tail of the neutrino distribution shows  the double bump structure due to kaon decay.}
	\label{Fig:neutrino_fluence_earth&ngamma_ICMART}
\end{figure}
For the model introduced in  Sec.~\ref{ICMART}, the typical radius necessary to make sure that runaway reconnection has enough time to grow is $R_{\mathrm{ICMART}} \simeq 10^{15}$~cm~\cite{Zhang:2010jt}, while the typical width of the reconnection region is $\Delta = c t_{v} $, where we adopt $ t_{v} \simeq 0.5$~s. The dissipation efficiency may be as high as $0.35$ in this model~\cite{Deng:2015xea} and this is the value we adopt for $\varepsilon_{d}$.

It has been shown that a Band-like spectrum may be reproduced in this scenario by considering an appropriate time dependent injection rate of particles in the emitting region~\cite{Liu:2020cvk}, and this is the spectrum we adopt for this model. If $\sigma$ is the magnetization parameter at $R_{\mathrm{ICMART}}$, the magnetic field in the bulk comoving frame can be expressed as~\cite{Zhang:2010jt}:
\begin{equation}
	B^\prime = \left(\frac{2\tilde{L}_{\mathrm{iso}}}{ \Gamma^{2} c R^{2}_{\mathrm{ICMART}}} \frac{\sigma}{\sigma + 1}\right)^{1/2}\ .
\end{equation}
We use as initial jet magnetization $\sigma_{0}=\sigma_{\rm in}=45$ (see Fig.~\ref{Fig:sketch ICMART}); this choice, as shown in the following,  allows for a consistent comparison with the results of  Sec.~\ref{subsubsec: Magnetic jet model with gradual dissipation}. By relying on the results from particle-in-cell simulations, we assume  $k_{p}(\sigma_{0})\simeq 2$ (see Eq.~\ref{eq: electron_slope_magnetic_model}). Furthermore,  we set  $\varepsilon_{p}= 0.5$ and $\varepsilon_{e}=0.5$ (see Eq.~\ref{eq:varepsilon_e_magnetic_model}). With this choice, the radiative efficiency of our benchmark GRB turns out to be $\eta_{\gamma}=\varepsilon_{d}\varepsilon_{e}\simeq 0.17$. The photon number density is normalized to $\eta_{\gamma}E^{\prime}_{\rm iso}$. Since $R_{\rm IS}\simeq R_{\rm ICMART}$, we obtain similar trends for the cooling times as in Fig.~\ref{Fig:t_p_cool_BAND}, except for the synchrotron loss that starts to dominate at $E^{\prime}_{p}\sim 10^{8}$~GeV and a slightly increased rate of $p\gamma$ interactions due to a larger photon number density in the dissipation region.

The neutrino fluence is displayed  in the right panel of Fig.~\ref{Fig:neutrino_fluence_earth&ngamma_ICMART}. The fluence for the muon flavor peaks at $ E^{\rm peak}_{\nu}= 1.3\times 10^{7}$~GeV. Note that  the double bump due to kaon decay is clearly visible in the high-energy tail of the energy distribution. This is due to   $E^{\prime}_{p,\rm max}\sim E^{\prime}_{K,\rm max}\sim 2\times 10^{9}$~GeV (while $E^{\prime}_{\pi,\rm max}~\sim 1.2\times 10^{7}$~GeV). This feature is determined by the stronger magnetic field in the acceleration region ($B^{\prime}\simeq 9$~kG, while for example $B^{\prime}\simeq 1$~kG in the  IS model).

\subsubsection{Magnetized jet model with gradual dissipation}
\label{subsubsec: Magnetic jet model with gradual dissipation}
For the model introduced in Sec.~\ref{sec:MAG-DISS}, we follow Ref.~\cite{Beniamini:2017fqh} and assume that the energy which is dissipated in the optically thick region  is reprocessed into  quasi-thermal emission, leading to a black-body-like emission from the photosphere. In the optically thin region, the synchrotron radiation from electrons is the dominant emission mechanism and it represents the non-thermal prompt emission. 
The energy emitted at the photosphere is obtained by integrating the energy dissipation rate  (Eq.~\ref{eq:E_B_diss_rate}) up to $R_{\mathrm{PH}}$ and considering that only the fraction $(R/R_{\mathrm{PH}})^{4/9}$ of the  energy dissipated at  $R$ remains thermal at $R_{\mathrm{PH}}$. In the optically thin region,  electrons are always in the fast cooling regime and 
$E^\prime_{\gamma, \mathrm{ssa}} \gg E^\prime_{\gamma,\mathrm{cool}}$ for $R_{\mathrm{ph}} < R <R_{\mathrm{sat}}$ with our choice of parameters. Here $E^\prime_{\gamma,\mathrm{ssa}}$ is the synchrotron self-absorption energy~\cite{Gill:2020oon}:
\begin{equation}
	E^\prime_{\gamma, \mathrm{ssa}} \sim \left(\frac{h^{3}}{8 \pi m_{p}} \frac{\xi \tilde{L}_{\mathrm{iso}}}{4 \pi \Gamma_{\rm sat}} \frac{1}{R^{2} \Gamma(R)}\right)^{1/3}\ .
\end{equation}
The shape of the synchrotron spectrum follows Eq.~\ref{eq:synch_spectrum_fast_cool}, but we replace $E^\prime_{\mathrm{cool}}$ with $E^\prime_{\gamma, \mathrm{ssa}}$ and use $\alpha _{\gamma}= 1$ for $E^\prime_{\gamma} < E^\prime_{\gamma, \mathrm{ssa}}$~\cite{Gill:2020oon}. Furthermore,  only a fraction $\xi= 0.2$~\cite{Beniamini:2017fqh} of electrons is accelerated.

For our reference GRB we adopt 
$ \lambda = 4\times 10^{8} $~cm~\cite{Beniamini:2017fqh}. The terminal Lorenz factor of the outflow is $\Gamma_{\rm sat} \simeq \Gamma_{0} \sigma_{0}$. We choose $\Gamma_{\rm sat} = 300$  and the initial jet magnetization is $\sigma_{0} = \Gamma^{2/3}_{\mathrm{sat}}\sim 45$.

For what concerns protons, we assume that they are accelerated into a power-law distribution starting from a minimum Lorentz factor
\begin{equation}
    \gamma^{\prime}_{p,\rm min}= \mathrm{max}\bigg[1,\frac{k_{p}-2}{k_{p}-1}\frac{\varepsilon_{p}}{2}\sigma( R ) \bigg]
\end{equation}
and extending up to a maximum value determined by balancing the  energy gain and loss rates, as described in Sec.~\ref{subsec:proton_cool_times}. We also assume that the power of the proton distribution is the same as  the one of electrons, namely $k_{p}=k_{e}$ (see Eq.~\ref{eq: electron_slope_magnetic_model}). The latter assumption is motivated by particle-in-cell simulations of magnetic reconnection for $\sigma \gg 1$~\cite{Guo2014}, but it has to be yet demonstrated for $\sigma\sim 1$~\cite{2019ApJ...880...37P}.

We compute the neutrino production assuming that dissipation and particle acceleration start at $R_{1}= a R_{\rm PH}$ with $a = 3$. Being an arbitrary choice for the starting radius, we explore the effects of $a$ on the neutrino fluence in  Appendix~\ref{MAG-DISS2}. For illustration purposes, we 
compute the  neutrino production rate at three radii ($R_{1}$, $R_{2}$, and $R_{3}$) equally distanced in logarithmic scale. 
We make this choice in order to establish the qualitative trend of the neutrino production during the evolution of the outflow in the optically thin region. The photon and proton distributions are normalized at each radius $R_{i}$ along the jet to the energy dissipated between $R_{i-1}$ and $R_{i}$, where $R_{0} = R_{\mathrm{PH}}$ and $R_{3} = R_{\mathrm{sat}}$. At each $R_{i}$, the photon field coming from $R_{j}$ is Lorentz transformed through the  relative Lorentz factor 
\begin{equation}
	\Gamma_{\mathrm{rel},ij} = \frac{1}{2} \left(\frac{\Gamma_{i}}{\Gamma_{j}} + \frac{\Gamma_{j}}{\Gamma_{i}}\right)\ ,
\end{equation}
that holds as long as $\Gamma_{i}, \Gamma_{j} \gg 1$. The total photon number density used as input at each radius $R_{i}$ for producing neutrinos is thus
\begin{equation}
\label{eq: n_gamma_at_R_i}
    n^{\prime}_{\rm tot}(E^{\prime}_{\gamma}, R_{i})= \sum_{j=0}^{i}\bigg(\frac{R_{j}}{R_{i}}\bigg)^{2} n^{\prime}_{j}\bigg(\frac{E^{\prime}_{\gamma}}{\Gamma_{\mathrm{rel},ij}}\bigg)\frac{1}{\Gamma_{\mathrm{rel},ij}}
\end{equation}
where $n^{\prime}_{j} (E^{\prime}_{\gamma})$ is the photon energy distribution at $R_{j}$ (in units of $\rm{GeV}^{-1} \rm{cm}^{-3}$).

Once the photon distributions are set, we evaluate the proton cooling times at each radius. In all the three cases,  dominant losses are due to  the adiabatic cooling up to $\simeq 10^{5}$~GeV, and $p\gamma$ interactions for $10^{5}\,\mathrm{GeV} \lesssim E^{\prime}_{p} \lesssim E^{\prime}_{p,\rm max}$. Synchrotron losses become relevant around $10^{7}$~GeV. Given the very strong magnetic field (see Table~\ref{Table: parameters at 3 Radii}), the secondaries  suffer strong synchrotron losses; this considerably affects the resulting neutrino spectrum, which is damped at energies much lower than in all the other models investigated so far in this work. A summary of the input parameters at the three $R_i$ is reported in Table~\ref{Table: parameters at 3 Radii}.

\begin{table}[t]
	\caption{\label{Table: parameters at 3 Radii} Summary table for the input parameters  adopted at the radii $R_{1}$, $R_{2}$ and $R_{3}$ in the magnetic model with gradual dissipation:  the radius ($R$), the comoving magnetic field ($B^{\prime}$), the Lorentz factor ($\Gamma$), the maximum energy of protons ($E^{\prime}_{p,\rm max}$), pions ($E^{\prime}_{\pi,\rm max}$), muons ($E^{\prime}_{\mu,\rm max}$), and kaons ($E^{\prime}_{K,\rm max}$), as well as the power-law slope ($k_{e}=k_{p}$) of electrons and protons.  }
	\begin{center}
		\begin{adjustbox}{width=1\textwidth}
			\begin{tabular}[c]{ccccccccc}
                \toprule
                \toprule
					&$R\,[\mathrm{cm}]$ & $B^{\prime}\,[\mathrm{kG}]$ & $\Gamma$ & $E^{\prime}_{p,\rm max}\,[\mathrm{GeV}]$ & $E^{\prime}_{\pi,\rm max}\,[\mathrm{GeV}]$ & $E^{\prime}_{\mu,\rm max}\,[\mathrm{GeV}]$ &
				$E^{\prime}_{K,\rm max}\,[\mathrm{GeV}]$ & $k_{e}$ \\
				\toprule

				$R_{1}$ & $7.1\times 10^{12}$ & $1.7\times 10^{3}$ & $176$ &$1.1\times 10^{8}$ & $6.5\times 10^{4}$ & $3.6\times 10^{3}$ & $7.5\times 10^{6}$& $2.4$\\
				$R_{2}$ & $1.6\times 10^{13}$ & $5.4\times 10^{2}$ & $230$ &$1.7\times 10^{8}$ & $2\times 10^{5}$ & $1.1\times 10^{4}$ & $2.3\times 10^{7}$ & $2.5$\\
				$R_{3}$ & $3.6\times 10^{13}$ & $1.7\times 10^{2}$ & $300$ &$2.5\times 10^{8}$ & $6.4\times 10^{5}$ & $3.5\times 10^{4}$ & $7.3\times 10^{7}$ & $2.6$\\
				\bottomrule
			\end{tabular}
		\end{adjustbox}
	\end{center}
\end{table}

\begin{figure}[]
	\centering
	\includegraphics[width=0.49\textwidth]{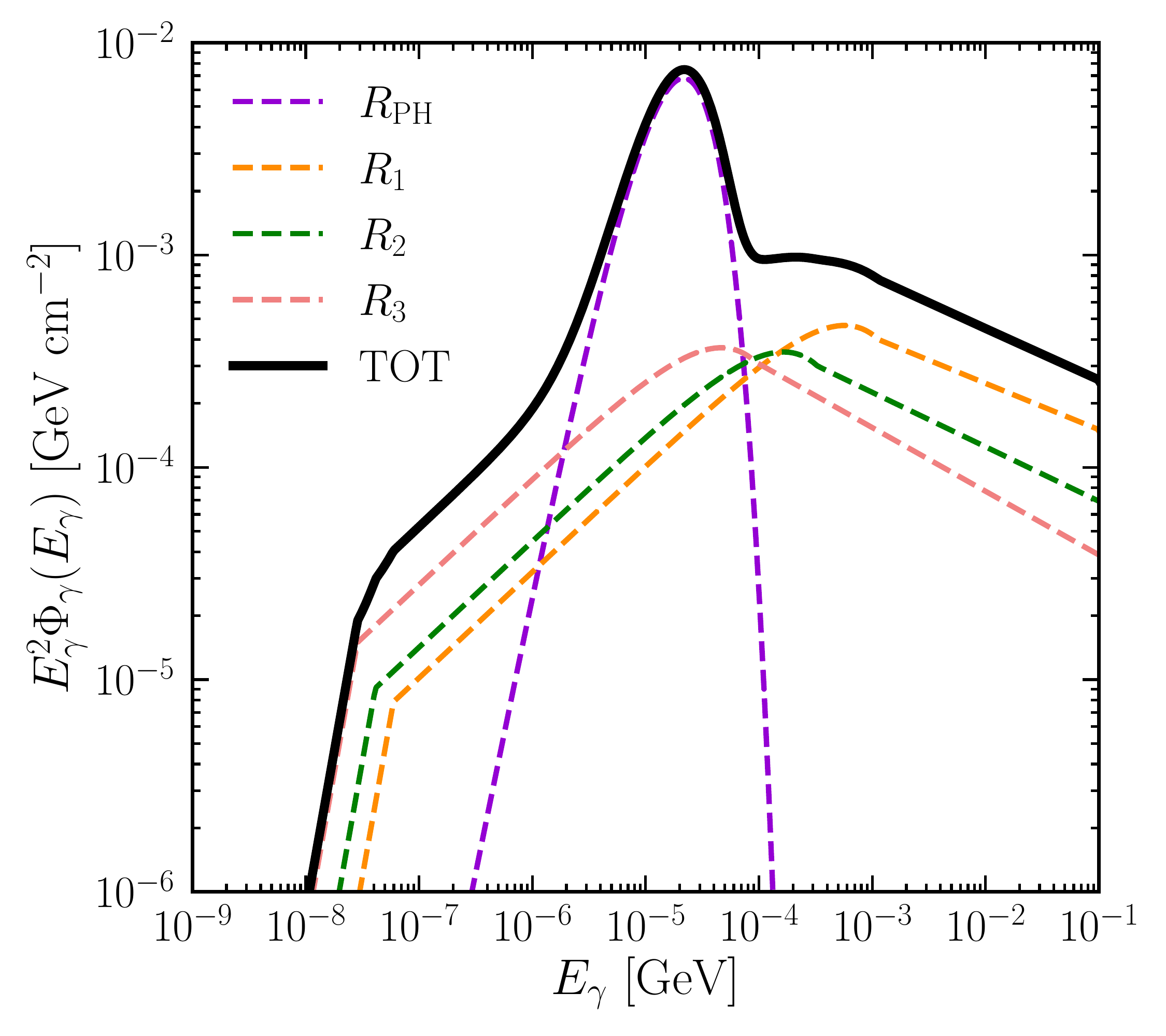}
	\includegraphics[width=0.49\textwidth]{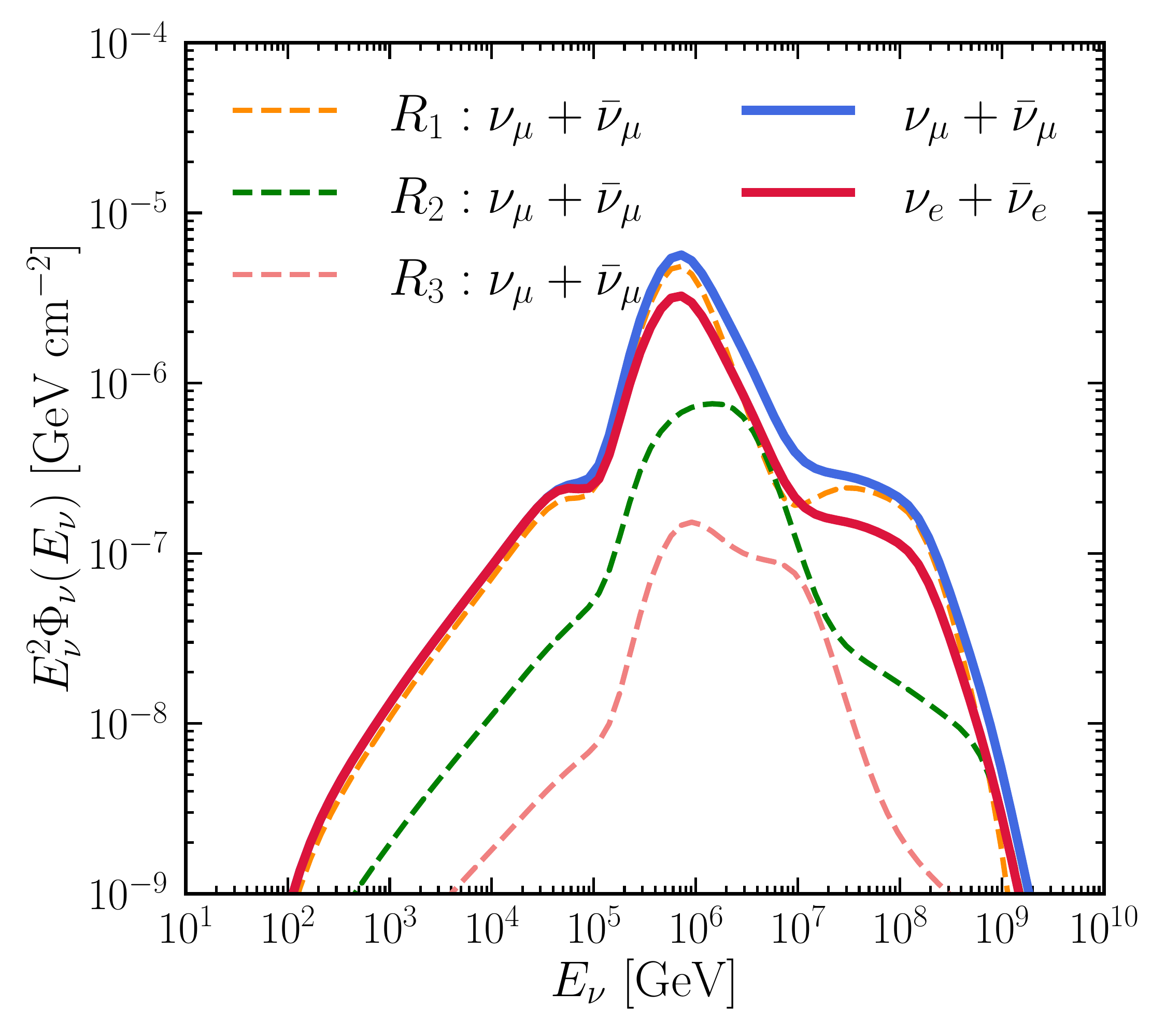}
	\caption{{\it Left}: Photon fluence in the observer frame for the GRB model invoking continuous magnetic dissipation for  the parameters reported in   Tables~\ref{Table: all model parameters} and \ref{Table: parameters at 3 Radii}. The total photon energy distribution is shown in black, and its components at $ R_{\rm PH}$, $R_{1}$, $R_{2}$, and $R_{3}$  are plotted in violet, orange, green, and coral respectively. {\it Right:} Correspondent $\nu_\alpha+\bar{\nu}_\alpha$ fluence at Earth after flavor oscillations   in the left panel (in blue for the muon flavor and in red for the electron one).  The  fluence for the muon flavor peaks at $E^{\rm peak}_{\nu}= 7.2\times 10^{5}$~GeV. An unusual spectral structure  is clearly visible.  
	}
	\label{Fig:PHOTON_MAGNETIC}
\end{figure}
The (photon) neutrino fluence at Earth is shown  in the (left) right panel of Fig.~\ref{Fig:PHOTON_MAGNETIC}.
The slope of the three non-thermal synchrotron components and their distribution peaks decrease as the  distance from the source increases. The high energy cut-off of each spectral component is given by Eq.~\ref{eq: E_cutoff}.
Notably, the dominant component comes from the smallest radius, while the contribution coming from larger radii gets lower and lower ($67\%$, $26\%$, and $7\%$ from $R_{1}, R_{2}$ and $R_{3}$, respectively).
The significant drop in the neutrino flux between $R_{1}$ and $R_{3}$ is mainly due to the 
decrease of the proton power slope (see Table~\ref{Table: parameters at 3 Radii}), which causes a more pronounced drop in proton number density in the energy range of interest.
This is a peculiar feature of this  model, which predicts parameters depending on the jet magnetization, and thus changing with the radius.

The  neutrino fluence for the muon flavor peaks at $E^{\rm peak}_{\nu}= 7.2\times 10^{5}$~GeV, which is about $\mathcal{O}(10-100)$~GeV smaller than in the models presented in the previous sections, although roughly comparable in intensity at peak. This is due to the fact that, in this case, the main contribution to the neutrino flux comes from the interaction of protons with thermal photons, whose energy peaks at $\sim 22$~keV. The second bump visible in the spectrum is instead represented by the kaon contribution. Apart from the ICMART model, this is the only other case out of the ones studied in this work in which this feature is clearly identified at higher energies. The reason is the very strong magnetic field in these two magnetic models. Another peculiar feature of this model is the low-energy tail of the neutrino distribution, which is higher than in  previous cases. This is  due to a combination of the larger number density [$\mathcal{O}(10^{3}-10^{4})$] of protons at low energies in the acceleration region and the extended photon field at higher energies.

\subsection{Proton synchrotron model}
\label{susubsec: P_SYNCH_result}

In order to estimate the neutrino production in the proton synchrotron model (see Sec.~\ref{sec:psync}), we need to evaluate  the fraction of the proton energy which goes into $p\gamma$ interactions. We  consider the photon spectral fit as in Eq.~\ref{eq:synch_spectrum_fast_cool} and follow Ref.~\cite{Oganesyan:2019fpa}, which provides the cooling energy $E_{\gamma, \mathrm{cool}}$, the peak energy (or minimum injection energy) $E_{\gamma, \mathrm{peak}} \equiv E_{\gamma, \mathrm{min}} $, and the energy flux at the cooling energy ($F_{\gamma, \mathrm{cool}}$). 

Another inferred quantity is the cooling timescale  of the radiating particles,  $t_{\mathrm{cool}} \sim 1$~s. The cooling time $ t_{\rm cool} $ is related, after Lorentz transforming, to the comoving magnetic field $B^{\prime}$ and $\gamma^{\prime}_{\rm cool}$ by means of  Eq.~\ref{cooling_gamma}. The  variability timescale is assumed to be $t_{v} = 0.5$~s; the duration of the burst, as well as the redshift information, is extracted from the GRB catalog~\cite{GRBweb}. These observables can be used to constrain  the source parameters, such as  $B^\prime$, $\gamma^\prime_{\mathrm{min}}$, $R_{\gamma}$, $\Gamma$, and $E^\prime_{\gamma, \mathrm{bol,iso}}$ through the following relations~\cite{Oganesyan:2019fpa}:
\begin{eqnarray}
	E_{\gamma, \mathrm{peak}} &=& \frac{3}{2} \frac{\hbar e B^\prime \gamma^{\prime 2}_{\mathrm{min}}}{m_{p} c} \frac{ \Gamma}{1 + z}\ ,\\
	\label{eq:epsilon_cool_proton}
	E_{\gamma, \mathrm{cool}} &=& \left(\frac{m_{p}}{m_{e}}\right)^{5} \frac{54 \pi^{2} \hbar e m_{e} c}{\sigma^{2}_{T} B^{\prime 3} t^{2}_{\mathrm{cool}}}\frac{1 + z}{ \Gamma}\ ,\\
	F_{\gamma} &=& F_{\gamma, \mathrm{cool}} \left(\frac{E_{\gamma, \mathrm{cool}}}{h}\right) \left[\frac{3}{4} + 2 \sqrt{\frac{E_{\gamma, \mathrm{peak}} }{E_{\gamma, \mathrm{cool}}} }- 2 + \frac{2}{k_{p} - 2} \sqrt{\frac{E_{\gamma, \mathrm{peak}}}{E_{\gamma, \mathrm{cool}}}}\right]\ , \\
	E^\prime_{\gamma, \mathrm{bol,iso}} &=& \frac{4 \pi d^{2}_{L}(z) F_{\gamma} t_{\mathrm{dur}}}{\Gamma (1+z)}\ , \\
	\label{eq:diss_radius}
	R_{\gamma} &=& \frac{2 c t_{v} \Gamma^{2}}{(1 + z)}\ .
\end{eqnarray}
where $ F_{\gamma} = \tilde{L}_{\gamma,\rm bol,iso}/ 4 \pi d_{L}^{2}(z)$ is the bolometric isotropic radiative flux (in units of $\rm GeV\, cm^{-2}\, s^{-1} $), $\tilde{L}_{\gamma,\rm bol, iso}$ being the bolometric isotropic luminosity of the burst over the whole energy range. Using these relations we can infer $ B^{\prime}, \gamma^{\prime}_{\rm{min}}, R_{\gamma} $ and $ E^\prime_{\gamma,\rm bol, iso} $ as functions of $ \Gamma$. 

\begin{figure}[]
	\centering
	\includegraphics[width=0.6\textwidth]{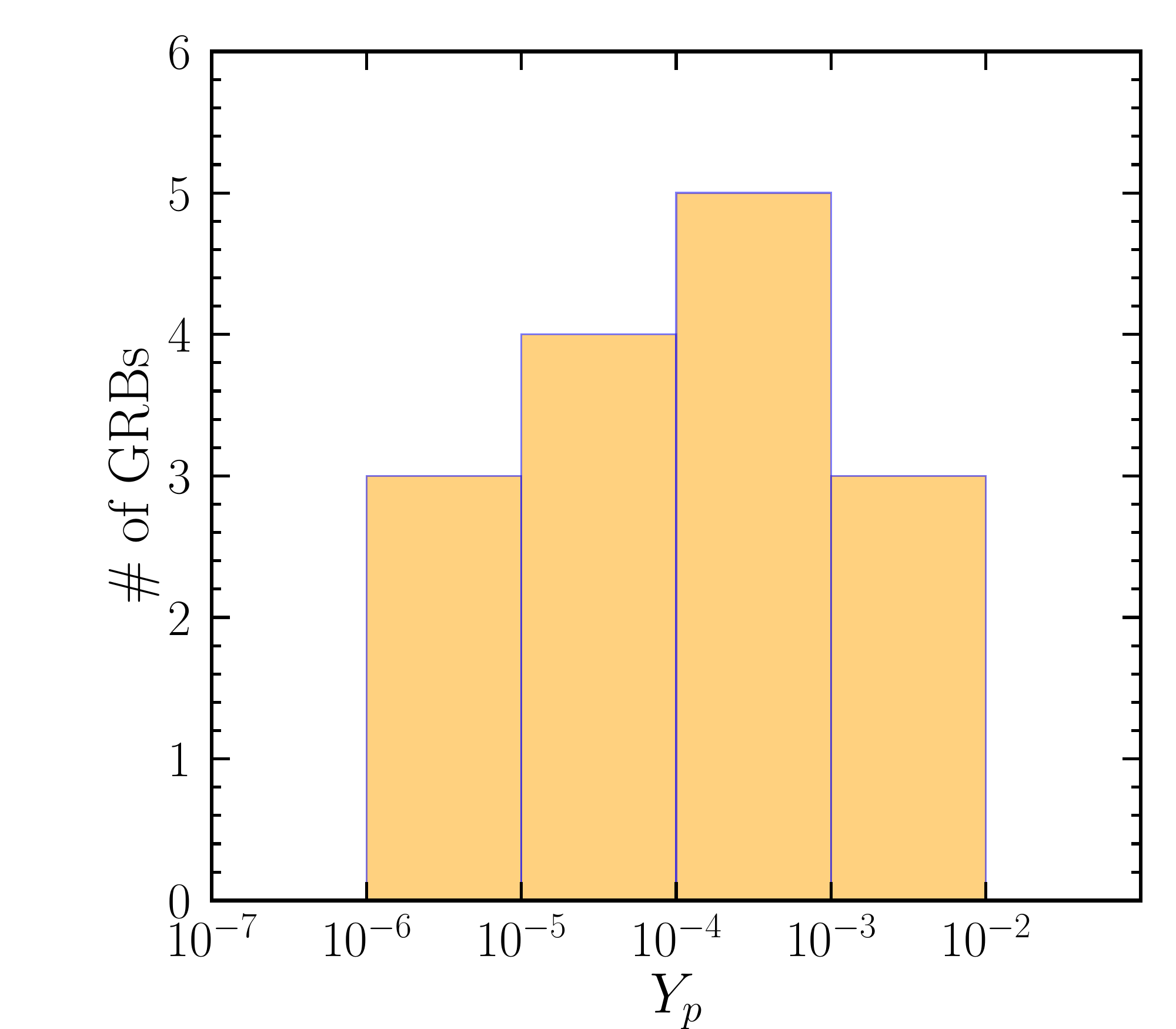}
	\caption{Histogram of $Y_{p}$ (see Eq.~\ref{eq: Y_p}) for a subset of GRBs analyzed in Ref.~\cite{Oganesyan:2019fpa} for which redshift information is available;  $ \Gamma =300 $ and $t_{v}= 0.5$~s are adopted.  The parameter $Y_{p}$ quantifies the relative importance between the proton synchrotron emission and $ p\gamma $ interactions.  The very low values of $ Y_{p} $ for most of GRBs in the sample suggest a negligible neutrino production of this class of GRBs.
	}
	\label{Fig:Y_p_proton_synch}
\end{figure}
In order to figure out the relative importance between proton synchrotron  and $p\gamma$ cooling for the sample of GRBs studied in Ref.~\cite{Oganesyan:2019fpa}, we introduce the following parameter~\cite{Gupta:2007yb}:
\begin{equation}
    \label{eq: Y_p}
    Y_{p} \equiv \frac{L^\prime_{\mathrm{p}, p\gamma}}{L^\prime_{\mathrm{p},\mathrm{syn}}} \approx \frac{\sigma_{p\gamma}}{\sigma_{\text{p,T}}} \frac{U^\prime_{p,\mathrm{syn}}}{U^\prime_{B}} = \frac{\sigma_{p\gamma}}{\sigma_{\text{p,T}}} \frac{E^\prime_{\gamma, \mathrm{tot, iso}}}{V^\prime_{\mathrm{iso}}} \frac{8 \pi}{B^{\prime 2}} = \frac{\sigma_{p\gamma}}{\sigma_{\text{p,T}}} \frac{8 \pi F_{\gamma} d^{2}_{L}(z)}{\Gamma^{2} R^{2}_{\gamma} c B^{\prime 2}}\ ,
\end{equation}
where  $L^\prime_{p, p\gamma}$ and $L^\prime_{p, \mathrm{sync}}$ are the proton energy loss rates for $p\gamma$ interactions and  synchrotron emission respectively, and $\sigma_{\text{p,T}}=\sigma_{\text{T}}(m_{e}/m_{p})^{2}$. By relying on Eqs.~\ref{eq:epsilon_cool_proton} and  \ref{eq:diss_radius},  $Y_{p}$ can be estimated as a function of the bulk Lorentz factor.

For our bencnhmark $ \Gamma = 300 $, we compute $ Y_{p} $  for the GRBs studied in Ref.~\cite{Oganesyan:2019fpa} for which redshift information is available. The histogram of $ Y_{p} $  is shown in Fig.~\ref{Fig:Y_p_proton_synch}. We can see that  $Y_{p}$ spreads over almost three orders of magnitude, with very low typical values.  Hence, assuming proton synchrotron radiation as the main emission mechanism, we expect this class of GRBs to be poor emitters of high energy neutrinos. To show this quantitatively, we compute the neutrino fluence for our representative GRB.

\begin{figure}[]
    \begin{center}
	\includegraphics[width=0.49\textwidth]{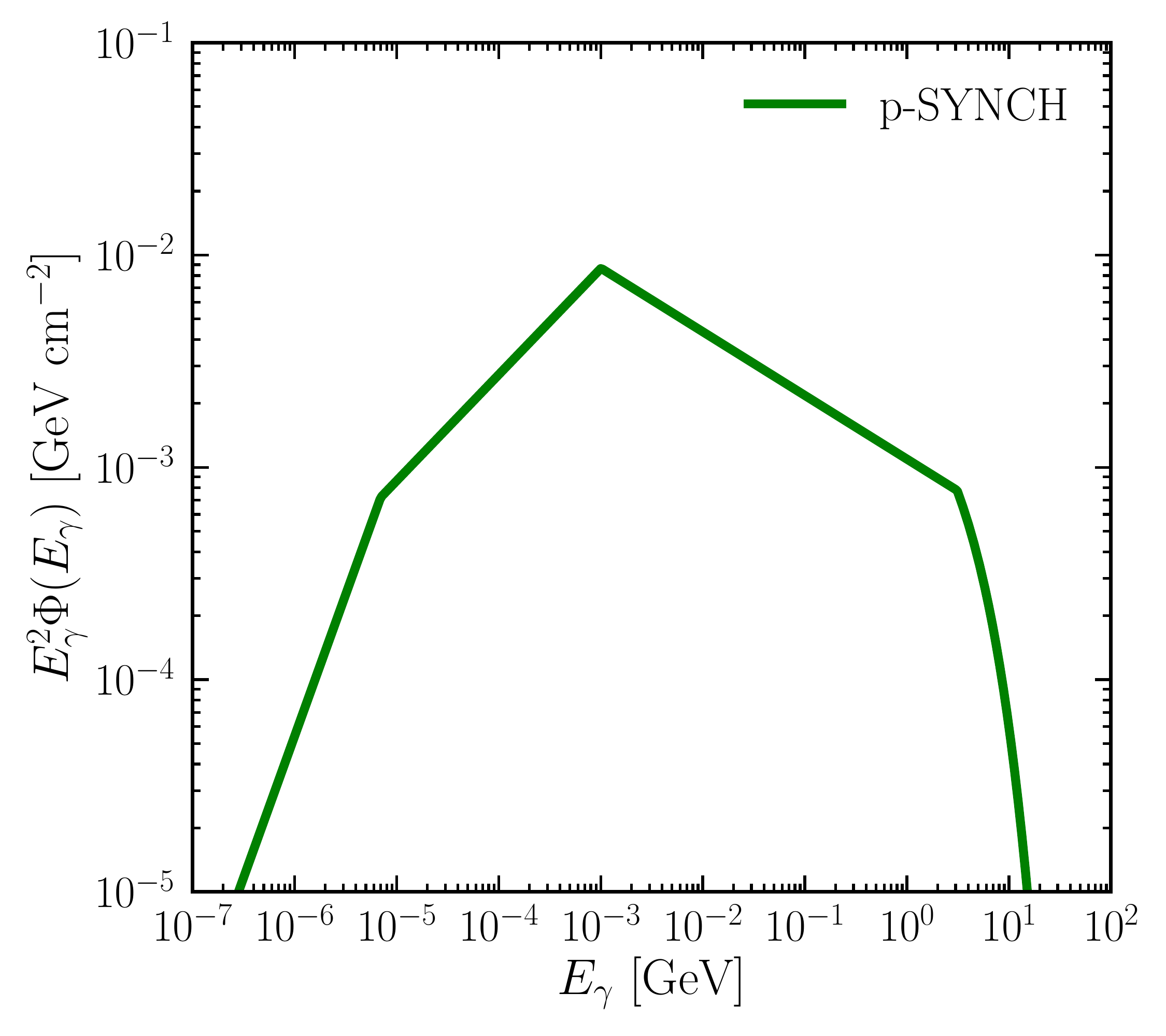}
	\includegraphics[width=0.49\textwidth]{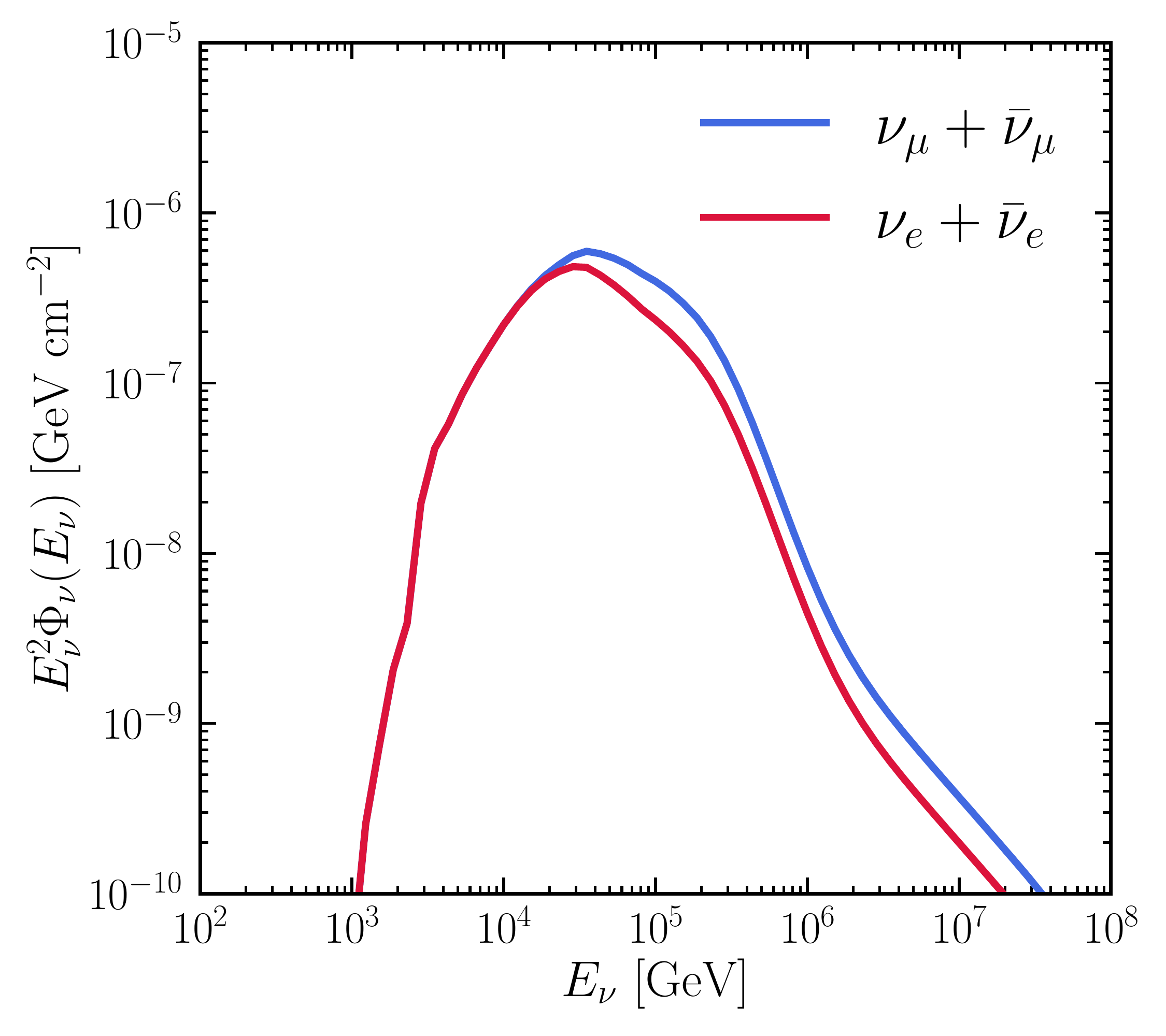}
	\caption{{\it Left}: Photon fluence in the observer frame for the  proton synchrotron  model; see  Table~\ref{Table: all model parameters} for the model parameters, in addition $\tilde{E}_{\mathrm{\gamma,\rm  iso}} = 5 \times 10^{53} $~ergs, $t_{\rm cool} = 0.5$~s, $\gamma_{\rm min}/\gamma_{\rm cool}=12$, $E_{\gamma,\mathrm{cool}} = 7$~keV. {\it Right:} Correspondent $\nu_\alpha+\bar{\nu}_\alpha$ fluence (in red and  blue for the electron and muon flavors, respectively).   The peak in the neutrino distribution ($E^{\rm peak}_{\nu} = 3.5\times 10^{4}$~GeV), due to the cooling energy break $E_{\gamma,\rm cool}$, is shifted to lower energies with respect to the other analized models. The damping at high energies is due to the very strong magnetic field  in the emitting region ($ B^{\prime} \simeq 8.5 \times 10^{6}$~G).}
	\label{Fig:PHOTON&NEUTRINO_P_SYNCH}
	\end{center}
\end{figure}

We adopt the following GRB parameters: 
${\gamma_{\rm min}/\gamma_{\rm cool} =12 }$, $E_{\gamma,\mathrm{cool}} = 7$~keV, $\Gamma=300$, $z=2$ $E_{\gamma,\mathrm{peak}} = (\gamma_{\rm min}/\gamma_{\rm cool} )^{2}\ E_{\gamma,\mathrm{cool}}$, which result in $\tilde{E}_{\gamma,\rm bol, iso}\simeq 7\times 10^{53}$~erg. These values are compatible with the average ones inferred from the sample considered in Ref.~\cite{Oganesyan:2019fpa}, namely $ \langle \tilde{E}_{\gamma,\rm bol, iso}\rangle \simeq  7.4\times 10^{53}$~erg, $ \langle E_{\gamma,\rm cool}\rangle\simeq 6.4 $~keV, and $ \langle\gamma_{\rm min}/\gamma_{\rm cool}\rangle\simeq 11.9 $. Furthermore, we choose $k_{p} = 2.6$ for the slope of the injection proton spectrum $Q^\prime(\gamma^\prime_{p})$, that reproduces the typical value of the high energy photon index $\beta \sim -2.3$; note that $k_p$ is almost never constrained for the sample  in Ref.~\cite{Oganesyan:2019fpa}. 

As for the radiated energy, this fiducial GRB is comparable to the ones analyzed in the previous sections, except for the total energetics. In fact, given the very high magnetic field, $ B^{\prime}\simeq 8.5\times 10^{6} $~G, the total isotropic energy is  $\tilde{E}_{B,\rm iso}\propto R^{2}_{\gamma}\Gamma^{2} B^{\prime 2} \sim \mathcal{O}(10^{60})$~erg, much larger than the typical energy that a GRB jets is able to release (spin down of magnetars or  through the Blandford-Znajek mechanism~\cite{Blandford:1977ds}). Since the synchrotron radiation dominates by many orders of magnitude over all the other proton energy loss mechanisms, we  assume $U^{\prime}_{p,\rm iso}\simeq E^{\prime}_{\gamma,\rm bol, iso}$, where $U^{\prime}_{p}$ is the total isotropic proton energy in the comoving frame. Such a jet turns out to be highly inefficient in radiating energy, given that $\tilde{E}_{\rm  iso}\sim \tilde{E}_{B,\rm iso}\gg \tilde{E}_{\gamma,\rm bol, iso}$.

By considering the proton energy distribution  as in Eq.~\ref{eq:synch_spectrum_fast_cool} and normalizing it to $U^{\prime}_{p}$, we compute the neutrino fluence and show it in the right panel of Fig.~\ref{Fig:PHOTON&NEUTRINO_P_SYNCH}. The left of the same figure shows the  total synchrotron photon fluence.
Analogously to the  model in Sec.~\ref{subsubsec: Magnetic jet model with gradual dissipation}, the peak in the neutrino distribution ($E^{\rm peak}_{\nu} = 3.5\times 10^{4}$~GeV) is  due to the cooling energy break $E_{\gamma,\rm cool}$ and it is  shifted to lower energies. The neutrino spectrum is furthermore strongly damped at high energies due to the synchrotron cooling of mesons in the jet. Our estimation of the neutrino emission results to be in agreement with the one reported in the independent work of Ref.~\cite{Florou2021}, for GRBs with similar parameters. 

The proton synchrotron model, besides requiring unreasonable total jet energies, predicts the smallest neutrino fluence among all models considered in this work. We note that with the choice made of parameters, our representative GRB has $Y_{p}\sim \mathcal{O}(10^{-4})$; hence, our estimation may be considered an optimistic one, given the distribution of $Y_{p}$ shown in Fig.~\ref{Fig:Y_p_proton_synch}.  We refer the interested reader to Ref.~\cite{Florou2021} for additional details and discussion on this model.

\section{Discussion}
\label{sec:discussion}
In this work, we have computed the neutrino fluence for a class of models adopted to describe the prompt phase of long GRBs, all having the same $\tilde{E}_{\mathrm{iso}}$. Because of the diversity of electromagnetic GRB data and the uncertainties inherent to the models (e.g., jet composition, energy dissipation mechanism, particle acceleration, and radiation mechanisms), an exhaustive theoretical explanation of the mechanism powering GRBs is still lacking. 
To  compare the neutrino production across models,  we have selected  fiducial input parameters for a benchmark GRB motivated by observations. In addition, the modeling of the dissipative and acceleration efficiencies, as well as the properties of the accelerated particle distributions have been guided by the most recent simulation findings. A summary of our input parameters  is reported in  Table~\ref{Table: all model parameters}. 
In this section, we compare the energetics of the GRB models explored in this work, discuss the the detection prospects of stacked neutrino fluxes as well as the variation of the quasi-diffuse neutrino flux due to the uncertainties in the jet parameters. 

\subsection{Energetics}

A summary of our findings is reported in Table~\ref{Table: Ratio_iso_gamma_nu}, where the radiative efficiency of the jet (Eq.~\ref{eq: radiation_efficiency}) is listed for the six GRB models investigated in this paper together with the isotropic photon and  neutrino (for six flavors) energies, as well as the ratio of the latter two. As already discussed in Sec.~\ref{sec:emission_models}, the least efficient model in converting $\tilde{E}_{\mathrm{iso}}$  in $\tilde{E}_{\gamma, \mathrm{iso}}$ is the proton synchrotron model, whilst the most efficient one is the model which considers a dissipative photosphere as the main source of  prompt emission. This is mainly due to the high dissipative efficiency suggested by recent  three-dimensional simulations~\cite{Gottlieb:2019aae}. Note that the radiative efficiency is an input parameter of each model, since  we do not compute  the radiation spectra self-consistently.
\begin{table}[b]
    	\caption{\label{Table: Ratio_iso_gamma_nu} Summary  of the derived quantities for the models considered in this work 	and our benchmark parameters value (see  Table~\ref{Table: all model parameters}).   The radiative efficiency of the jet (Eq.~\ref{eq: radiation_efficiency}),  the isotropic photon energy in the  $1\,\mathrm{keV}$--$10\,\mathrm{MeV}$ energy range, the isotropic  neutrino energy for neutrinos and antineutrinos of all flavors, the ratio between the isotropic total neutrino and photon energies, the neutrino energy at the fluence peak, and the maximum proton energy are listed. The model with the smallest radiative efficiency is the proton synchrotron model; this model has also the smallest $\tilde{E}_{\nu, \mathrm{iso}}$. The most radiatively efficient model is the one with a dissipative photosphere.
	    	}
    \begin{center}
    \begin{adjustbox}{width=0.99\textwidth}
	\begin{tabular}[c]{ccccccc}
		\toprule
		\toprule
		\it{Model} & $\eta_{\gamma}\,(\%)$ & $ \tilde{E}_{\gamma,\rm iso}\, [\rm erg]$ & $\tilde{E}_{\nu, \rm iso}\, [\rm erg]$ & $\tilde{E}_{\nu,\rm iso}/{\tilde{E}_{\gamma, \rm iso}} $& $E^{\rm peak}_{\nu_{\mu}}\, [\rm GeV]$ & $E_{p,\rm max}\,[\rm GeV]$ \\
		\toprule
		IS & $0.2$& $ 6.8\times 10^{51} $ & $ 2.3\times 10^{48} $ & $3.4\times 10^{-4}$& $4\times 10^{7}$ & $1.2\times 10^{11}$\\
		PH-IS & $20$ & $ 6.9\times 10^{53} $& $ 7.2\times 10^{49} $ & $1.1\times 10^{-4}$& $3.2\times 10^{7}$& $7.5\times 10^{10}$\\
		3-COMP & $0.3$& $8.7\times 10^{51}$ & $5.2\times 10^{48}$ & $6\times 10^{-4}$& $4.2\times 10^{7}$ & $1.2\times 10^{11}$\\
		ICMART & $17.5$ & $ 6\times 10^{53} $ & $ 1.8\times 10^{51} $ & $3\times 10^{-3}$& $1.3\times 10^{7}$ & $1.7\times 10^{11}$\\
		MAG-DISS & $8$ & $ 2.7\times 10^{53} $ & $ 5.2\times 10^{50} $ & $2\times 10^{-3}$& $7.2\times 10^{5}$ & $2.5\times 10^{10}$\\
		p-SYNCH & $2\times 10^{-5}$ & $ 4.8\times 10^{53} $ & $ 7.2\times 10^{49} $ & $1.4\times 10^{-4}$ & $3.5\times 10^{4}$ & $6.9\times 10^{9}$\\
		\bottomrule
	\end{tabular}
		\end{adjustbox}
		\end{center}
	\end{table}

Among the models considered in this work, all with identical $\tilde{E}_{\mathrm{iso}}$,   neutrinos carry the largest amount of energy in the ICMART model, followed by the  model invoking magnetic dissipation; among the kinetic dominated jet scenarios, the case with a dissipative photosphere is the most efficient one in terms of neutrino production.
It is worth noting that, although in the model with three components $E_{\gamma,\rm iso}$ is just $30\%$ higher than in the IS one, $E_{\nu,\rm iso}$ is a factor $2.3$ larger. The reason for this lies in the fact that protons interact with a high-energy photon component comparable in intensity to the one in the $\gamma$-ray range (i.e., $1$~keV--$10$~MeV) in the three component model, while  the number density of photons above $10$~MeV is negligible in the IS model.
This also explains  the trend for $\tilde{E}_{\nu, \mathrm{iso}}/\tilde{E}_{\gamma, \mathrm{iso}}$  reported in Table~\ref{Table: Ratio_iso_gamma_nu} (note that $\tilde{E}_{\gamma,\rm iso}$ in Table~\ref{Table: Ratio_iso_gamma_nu} is estimated over the energy range $1$~keV--$10$~MeV;  hence,  this ratio, when defined with the bolometric photon energy used for neutrino production, should  be  slightly smaller than the one reported for the IS model with a dissipiative photosphere, the model with three components, the magnetic one with gradual dissipation, and the proton synchrotron model).

One last remark should be done on our results for the IS and ICMART models. In Ref.~\cite{Zhang:2012qy}, the ICMART scenario predicts  the least neutrino flux, given the larger emission radius than $R_{\rm IS}$. This is not the case in our work for two  reasons: first, the chosen representative variability timescale $t_{v}$ provides emission radii comparable in the two scenarios; second, the microphysics parameter that we adopt for the IS case  are less favorable in terms of radiative efficiency and neutrino production efficiency, if compared to the parameters adopted in Ref.~\cite{Zhang:2012qy}, which result to be the same for all their cases of study.

\subsection{Detection perspectives}
In order to compare the neutrino detection perspectives for our six models, we compute the all-sky quasi-diffuse   flux for  neutrinos and antineutrinos. We assume that our benchmark GRB at $z=2$ yields a neutrino emission that is  representative of the entire GRB population. For  ${ \rm \dot{N}\simeq 667}$~yr$ ^{-1} $  long GRBs per year~\cite{Aartsen:2017wea}, the stacking flux for the muon flavor over the whole sky is defined as
\begin{equation}
	F_{\nu_{\mu}}(E_{\nu})=\frac{1}{4\pi}{\rm {\dot{N}}} \Phi_{\nu_{\mu}} (E_{\nu},z=2)\ .
	\label{eq:stacking}
\end{equation}
\begin{figure}[t]
	\centering
	\includegraphics[width=0.8\textwidth]{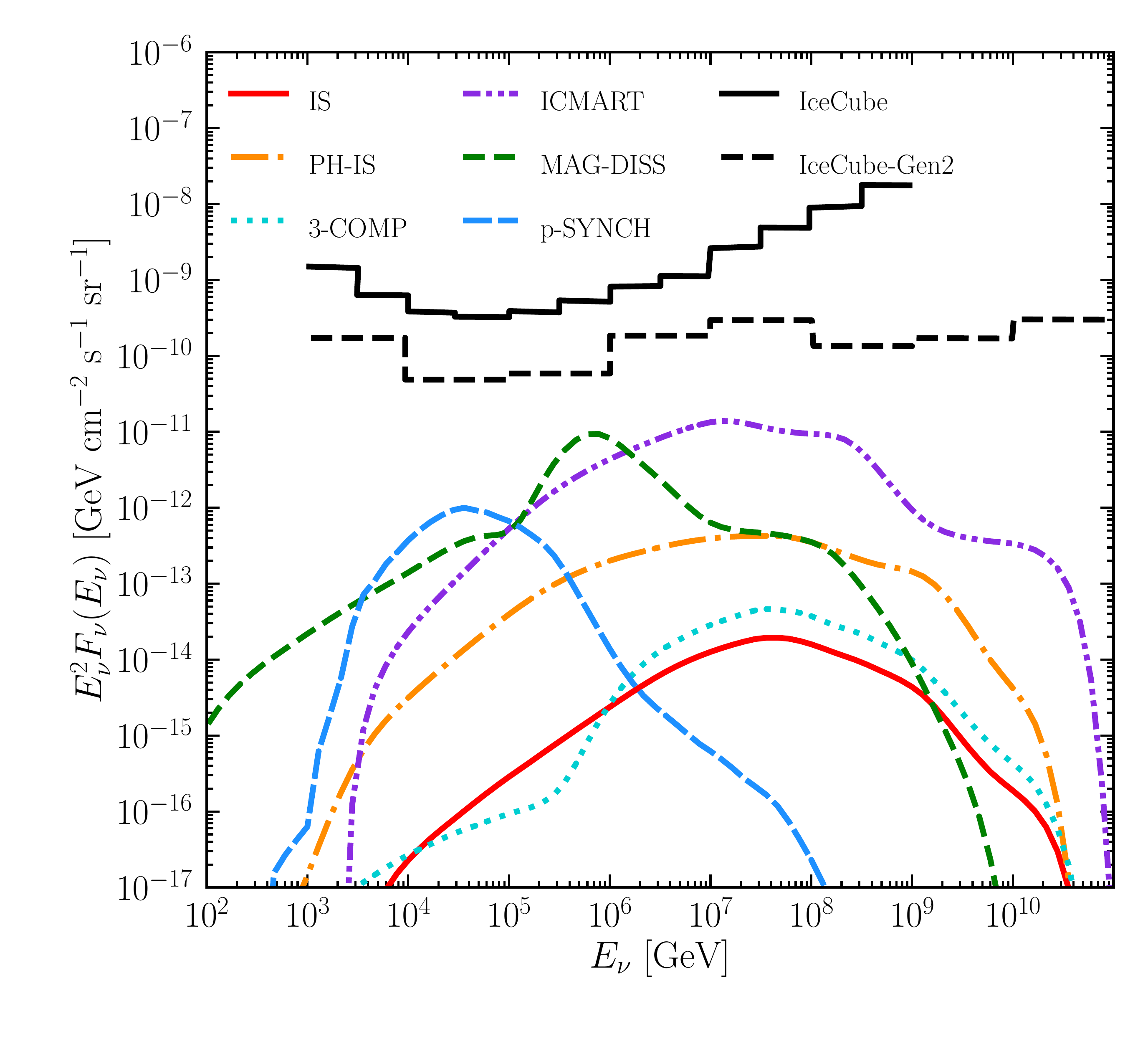}
	\caption{Model comparison of the expected all-sky quasi-diffuse   fluxes for the six GRB models considered in this work for the benchmark jet parameters listed in Table~\ref{Table: all model parameters}. The quasi-diffuse flux has been computed by relying on Eq.~\ref{eq:stacking} for $\nu_\mu+\bar\nu_\mu$; all models have identical $\tilde{E}_{\mathrm{iso}}$. For comparison, the IceCube staking limits (combined analysis for $1172$ GRBs)~\cite{Aartsen:2017wea} and the expected sensitivity for IceCube-Gen2 (based on a sample of $1000$ GRBs)~\cite{Aartsen:2020fgd}  are reported (solid and dashed black lines). By relying on the most up-to-date best-fit  GRB parameters, all models predict a quasi-diffuse flux that lies below the sensitivity curves; however, a  large spread in energy and shape of the expected neutrino fluxes is expected for different jet models.}
	\label{Fig:ALL_MODELS}
\end{figure}

Figure~\ref{Fig:ALL_MODELS} shows the resultant all-sky quasi-diffuse  fluxes for the muon flavor for the six GRB models as  functions of the neutrino energy (colored curves). For comparison, we also show the GRB staking limits of IceCube~\cite{Aartsen:2017wea} and the projected ones for IceCube-Gen2~\cite{Aartsen:2020fgd}
(black curves).  
In agreement with the non-detection of high-energy neutrinos from targeted GRB searches~\cite{Aartsen:2017wea}, our forecast for the neutrino fluxes lies below the experimental limits and is in agreement with the upper limits reported by the ANTARES Collaboration~\cite{Albert:2020lvs} and  with  the  ones expected for KM3NeT~\cite{Adrian-Martinez:2016fdl}.
The ICMART and the magnetic model with gradual dissipation predict comparable neutrino flux at peak energy. The models invoking ISs  (IS, PH-IS, 3-COMP) are the ones with the lowest photon and neutrino yield. This is due to the microphysics parameters adopted in this work;  we refer the reader to   Appendix~\ref{Appendix: Standard shock parmeters} for details on the  differences with respect to standard assumptions commonly used in the literature.

An important aspect to consider in targeted GRB searches is the large spread in energy and shape of the expected neutrino fluxes for different jet models. It is evident from Fig.~\ref{Fig:ALL_MODELS} that the neutrino flux  peak  energy ranges from $\mathcal{O}(10^4)$~GeV for the proton synchrotron model to $\mathcal{O}(10^8-10^9)$~GeV for the IS model with a dissipative photosphere. As such, targeted searches assuming one specific GRB model, such as the  IS one, as benchmark case for the GRB neutrino emission may lead to biased results. 

Another caveat of our modeling  is that the spectral energy distributions of photons and  the ones of the secondary particles  produced through $p\gamma$ interactions are not computed self-consistently; this may affect the overall expected emission, see e.g.~Refs.~\cite{Petropoulou:2014awa,Petropoulou:2014sja,Murase:2011cx,Asano:2013jea,Mastichiadis:2020yjo} for dedicated discussions. In addition, since none of the considered jet models  can account for all observational constraints, a population study~\cite{Acuner:2020zvi} may further affect the expected quasi-diffuse emission.

\subsection{Uncertainties in the jet parameters}

In this work, we have chosen one benchmark GRB as representative of the whole population and have relied on  the best fit values of the input parameters.   However, there are intrinsic uncertainties  of  the characteristic jet parameters, which can lead to variations of  the expected neutrino flux, see e.g.~Refs.~\cite{Tamborra:2015qza,Rudolph:2019ccl}. Hence, we now  investigate the impact  of the variation of two of the most uncertain parameters,  the variability timescale $t_{v}$ and the Lorentz boost factor $\Gamma$, on the quasi-diffuse neutrino flux.

\begin{figure}[h!]
	\begin{center}
		\includegraphics[width=0.68\textwidth]{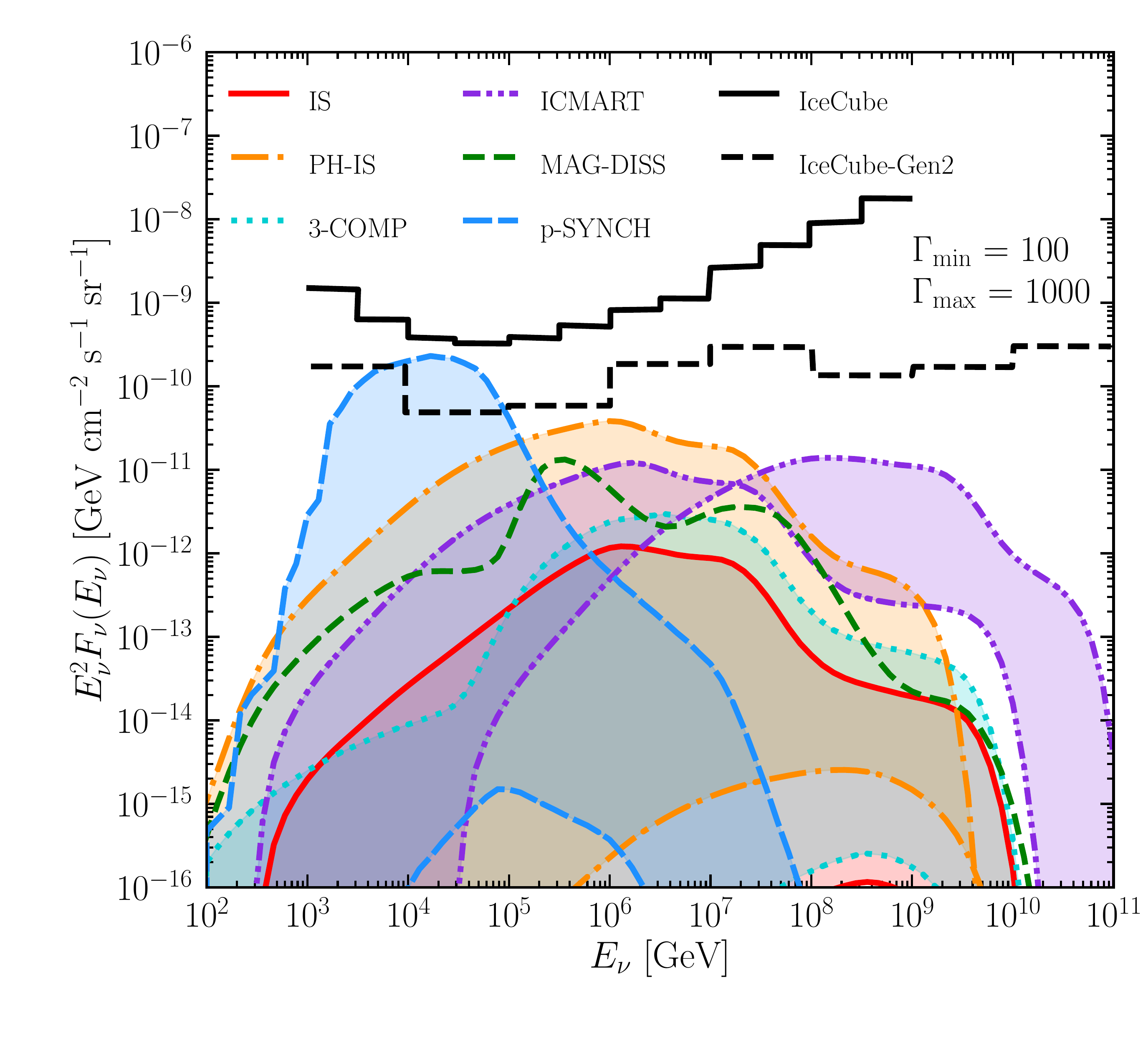}
		\includegraphics[width=0.68\textwidth]{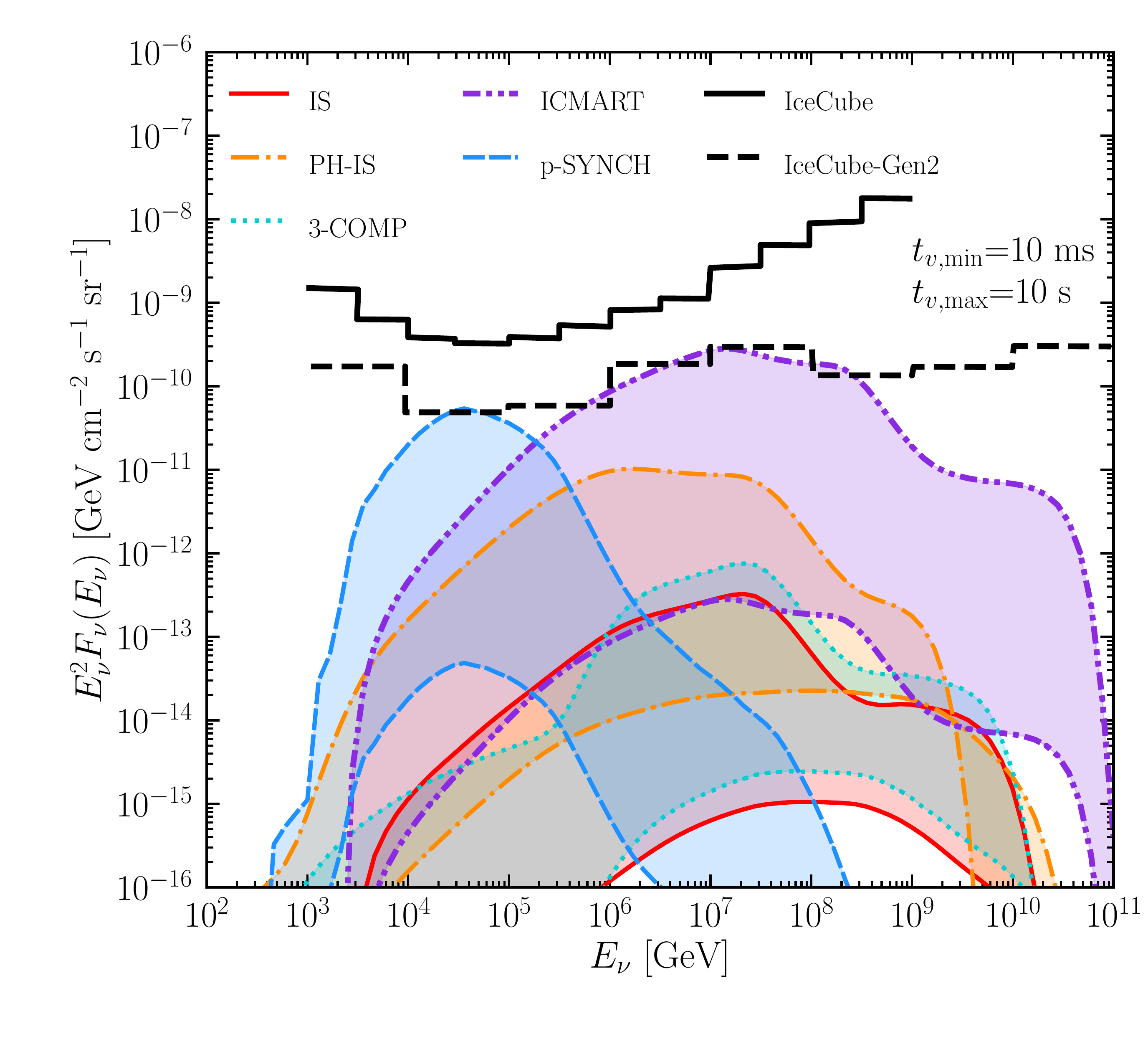}
		\caption{Same as Fig.~\ref{Fig:ALL_MODELS}, but for extreme values of boost factor $\Gamma$ and the variability timescale $t_v$. {\it Top}: The bands for the quasi-diffuse neutrino flux  are displayed for $\Gamma_{\rm min} = 100$ and $\Gamma_{\rm max}=1000$.  {\it Bottom}: The bands for the quasi-diffuse neutrino flux  are displayed for   $t_{v}= 10$~ms and $t_{v}=10$~s.}
		\label{Fig:ALL_models_variation_of  parameters}
	\end{center}
\end{figure}

Consistently with dedicated analysis~\cite{Racusin:2011jf,Ghirlanda:2017opl}, we adopt $\Gamma_{\rm min} = 100$ and $\Gamma_{\rm max} = 1000$ as the minimum and maximum values of the Lorentz factor, respectively. The resulting neutrino fluxes are displayed in the top panel of Fig.~\ref{Fig:ALL_models_variation_of  parameters}.  A variation up to five orders of magnitude for the proton synchrotron case is observable.  We note that  a band for the model with magnetic dissipation case is missing. This is due to the fact that  the outflow saturates below the photosphere for $\Gamma_{\rm sat}=100$ ($R_{\rm sat}> R_{\rm PH}$ for $\Gamma_{\rm sat}\gtrsim 121$;  since  we focus on the production of neutrinos in the optically thin region above the photosphere, we do not provide information about the  $\Gamma_{\rm sat}=100$ case); we instead show the case with $\Gamma=1000$ (see Appendix~\ref{MAG-DISS2} for a discussion on the dependence of the neutrino emission on the input parameters in the magnetized model with gradual dissipation). The neutrino flux for the ICMART case is  shifted to higher energies for larger boost factors; this is determined by  a compensation effect due to the fact that  the distance of the emitting region from the central engine is assumed to be constant and around $\sim 10^{15}$~cm in this model, thus being  completely independent on the bulk Lorentz factor. 

In the bottom panel of Fig.~\ref{Fig:ALL_models_variation_of  parameters},  the variation of the quasi-diffuse neutrino flux is shown as a function of  the  variability timescale. According to the analysis performed on a wide sample of GRB lightcurves~\cite{Albert:2020lvs}, we choose  $t_{v,\rm min}=10$~ms and $t_{v,\rm max}=10$~s as representative extreme values for the variability time. For the simple IS model, the model with three components, the proton-synchrotron model, and the ICMART model, $t_{v}$  is directly connected to the size of the emitting region. For the IS model with dissipative photosphere and the magnetic model with gradual dissipation instead, $t_{v}$ is not related to any observable erratic behavior in the electromagnetic signal;
this explains why no band is considered for the  magnetic model with gradual dissipation, while the band in the case of the IS model with dissipative photosphere comes from simply varying $R_{\rm IS}$.

It is worth noting that while the quasi-diffuse fluxes shown in Fig.~\ref{Fig:ALL_MODELS} sit below the IceCube stacking limits and expected sensitivity of IceCube-Gen2,  once taking into account the variability ranges of $t_{v}$ and $
\Gamma$,  the quasi-diffuse fluxes for the different models can  hit the expected sensitivity of IceCube-Gen2; this hints that it may be possible to  constrain  extreme configurations responsible for the prompt neutrino emission.

\section{Conclusions}
\label{sec:conclusions}
Long duration gamma-ray bursts (GRBs) are subject of investigation since long time, being among the most mysterious transients occurring in our universe. In the attempt of explaining the observed electromagnetic GRB emission, various models have been proposed. The main goal of this work is to show that the neutrino emission strongly depends on the chosen jet model, despite the fact that different jet models may be equally successful in fitting the observed electromagnetic spectral energy distributions. 

To this purpose, we choose a benchmark GRB and compute the neutrino emission for kinetic dominated jets, i.e.~in the  internal shock model, also including a dissipative photosphere as well as three spectral components. We also consider Poynting flux dominated jets: a jet model invoking internal-collision-induced magnetic reconnection and turbulence (ICMART) and a magnetic jet model with gradual dissipation. A jet  model with dominant proton synchrotron radiation in the keV-MeV energy range is also taken into consideration. In particular, the neutrino production for the latter two models has been investigated for the first time in this work. 

Defining the radiative efficiency as the ratio of isotropic gamma-ray energy to the total isotropic energy of the jet, we find that the least radiatively efficient model is  the proton synchrotron one, while the most efficient one is the  model with a dissipative photosphere.  However, the model predicting the largest amount of isotropic-equivalent energy going into neutrinos is the ICMART one. 

In the context of targeted searches, it should be noted that the expected quasi-diffuse neutrino flux can vary up to $3$ orders of magnitude in amplitude and  peak at energies ranging from $10^4$ to $10^8$~GeV. The predicted spectral shape of the neutrino distribution is also strongly dependent on the adopted jet model. A summary of our findings is reported in Table~\ref{Table: Ratio_iso_gamma_nu} and Fig.~\ref{Fig:ALL_MODELS}. 

This work highlights the great potential of neutrinos in pinpointing the GRB emission mechanism in the case of successful neutrino detection. In particular, it suggests the need to rely on a wide range of jet models in targeted stacking searches.

\vspace{1cm}
\noindent{\it Note added}: The modeling of the neutrino emission for the proton synchrotron model  is also presented in the independent
work of Ref.~\cite{Florou2021}. Our paper focuses on the comparison of the neutrino production across different GRB models for the prompt emission, while Ref.~\cite{Florou2021} investigates the plausibility of the proton synchrotron interpretation.

\acknowledgments
We are grateful to Daniele Caprioli, Jochen Greiner, Gor Oganesyan  for insightful discussions and Damien B{\'e}gu{\'e}, Kohta Murase and Walter Winter for useful comments on the manuscript. This project has received funding from the  Villum Foundation (Project No.~13164), the Carlsberg Foundation (CF18-0183), the Knud H\o jgaard Foundation, the Deutsche Forschungsgemeinschaft through Sonderforschungbereich
SFB~1258 ``Neutrinos and Dark Matter in Astro- and Particle Physics'' (NDM), and the MERAC Foundation.

\appendix
\section{Spectral energy distributions of photons: fitting functions}
\label{sec:fitting_models}
In order to describe the electromagnetic emission, in this appendix we introduce  the  main spectral functions  used to fit the electromagnetic data: the Band function, the cut-off power-law, a simple power-law, and a double broken power-law usually representing the synchrotron emission from a marginally fast cooling particle population. 
The various spectral functions introduced here are then employed to model the GRB emission in Sec.~\ref{sec:emission_models}. 

When the energy distribution does not present an intrinsic cut-off at high energies (e.g., Band function and synchrotron spectrum), we define $E^{\prime}_{\gamma,\rm cutoff}$ as the energy at which the opacity to photon-photon pair production becomes unity
\begin{equation}
    \label{eq: E_cutoff}
    \tau_{\gamma\gamma}(E^{\prime}_{\gamma,\rm cutoff})\simeq 0.1 \sigma_{T} E^{\prime}_{\ast} n^{\prime}_{\gamma} (E^{\prime}_{\ast}) \frac{R_{\gamma}}{2\Gamma} = 1 
\end{equation}
where $ E^{\prime}_{\ast}= m^{2}_{e} c^{4}/E^{\prime}_{\gamma,\rm cutoff}$ and $\tau_{\gamma\gamma}(E^{\prime}_{\gamma,\rm cutoff})$ is the opacity for the photons with energy $E^{\prime}_{\gamma,\rm cutoff}$ and number density distribution $n^{\prime}_{\gamma}(E^{\prime}_{\gamma})$.

\subsection{Band function}
\label{subsec:Band_function}

The Band function~\cite{Band:1993eg} is the most used empirical function  to fit the time-integrated electromagnetic spectra. Despite fitting well the data,  it is still lacking a clear physical meaning. It consists of a smoothly joint broken power-law:
\begin{equation}
	n^{\rm{Band}}_{\gamma}(E_{\gamma}) = C
	\begin{cases}
		\left(\frac{E_{\gamma}}{100\ \mathrm{keV}}\right)^{\alpha_{\gamma}} \exp\left[- \frac{(\alpha_{\gamma} + 2) E_{\gamma}}{E_{\gamma,\rm{peak}}}\right] \quad &E_{\gamma} < E_{\gamma,c}\\
		\left(\frac{E_{\gamma}}{100\ \mathrm{keV}}\right)^{\beta_{\gamma}} \exp(\beta_{\gamma} - \alpha_{\gamma}) \left(\frac{E_{\gamma,c}}{100\ \mathrm{KeV}}\right)^{\alpha_{\gamma} - \beta_{\gamma}} \quad &E_{\gamma} \geq E_{\gamma,c}    
	\end{cases}
	\label{eq:BAND function}
\end{equation}
where 
\begin{equation}
	E_{\gamma,c} = \left(\frac{\alpha_{\gamma} - \beta_{\gamma}}{\alpha_{\gamma} + 2}\right) E_{\gamma,\rm peak}\ ,
\end{equation}
$C$ is a normalization constant (in units of $\mathrm{GeV}^{-1} \mathrm{cm}^{-3}$), $\alpha_{\gamma}$ and $\beta_{\gamma}$ are the low-energy and high-energy power-law photon indices, $E_{\gamma,c}$ represents the energy  where the low-energy
power-law with an exponential cutoff ends and the pure high energy
power-law begins.  The peak energy $E_{\gamma,\rm peak}$ is chosen to satisfy the Amati relation~\cite{Amati:2006ky}: 
\begin{equation}
	\label{eq: Amati_peak}
	\tilde{E}_{\gamma,\rm peak} = 80 \left(\frac{\tilde{E}_{\gamma,\mathrm{iso}}}{10^{52}\ \mathrm{erg}}\right)^{0.57}\ \mathrm{keV}\ .
\end{equation}
The typical spectral parameters inferred from  observations are: $\alpha_{\gamma} \simeq -1.1$, $\beta_{\gamma} \simeq -2.2$, and $E_{\gamma,\rm peak} \simeq 300$~keV~\cite{Gruber:2014iza}.

\subsection{Cut-off power-law}
\label{subsec:cutoff_PL}

Although  the Band spectrum is the best fitting function for most GRBs, it has been shown that in some cases a cut-off power-law (CPL) can represent the preferred model~\cite{Yu:2016epf, Yu:2018vfp,Asano:2015oia}. 
The CPL   is a power-law model with a high energy exponential cut-off :
\begin{equation}
	n^{\rm{CPL}}_{\gamma}(E_{\gamma}) = C \left(\frac{E_{\gamma}}{100\ \mathrm{keV}}\right)^{\alpha_{\gamma}} \exp\left[-\frac{(\alpha_{\gamma}+2) E_{\gamma}}{E_{\gamma,\rm peak}}\right]\ ,
	\label{cutoff_PL}
\end{equation}
where  $\alpha_{\gamma}$ is the photon index and $E_{\gamma, \rm peak}$ the peak energy, whose value will be specified later.
In an optically thick thermal scenario, $\alpha_{\gamma} = 1$ in the Rayleagh-Jeans limit, $\alpha_{\gamma} = 2$ in the Wien limit, $\alpha_{\gamma} = 0.4$ for a non-dissipative photosphere in the coasting phase and $\alpha_{\gamma} < 0$ for all non-thermal emissions~\cite{Acuner:2019rif}.

\subsection{Power law}
\label{subsec:PL}
In the cases of faint bursts or narrow detector bandpass, the whole GRB spectrum, or one of its components, can be fitted with a simple power-law~\cite{Sakamoto:2007yi} defined as 
\begin{equation}
	\label{simple_PL}
	n^{\mathrm{PL}}_{\gamma}(E_{\gamma}) = C \left(\frac{E_{\gamma}}{100\ \mathrm{keV}}\right)^{\alpha_{\gamma}}\ ,
\end{equation}
where $C$ is the normalization and $\alpha_{\gamma}$ is the power-law photon index. 

\subsection{Double broken power law}
\label{subsec:synch_spectrum}
This is a spectral model that is commonly adopted to describe the synchrotron emission of a fast cooling population of particles that are being injected into the emitting region with a power-law distribution at a rate
$Q(\gamma) \propto \gamma^{-k}$ with $\gamma_{\mathrm{min}} < \gamma < \gamma_{\mathrm{max}}$. 
During an emission period $t$, the charged particles of mass $m$ loose most of their energy above the characteristic value $\gamma_{\rm cool}$:
\begin{equation}
	\label{cooling_gamma}
	\gamma_{\mathrm{cool}}(t) = \frac{6 \pi mc}{\sigma_{\mathrm{T}} \beta^{2} B^{2} t}\left(\frac{m}{m_{e}}\right)^{2}\ ,
\end{equation}
where $ m_{e} $ is the electron mass. Considering a constant injection rate of particles in the emitting region which radiate in the fast cooling regime ($\gamma_{\mathrm{min}} > \gamma_{\mathrm{cool}}$) at a rate 
$\propto \gamma^{2}$, after a time $t$ the emitting particle distribution has the following shape~\cite{Zhang:2018ond}:
\begin{equation}
	n(\gamma, t) \propto \begin{cases}
		0 & \gamma < \gamma_{\mathrm{cool}}\ \  \rm{and}\ \  \gamma > \gamma_{\mathrm{\mathrm{max}}}\\
		\gamma^{-2} & \gamma_{\mathrm{cool}} < \gamma < \gamma_{\mathrm{min}}\\
		\gamma^{-(k+1)} & \gamma_{\mathrm{min}} < \gamma < \gamma_{\mathrm{max}}\ .
	\end{cases}
	\label{N_part_synch}
\end{equation}

Given that each particle  radiates photons with a characteristic synchrotron energy
\begin{equation}
\label{eq: characteristic_sycnh_energy}
E_{\gamma}(\gamma) = \frac{3}{2} \frac{\hbar e}{m c} \gamma^{2} B, 
\end{equation}
the particle distribution  in Eq.~\ref{N_part_synch} emits the following synchrotron spectrum:
\begin{equation}
	\label{eq:synch_spectrum_fast_cool}
	n^{\mathrm{sync}}_{\gamma}(E_{\gamma}) = C
	\begin{cases}
		\left(\frac{E_{\gamma}}{E_{\gamma,\mathrm{cool}}}\right)^{-\frac{2}{3}} \quad &E_{\gamma} < E_{\gamma,\mathrm{cool}}\\
		\left(\frac{E_{\gamma}}{E_{\gamma,\mathrm{cool}}}\right)^{-\frac{3}{2}} \quad &E_{\gamma,\mathrm{cool}} < E_{\gamma} < E_{\gamma,\mathrm{min}}\\  
		\left(\frac{E_{\gamma,\mathrm{min}}}{E_{\gamma,\mathrm{cool}}}\right)^{-\frac{3}{2}} \left(\frac{E_{\gamma}}{E_{\gamma,\mathrm{min}}}\right)^{- \frac{k + 2}{2}} &E_{\gamma,\mathrm{min}} <E_{\gamma} < E_{\gamma,\mathrm{max}}\ ,\\  
	\end{cases}
\end{equation}
where $E_{\gamma,\mathrm{cool}}$, $E_{\gamma,\mathrm{min}}$, and $E_{\gamma,\mathrm{max}}$ correspond to the characteristic photon energies mainly emitted by particles with gamma factors $\gamma_{\mathrm{cool}}$, $\gamma_{\mathrm{min}}$, and $\gamma_{\mathrm{max}}$, respectively.

\section{Magnetized jet model with gradual dissipation: dependence of the neutrino emission on the input parameters}
\label{MAG-DISS2}
One of the main, but less certain, parameters of the jet model with gradual magnetic dissipation is the initial magnetization $\sigma_0$. This, in turn, determines the photospheric radius, the saturation Lorentz factor, the energy dissipation rate, and other parameters. For this reason, we investigate the impact of $\sigma_0$ on the photon and neutrino fluences by considering a case with $\sigma_0=100$.
All the other parameters, like $\tilde{E}_{\rm iso}$, are identical to the ones adopted in Sec.~\ref{subsubsec: Magnetic jet model with gradual dissipation}. We follow the same procedure to calculate the neutrino flux as outlined in Sec.~\ref{subsubsec: Magnetic jet model with gradual dissipation}. 

\begin{figure}[]
	\centering
	\includegraphics[width=0.49\textwidth]{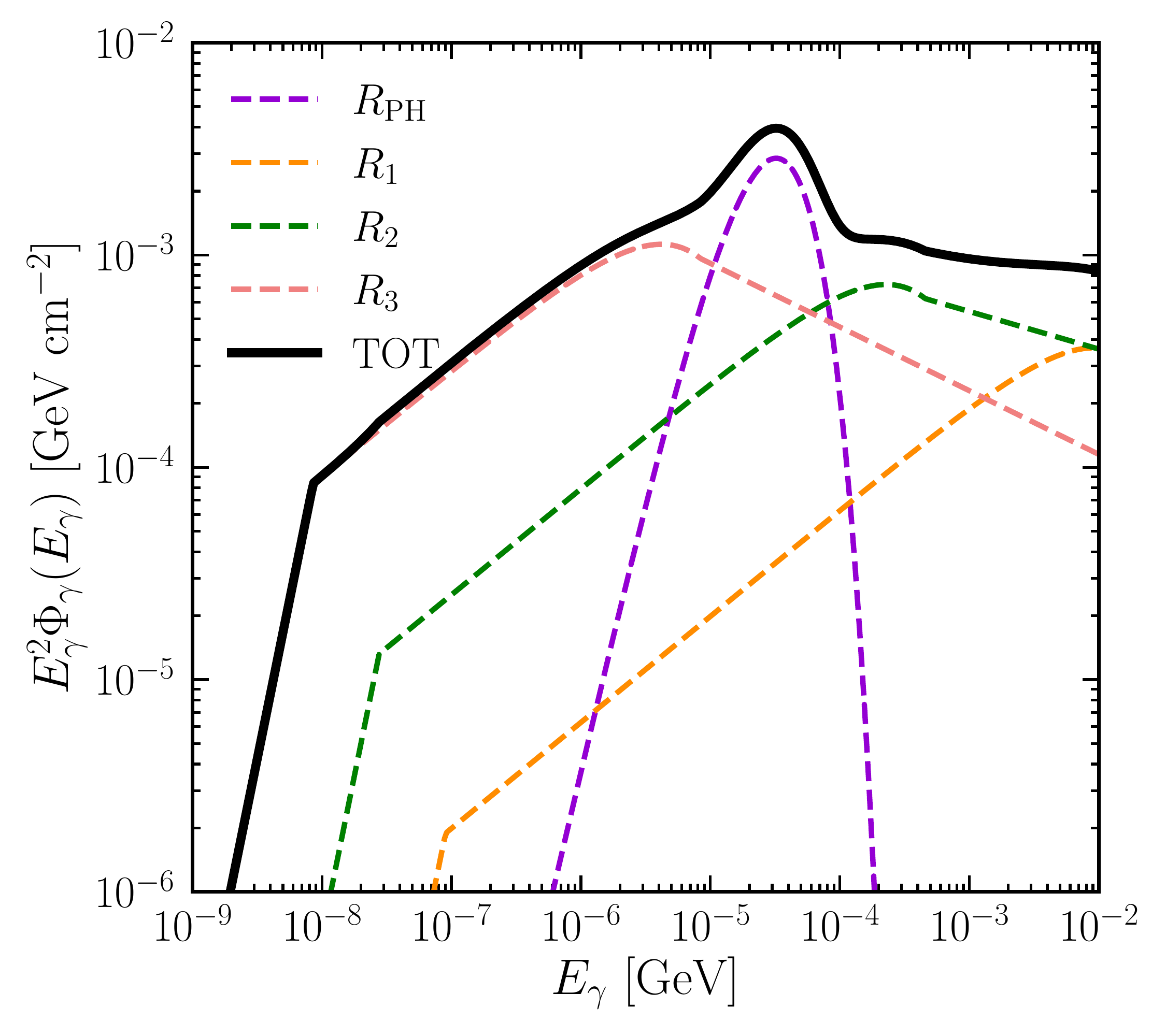}
	\includegraphics[width=0.49\textwidth]{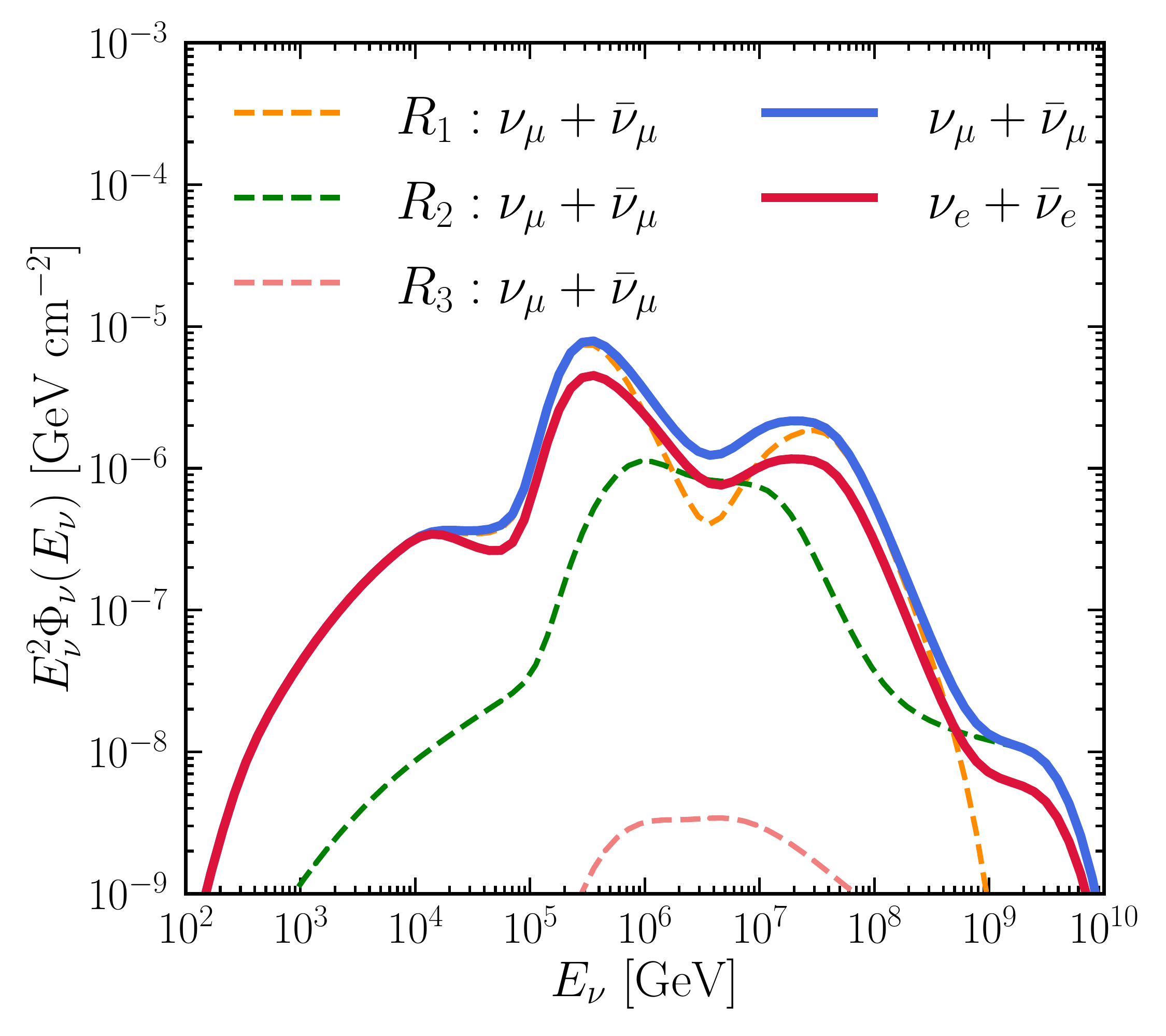}
	\caption{Similar to
	Fig.~\ref{Fig:PHOTON_MAGNETIC}, but for $\sigma_{0} = 100$. 
	The  fluence for the muon flavor peaks at $E^{\rm peak}_{\nu}= 3.3\times 10^{5}$~GeV. In addition, $E_{\nu,\rm iso}=8.6\times 10^{50}$~erg, $E_{\gamma,\rm iso}=2.3\times 10^{53}$~erg, $\Gamma_{\rm sat}=1000$, $\sigma_{0}=100$, $R_{\rm PH}=7.1\times 10^{11}$~cm, $R_{1}=2.1\times 10^{12}$~cm, $R_{2}=3\times 10^{13}$~cm, $R_{3}=4\times 10^{14}$~cm, $\Gamma_{1}=180, \Gamma_{2}=422, \Gamma_{3}=1000$, $\tilde{E}_{\nu,\rm iso}/\tilde{E}_{\gamma,\rm iso}=3.7\times 10^{-3}$, $\eta_{\gamma}=7\%$.
	}
	\label{Fig:PHOTON_MAGNETIC_sigma_100}
\end{figure}

\begin{figure}[]
	\centering
	\includegraphics[width=0.49\textwidth]{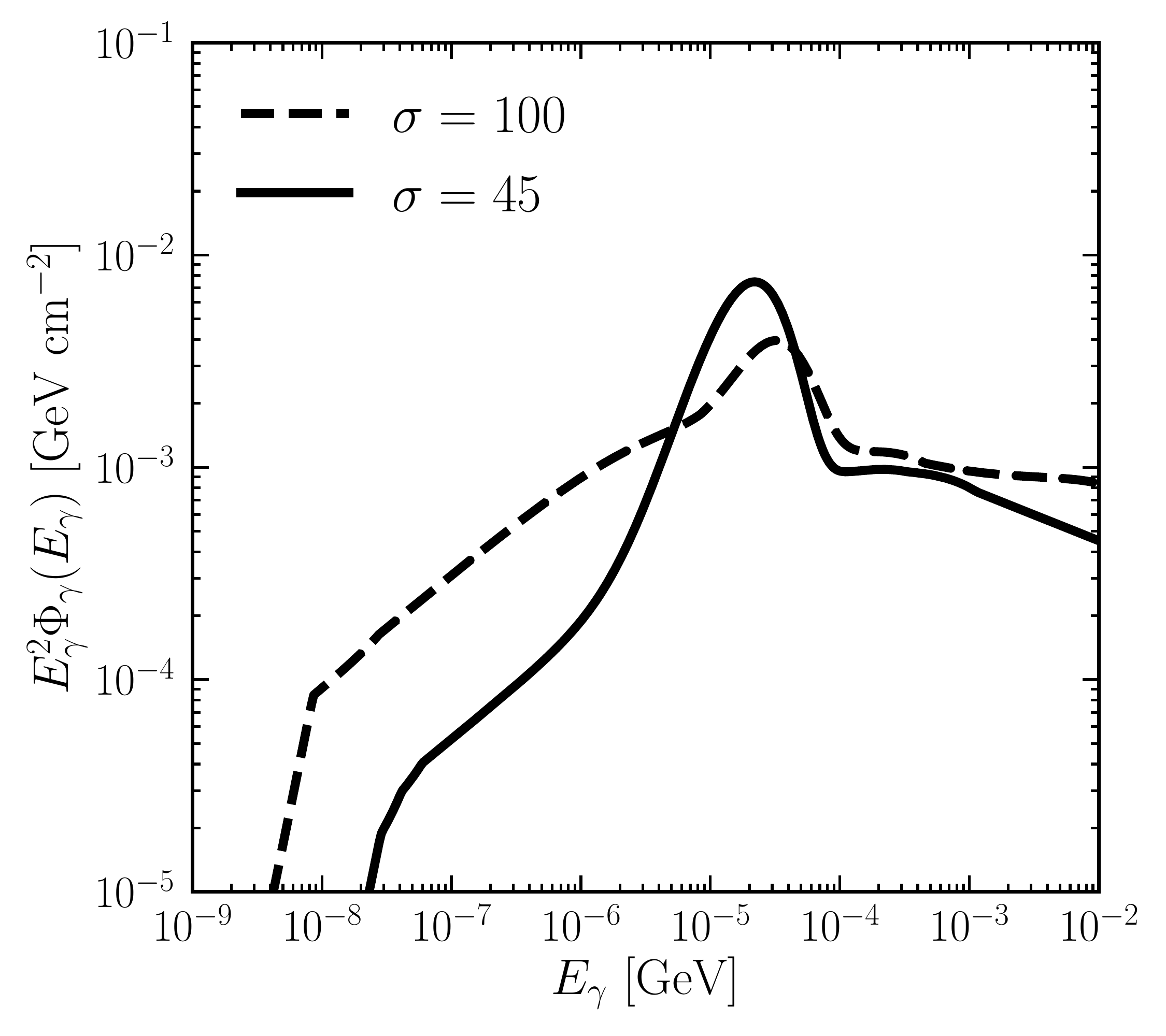}
	\includegraphics[width=0.49\textwidth]{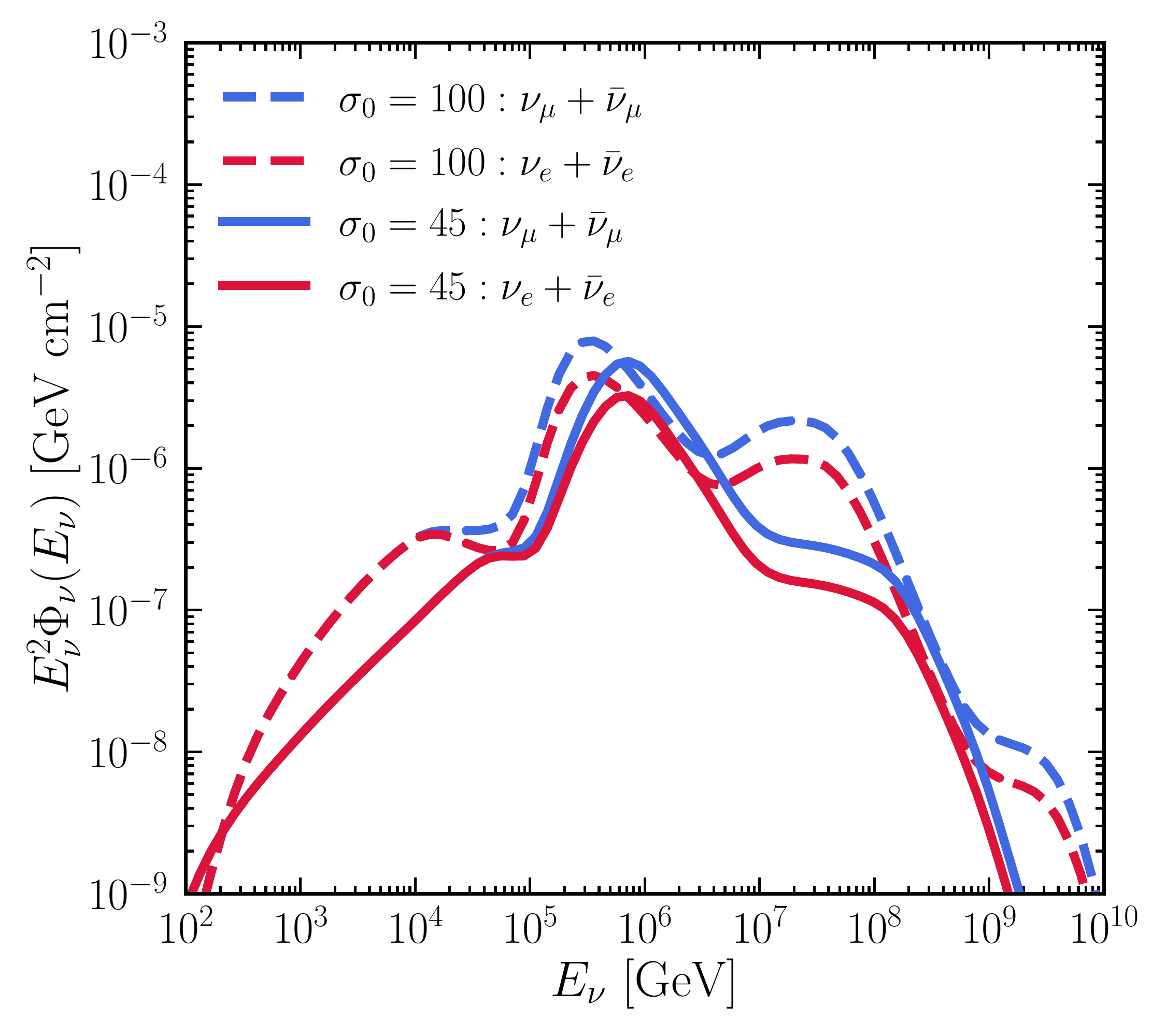}
	\caption{\textit{Left}: Total photon fluence in the observer reference frame, obtained as the sum of the components produced at $R_{\rm{PH}},\,R_{1},\,R_{2}$ and $R_{3}$ for the $\sigma_{0}=45$ (solid line) and $\sigma_{0}=100$ (dashed line) cases, respectively. \textit{Right}: Correspondent $\nu_{\alpha}+\bar{\nu}_{\alpha}$ fluence (in red and in blue for the electron and muon flavors, respectively). For parameters used, see captions of Figs.~\ref{Fig:PHOTON_MAGNETIC} and~\ref{Fig:PHOTON_MAGNETIC_sigma_100}.
	}
	\label{Fig:PHOTON_MAGNETIC_sigma_45_100}
\end{figure}

In Fig.~\ref{Fig:PHOTON_MAGNETIC_sigma_100}, we show snapshots of the photon fluence (left panel) and neutrino fluence (right panel) for $\sigma_0=100$ at three indicative radii. A comparison between the  $\sigma_{0}=45$ and $\sigma_{0}=100$ cases is shown in  Fig.~\ref{Fig:PHOTON_MAGNETIC_sigma_45_100}.
A noticeable difference is appreciable between the photon spectral energy distributions. As the initial magnetization increases, the saturation Lorentz factor increases, namely  $\Gamma_{\rm sat} = \sigma^{3/2}_{0} = 1000$ for $\sigma_0=100$.  The energy is dissipated at a rate $\dot{E}\propto R^{1/3}$, while 
the photosphere occurs at a smaller distance from the source ($R_{\rm PH} \propto 1/\Gamma_{\rm sat}$).  As a result, less energy is dissipated during the optically thick regime (most of the energy is dissipated at $R>R_{\rm PH}$) and the photospheric emission becomes dimmer (compare dashed and solid lines at $E_{\gamma}\sim 10^{-5}$--$10^{-4}$~GeV in Fig.~\ref{Fig:PHOTON_MAGNETIC_sigma_45_100}). The characteristic synchrotron energy $E_{\gamma,\rm min} \propto \Gamma \gamma^{\prime 2}_{\rm min} B^{\prime}$ (Eq.~\ref{eq: characteristic_sycnh_energy}) decreases with the radius (see, e.g., dashed colored curves in the left panel of Fig.~\ref{Fig:PHOTON_MAGNETIC_sigma_100}), while the normalization of the synchrotron photon spectra increases with respect to the case of $\sigma_0=45$ (see dashed curve in the left panel of Fig.~\ref{Fig:PHOTON_MAGNETIC_sigma_45_100}) because of the higher dissipation rate (Eq.~\ref{eq:E_B_diss_rate}).
For a higher $\sigma_0$, particle acceleration begins at smaller radii and so does the production of neutrinos. Moreover, the power slopes of the electron and proton distributions (accelerated via reconnection) are harder~\cite{Sironi:2014jfa, Guo2014} because of the higher magnetization in the acceleration region. Because of the larger saturation radius ($R_{\rm sat}\propto \Gamma_{\rm sat}^2$) found for higher $\sigma_0$, the dissipated energy up to $R_1=aR_{\rm ph}\propto \Gamma_{\rm sat}^{-1}$ that is available for relativistic particles is less than in the case of lower initial magnetizations.
\begin{figure}[]
	\centering
	\includegraphics[width=0.49\textwidth]{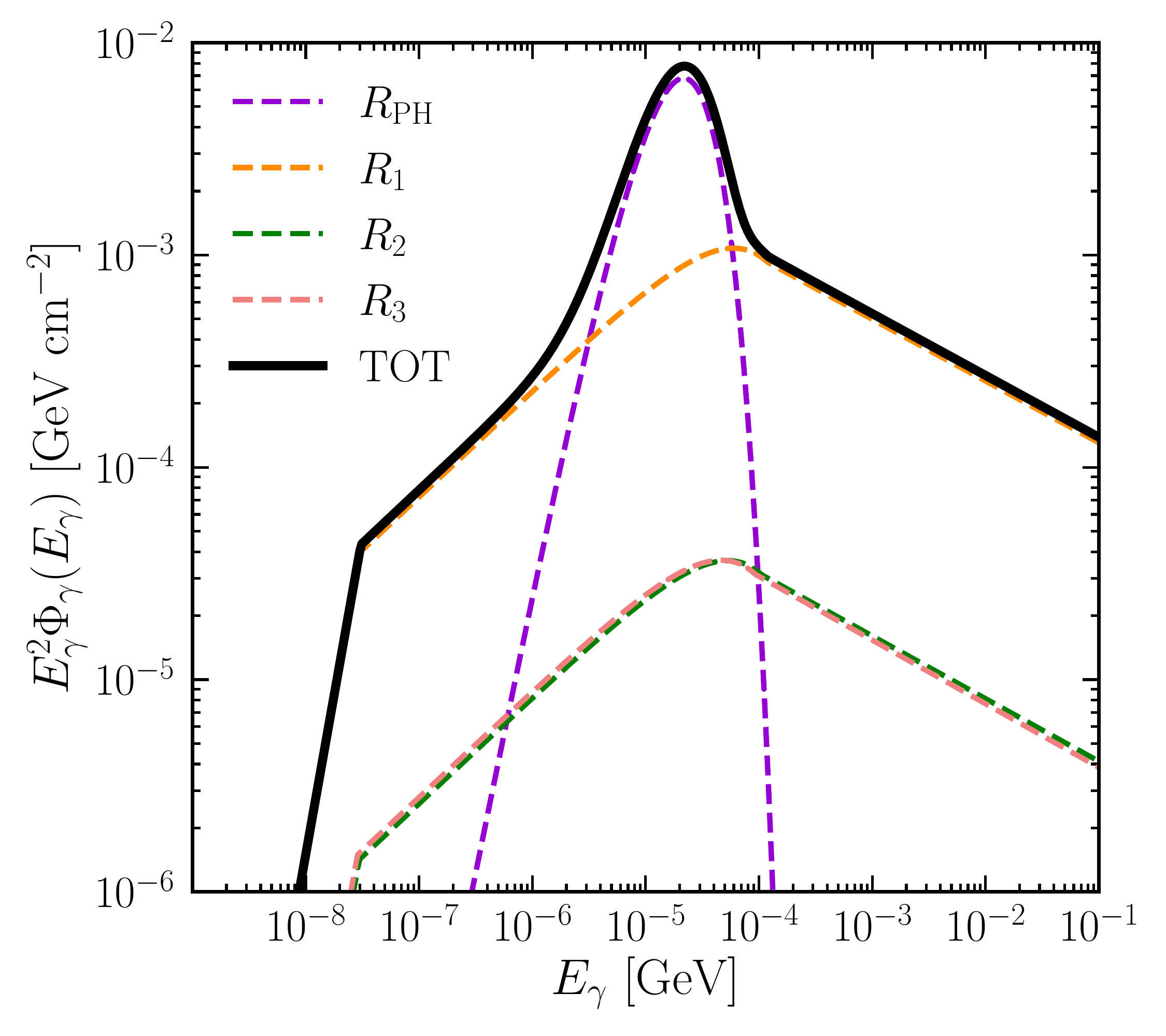}
	\includegraphics[width=0.49\textwidth]{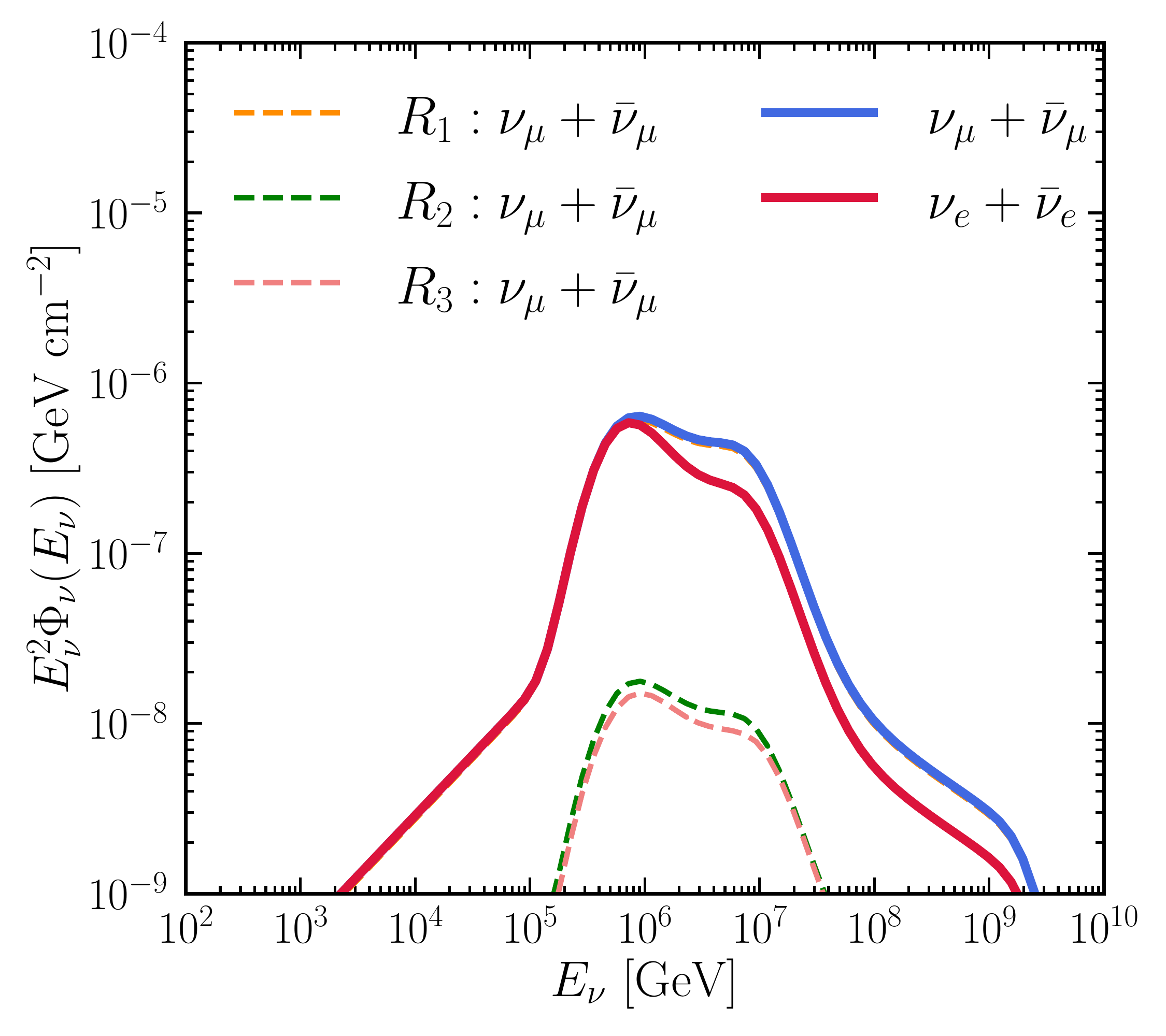}
	\caption{\textit{Left}: Total photon fluence in the observer reference frame, obtained as the sum of the components produced at $R_{\rm{PH}},\,R_{1},\,R_{2}$ and $R_{3}$ for the $\sigma_{0}=45$ and $a=13$ case. \textit{Right}: Correspondent $\nu_{\alpha}+\bar{\nu}_{\alpha}$ fluence (in red and in blue for the electron and muon flavors, respectively).
	}
	\label{Fig:PHOTON_MAGNETIC_a=13}
\end{figure}
The combination of 
a smaller amount of dissipated energy up to a given radius, smaller volume and harder proton power slope leads to a  neutrino flux at  peak (whose main contribution comes from $R_{1}$)  comparable to the one with $\sigma_{0} = 45$ (see right panel of Fig.~\ref{Fig:PHOTON_MAGNETIC_sigma_45_100}). 

For the case of $\sigma_{0} = 100$, the second bump in the neutrino spectrum has a fluence that is comparable to the one of the first bump at $\sim 10^6$~GeV. On the contrary, the second bump is barely visible in the neutrino energy distribution with $\sigma_0=45$ (compare solid and dashed lines in the right panel of Fig.~\ref{Fig:PHOTON_MAGNETIC_sigma_45_100}). This is because in the  $\sigma_{0}=100$ case, pions suffer  stronger synchrotron losses, hence the neutrino intensity  resulting from the decay of pions  decreases to the level of the one produced by kaons. This is also the reason for a slight shift in the neutrino flux peak to lower energies. Another noticeable feature is the low energy tail. The latter turns out to be higher in the $\sigma_{0}=100$ case, given the higher number density of photons at higher energies.

Finally, in order to explore the effects of the arbitrary choice of the parameter $a$, we considered the case with $\sigma_{0}=45$ and $a=13$, where $R_{1}\sim R_{\rm sat}$, see Fig.~\ref{Fig:PHOTON_MAGNETIC_a=13}. Since most of the energy is dissipated within $R_{1}$, the neutrino contribution from $R_{1}$ is dominant, although lower by a factor $\mathcal{O}(10)$ if compared to the case with $a=3$. However, since the case $a=13$ represents an extreme case, such that all the energy is locally dissipated near the saturation radius, the fluence  of $\mathcal{O}(10^{-6})\,\rm{GeV cm}^{-2}$ should be considered as  the lower limit for the neutrino production from a GRB described by the magnetic model with gradual dissipation for the specific set of parameters adopted in this work. 

\section{Quasi-diffuse neutrino flux for standard internal shock parameters}
\label{Appendix: Standard shock parmeters}

\begin{figure}[h]
	\centering
	\includegraphics[width=0.7\textwidth]{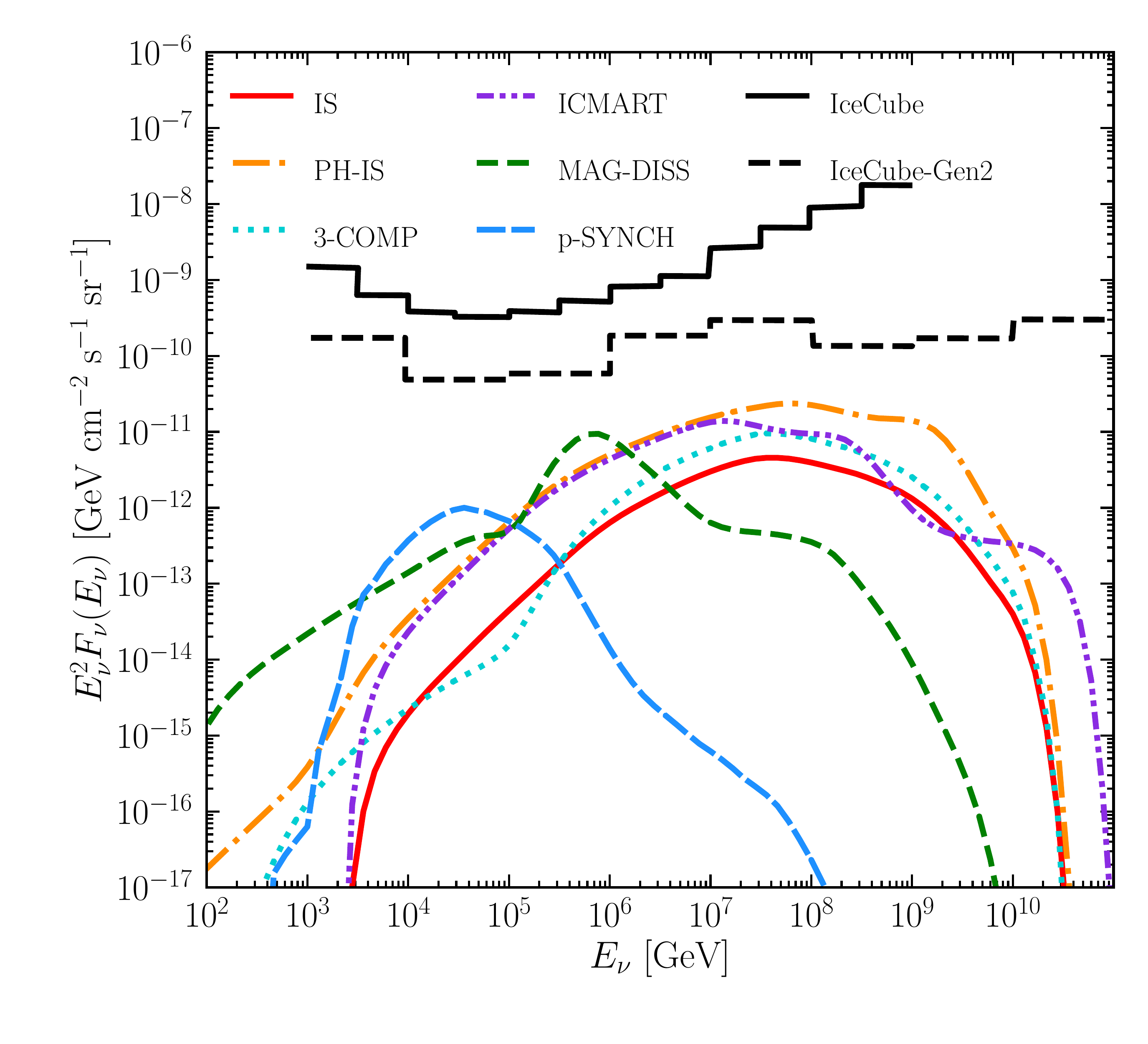}
	\caption{Quasi-diffuse neutrino flux for the six models considered in this work and computed as in Fig.~\ref{Fig:ALL_MODELS}, but with the classically adopted microphysics parameters for the Internal shocks: ${\varepsilon_{p}=10/12, \varepsilon_{e}=\varepsilon_{B}=1/12}$, and $k_{p}=2$.}
	\label{Fig:ALL_MODELS_0}
\end{figure}

In this work,  we have adopted input parameters inspired by the results of recent PIC simulations of mildly relativistic shocks for the IS models (see  Sec.~\ref{subsec: Reference jet parameters}). However, in the literature, under the assumption that GRBs are the main sources of ultra-high-energy cosmic rays, the following parameters are often adopted: $\varepsilon_{p}=10/12$, $\varepsilon_{e}=\varepsilon_{B}=1/12$, and $k_{p}=2$, see e.g.~\cite{Zhang:2013ycn}. Fig.~\ref{Fig:ALL_MODELS_0} shows the quasi-diffuse neutrino emission for these input parameters, in order to facilitate a comparison with the existing literature. One can see that the flux normalization of the simple IS model, the IS model with dissipative photosphere and the IS model with three components is indeed larger than what is shown in Fig.~\ref{Fig:ALL_MODELS}, and roughly at the same level of the ICMART model, the proton synchrotron model and magnetized jet model with gradual dissipation.

\bibliographystyle{JHEP}
\bibliography{myreferences}

\providecommand{\href}[2]{#2}\begingroup\raggedright\begin{thebibliography}{100}

\bibitem{Klebesadel:1973iq}
R.~W. Klebesadel, I.~B. Strong and R.~A. Olson, \emph{{Observations of
  Gamma-Ray Bursts of Cosmic Origin}},
  \href{https://doi.org/10.1086/181225}{\emph{Astrophys. J. Lett.} {\bfseries
  182} (1973) L85--L88}.

\bibitem{Gruber:2014iza}
D.~Gruber et~al., \emph{{The Fermi GBM Gamma-Ray Burst Spectral Catalog: Four
  Years Of Data}},
  \href{https://doi.org/10.1088/0067-0049/211/1/12}{\emph{Astrophys. J. Suppl.}
  {\bfseries 211} (2014) 12},
  [\href{https://arxiv.org/abs/1401.5069}{{\ttfamily 1401.5069}}].

\bibitem{Woosley:1993wj}
S.~E. Woosley, \emph{{Gamma-ray bursts from stellar mass accretion disks around
  black holes}}, \href{https://doi.org/10.1086/172359}{\emph{Astrophys. J.}
  {\bfseries 405} (1993) 273}.

\bibitem{MacFadyen:1998vz}
A.~MacFadyen and S.~Woosley, \emph{{Collapsars: Gamma-ray bursts and explosions
  in 'failed supernovae'}},
  \href{https://doi.org/10.1086/307790}{\emph{Astrophys. J.} {\bfseries 524}
  (1999) 262}, [\href{https://arxiv.org/abs/astro-ph/9810274}{{\ttfamily
  astro-ph/9810274}}].

\bibitem{Woosley:2006fn}
S.~Woosley and J.~Bloom, \emph{{The Supernova Gamma-Ray Burst Connection}},
  \href{https://doi.org/10.1146/annurev.astro.43.072103.150558}{\emph{Ann. Rev.
  Astron. Astrophys.} {\bfseries 44} (2006) 507--556},
  [\href{https://arxiv.org/abs/astro-ph/0609142}{{\ttfamily
  astro-ph/0609142}}].

\bibitem{Atteia:2017dcj}
J.~Atteia, V.~Heussaff, J.~P. Dezalay, A.~Klotz, D.~Turpin, A.~Tsvetkova
  et~al., \emph{{The maximum isotropic energy of gamma-ray bursts}},
  \href{https://doi.org/10.3847/1538-4357/aa5ffa}{\emph{Astrophys. J.}
  {\bfseries 837} (2017) 119},
  [\href{https://arxiv.org/abs/1702.02961}{{\ttfamily 1702.02961}}].

\bibitem{2020MNRAS.496.2910P}
M.~{Petropoulou}, P.~{Beniamini}, G.~{Vasilopoulos}, D.~{Giannios} and
  R.~{Barniol Duran}, \emph{{Deciphering the properties of the central engine
  in GRB collapsars}}, \href{https://doi.org/10.1093/mnras/staa1695}{\emph{Mon.
  Not. Roy. Astron. Soc.} {\bfseries 496} (Aug., 2020) 2910--2921},
  [\href{https://arxiv.org/abs/2006.07482}{{\ttfamily 2006.07482}}].

\bibitem{Chen:2006rra}
W.-X. Chen and A.~M. Beloborodov, \emph{{Neutrino-Cooled Accretion Disks around
  Spinning Black Hole}}, \href{https://doi.org/10.1086/508923}{\emph{Astrophys.
  J.} {\bfseries 657} (2007) 383--399},
  [\href{https://arxiv.org/abs/astro-ph/0607145}{{\ttfamily
  astro-ph/0607145}}].

\bibitem{Lei:2017zro}
W.-H. Lei, B.~Zhang, X.-F. Wu and E.-W. Liang, \emph{{Hyperaccreting Black Hole
  as Gamma-Ray Burst Central Engine. II. Temporal Evolution of the Central
  Engine Parameters during the Prompt and Afterglow Phases}},
  \href{https://doi.org/10.3847/1538-4357/aa9074}{\emph{Astrophys. J.}
  {\bfseries 849} (2017) 47},
  [\href{https://arxiv.org/abs/1708.05043}{{\ttfamily 1708.05043}}].

\bibitem{Blandford:1977ds}
R.~D. Blandford and R.~L. Znajek, \emph{{Electromagnetic extractions of energy
  from Kerr black holes}},
  \href{https://doi.org/10.1093/mnras/179.3.433}{\emph{Mon. Not. Roy. Astron.
  Soc.} {\bfseries 179} (1977) 433--456}.

\bibitem{Usov:1993qm}
V.~V. Usov, \emph{{On the nature of nonthermal radiation from cosmological
  gamma-ray bursters}},
  \href{https://doi.org/10.1093/mnras/267.4.1035}{\emph{Mon. Not. Roy. Astron.
  Soc.} {\bfseries 267} (1994) 1035},
  [\href{https://arxiv.org/abs/astro-ph/9312024}{{\ttfamily
  astro-ph/9312024}}].

\bibitem{Paczynski:1986px}
B.~Paczynski, \emph{{Gamma-ray bursters at cosmological distances}},
  \href{https://doi.org/10.1086/184740}{\emph{Astrophys. J. Lett.} {\bfseries
  308} (1986) L43--L46}.

\bibitem{Drenkhahn:2002ug}
G.~Drenkhahn and H.~Spruit, \emph{{Efficient acceleration and radiation in
  Poynting flux powered GRB outflows}},
  \href{https://doi.org/10.1051/0004-6361:20020839}{\emph{Astron. Astrophys.}
  {\bfseries 391} (2002) 1141},
  [\href{https://arxiv.org/abs/astro-ph/0202387}{{\ttfamily
  astro-ph/0202387}}].

\bibitem{Meszaros:2017fcs}
P.~M{\'e}sz{\'a}ros, \emph{{Astrophysical Sources of High Energy Neutrinos in
  the IceCube Era}},
  \href{https://doi.org/10.1146/annurev-nucl-101916-123304}{\emph{Ann. Rev.
  Nucl. Part. Sci.} {\bfseries 67} (2017) 45--67},
  [\href{https://arxiv.org/abs/1708.03577}{{\ttfamily 1708.03577}}].

\bibitem{Heinze:2020zqb}
J.~Heinze, D.~Biehl, A.~Fedynitch, D.~Boncioli, A.~Rudolph and W.~Winter,
  \emph{{Systematic parameter space study for the UHECR origin from GRBs in
  models with multiple internal shocks}},
  \href{https://doi.org/10.1093/mnras/staa2751}{\emph{Mon. Not. Roy. Astron.
  Soc.} {\bfseries 498} (2020) 5990--6004},
  [\href{https://arxiv.org/abs/2006.14301}{{\ttfamily 2006.14301}}].

\bibitem{Waxman:1997ti}
E.~Waxman and J.~N. Bahcall, \emph{{High-energy neutrinos from cosmological
  gamma-ray burst fireballs}},
  \href{https://doi.org/10.1103/PhysRevLett.78.2292}{\emph{Phys. Rev. Lett.}
  {\bfseries 78} (1997) 2292--2295},
  [\href{https://arxiv.org/abs/astro-ph/9701231}{{\ttfamily
  astro-ph/9701231}}].

\bibitem{Guetta:2003wi}
D.~Guetta, D.~Hooper, J.~Alvarez-Muniz, F.~Halzen and E.~Reuveni,
  \emph{{Neutrinos from individual gamma-ray bursts in the BATSE catalog}},
  \href{https://doi.org/10.1016/S0927-6505(03)00211-1}{\emph{Astropart. Phys.}
  {\bfseries 20} (2004) 429--455},
  [\href{https://arxiv.org/abs/astro-ph/0302524}{{\ttfamily
  astro-ph/0302524}}].

\bibitem{Wang:2018xkp}
K.~Wang, R.-Y. Liu, Z.-G. Dai and K.~Asano, \emph{{Hadronic origin of prompt
  high-energy emission of gamma-ray bursts revisited: in the case of a limited
  maximum proton energy}},
  \href{https://doi.org/10.3847/1538-4357/aab667}{\emph{Astrophys. J.}
  {\bfseries 857} (2018) 24},
  [\href{https://arxiv.org/abs/1803.04112}{{\ttfamily 1803.04112}}].

\bibitem{Razzaque:2003uv}
S.~Razzaque, P.~M{\'e}sz{\'a}ros and E.~Waxman, \emph{{Neutrino tomography of
  gamma-ray bursts and massive stellar collapses}},
  \href{https://doi.org/10.1103/PhysRevD.68.083001}{\emph{Phys. Rev. D}
  {\bfseries 68} (2003) 083001},
  [\href{https://arxiv.org/abs/astro-ph/0303505}{{\ttfamily
  astro-ph/0303505}}].

\bibitem{Murase:2013ffa}
K.~Murase and K.~Ioka, \emph{{TeV--PeV Neutrinos from Low-Power Gamma-Ray Burst
  Jets inside Stars}},
  \href{https://doi.org/10.1103/PhysRevLett.111.121102}{\emph{Phys. Rev. Lett.}
  {\bfseries 111} (2013) 121102},
  [\href{https://arxiv.org/abs/1306.2274}{{\ttfamily 1306.2274}}].

\bibitem{Metzger:2011xs}
B.~D. Metzger, D.~Giannios and S.~Horiuchi, \emph{{Heavy Nuclei Synthesized in
  Gamma-Ray Burst Outflows as the Source of UHECRs}},
  \href{https://doi.org/10.1111/j.1365-2966.2011.18873.x}{\emph{Mon. Not. Roy.
  Astron. Soc.} {\bfseries 415} (2011) 2495},
  [\href{https://arxiv.org/abs/1101.4019}{{\ttfamily 1101.4019}}].

\bibitem{Meszaros:2015krr}
P.~M{\'e}sz{\'a}ros, \emph{{Gamma Ray Bursts as Neutrino Sources}},
  \href{https://arxiv.org/abs/1511.01396}{{\ttfamily 1511.01396}}.

\bibitem{Waxman:2015ues}
E.~Waxman, \emph{{The Origin of IceCube\textquoteright{}s Neutrinos: Cosmic Ray
  Accelerators Embedded in Star Forming Calorimeters}},
  \href{https://arxiv.org/abs/1511.00815}{{\ttfamily 1511.00815}}.

\bibitem{Murase:2015ndr}
K.~Murase, \emph{{Active Galactic Nuclei as High-Energy Neutrino Sources}},
  \href{https://arxiv.org/abs/1511.01590}{{\ttfamily 1511.01590}}.

\bibitem{Ahlers:2018fkn}
M.~Ahlers and F.~Halzen, \emph{{Opening a New Window onto the Universe with
  IceCube}}, \href{https://doi.org/10.1016/j.ppnp.2018.05.001}{\emph{Prog.
  Part. Nucl. Phys.} {\bfseries 102} (2018) 73--88},
  [\href{https://arxiv.org/abs/1805.11112}{{\ttfamily 1805.11112}}].

\bibitem{Ahlers:2015lln}
M.~Ahlers and F.~Halzen, \emph{{High-energy cosmic neutrino puzzle: a review}},
  \href{https://doi.org/10.1088/0034-4885/78/12/126901}{\emph{Rept. Prog.
  Phys.} {\bfseries 78} (2015) 126901}.

\bibitem{Abbasi:2020jmh}
{\scshape IceCube} collaboration, R.~Abbasi et~al., \emph{{The IceCube
  high-energy starting event sample: Description and flux characterization with
  7.5 years of data}},  \href{https://arxiv.org/abs/2011.03545}{{\ttfamily
  2011.03545}}.

\bibitem{Abbasi:2020zmr}
{\scshape IceCube} collaboration, R.~Abbasi et~al., \emph{{Measurement of
  Astrophysical Tau Neutrinos in IceCube's High-Energy Starting Events}},
  \href{https://arxiv.org/abs/2011.03561}{{\ttfamily 2011.03561}}.

\bibitem{Anchordoqui:2013dnh}
L.~A. Anchordoqui et~al., \emph{{Cosmic Neutrino Pevatrons: A Brand New Pathway
  to Astronomy, Astrophysics, and Particle Physics}},
  \href{https://doi.org/10.1016/j.jheap.2014.01.001}{\emph{JHEAp} {\bfseries
  1-2} (2014) 1--30}, [\href{https://arxiv.org/abs/1312.6587}{{\ttfamily
  1312.6587}}].

\bibitem{Vitagliano:2019yzm}
E.~Vitagliano, I.~Tamborra and G.~Raffelt, \emph{{Grand Unified Neutrino
  Spectrum at Earth: Sources and Spectral Components}},
  \href{https://doi.org/10.1103/RevModPhys.92.045006}{\emph{Rev. Mod. Phys.}
  {\bfseries 92} (2020) 45006},
  [\href{https://arxiv.org/abs/1910.11878}{{\ttfamily 1910.11878}}].

\bibitem{Albert:2020lvs}
{\scshape ANTARES} collaboration, A.~Albert et~al., \emph{{Constraining the
  contribution of Gamma-Ray Bursts to the high-energy diffuse neutrino flux
  with 10 years of ANTARES data}},
  \href{https://doi.org/10.1093/mnras/staa3503}{\emph{Mon. Not. Roy. Astron.
  Soc.} {\bfseries 500} (2020) 5614--5628},
  [\href{https://arxiv.org/abs/2008.02127}{{\ttfamily 2008.02127}}].

\bibitem{Aartsen:2017wea}
{\scshape IceCube} collaboration, M.~Aartsen et~al., \emph{{Extending the
  search for muon neutrinos coincident with gamma-ray bursts in IceCube data}},
  \href{https://doi.org/10.3847/1538-4357/aa7569}{\emph{Astrophys. J.}
  {\bfseries 843} (2017) 112},
  [\href{https://arxiv.org/abs/1702.06868}{{\ttfamily 1702.06868}}].

\bibitem{Aartsen:2016qcr}
{\scshape IceCube} collaboration, M.~Aartsen et~al., \emph{{An All-Sky Search
  for Three Flavors of Neutrinos from Gamma-Ray Bursts with the IceCube
  Neutrino Observatory}},
  \href{https://doi.org/10.3847/0004-637X/824/2/115}{\emph{Astrophys. J.}
  {\bfseries 824} (2016) 115},
  [\href{https://arxiv.org/abs/1601.06484}{{\ttfamily 1601.06484}}].

\bibitem{Li:2012gf}
Z.~Li, \emph{{Fermi Limit on the Neutrino Flux from Gamma-Ray Bursts}},
  \href{https://doi.org/10.1088/2041-8205/770/2/L40}{\emph{Astrophys. J. Lett.}
  {\bfseries 770} (2013) L40},
  [\href{https://arxiv.org/abs/1210.6594}{{\ttfamily 1210.6594}}].

\bibitem{Hummer:2011ms}
S.~H{\"u}mmer, P.~Baerwald and W.~Winter, \emph{{Neutrino Emission from
  Gamma-Ray Burst Fireballs, Revised}},
  \href{https://doi.org/10.1103/PhysRevLett.108.231101}{\emph{Phys. Rev. Lett.}
  {\bfseries 108} (2012) 231101},
  [\href{https://arxiv.org/abs/1112.1076}{{\ttfamily 1112.1076}}].

\bibitem{Tamborra:2015qza}
I.~Tamborra and S.~Ando, \emph{{Diffuse emission of high-energy neutrinos from
  gamma-ray burst fireballs}},
  \href{https://doi.org/10.1088/1475-7516/2015/9/036}{\emph{JCAP} {\bfseries
  09} (2015) 036}, [\href{https://arxiv.org/abs/1504.00107}{{\ttfamily
  1504.00107}}].

\bibitem{Tamborra:2015fzv}
I.~Tamborra and S.~Ando, \emph{{Inspecting the
  supernova\textendash{}gamma-ray-burst connection with high-energy
  neutrinos}}, \href{https://doi.org/10.1103/PhysRevD.93.053010}{\emph{Phys.
  Rev. D} {\bfseries 93} (2016) 053010},
  [\href{https://arxiv.org/abs/1512.01559}{{\ttfamily 1512.01559}}].

\bibitem{Bustamante:2014oka}
M.~Bustamante, P.~Baerwald, K.~Murase and W.~Winter, \emph{{Neutrino and
  cosmic-ray emission from multiple internal shocks in gamma-ray bursts}},
  \href{https://doi.org/10.1038/ncomms7783}{\emph{Nature Commun.} {\bfseries 6}
  (2015) 6783}, [\href{https://arxiv.org/abs/1409.2874}{{\ttfamily
  1409.2874}}].

\bibitem{He:2012tq}
H.-N. He, R.-Y. Liu, X.-Y. Wang, S.~Nagataki, K.~Murase and Z.-G. Dai,
  \emph{{Icecube non-detection of GRBs: Constraints on the fireball
  properties}},
  \href{https://doi.org/10.1088/0004-637X/752/1/29}{\emph{Astrophys. J.}
  {\bfseries 752} (2012) 29},
  [\href{https://arxiv.org/abs/1204.0857}{{\ttfamily 1204.0857}}].

\bibitem{Oganesyan:2019fpa}
G.~Oganesyan, L.~Nava, G.~Ghirlanda, A.~Melandri and A.~Celotti, \emph{{Prompt
  optical emission as a signature of synchrotron radiation in gamma-ray
  bursts}}, \href{https://doi.org/10.1051/0004-6361/201935766}{\emph{Astron.
  Astrophys.} {\bfseries 628} (2019) A59},
  [\href{https://arxiv.org/abs/1904.11086}{{\ttfamily 1904.11086}}].

\bibitem{Zhang:2015bsa}
B.-B. Zhang, Z.~L. Uhm, V.~Connaughton, M.~S. Briggs and B.~Zhang,
  \emph{{Synchrotron Origin of the Typical GRB Band Function\textemdash{}a Case
  Study of GRB 130606b}},
  \href{https://doi.org/10.3847/0004-637X/816/2/72}{\emph{Astrophys. J.}
  {\bfseries 816} (2016) 72},
  [\href{https://arxiv.org/abs/1505.05858}{{\ttfamily 1505.05858}}].

\bibitem{Acuner:2019rif}
Z.~Acuner, F.~Ryde and H.-F. Yu, \emph{{Non-dissipative photospheres in GRBs:
  Spectral appearance in the Fermi/GBM catalogue}},
  \href{https://doi.org/10.1093/mnras/stz1356}{\emph{Mon. Not. Roy. Astron.
  Soc.} {\bfseries 487} (2019) 5508--5519},
  [\href{https://arxiv.org/abs/1906.01318}{{\ttfamily 1906.01318}}].

\bibitem{Acuner:2020zvi}
Z.~{Acuner}, F.~{Ryde}, A.~{Pe'er}, D.~{Mortlock} and B.~{Ahlgren}, \emph{{The
  Fraction of Gamma-Ray Bursts with an Observed Photospheric Emission
  Episode}}, \href{https://doi.org/10.3847/1538-4357/ab80c7}{\emph{Astrophys.
  J.} {\bfseries 893} (2020) 128},
  [\href{https://arxiv.org/abs/2003.06223}{{\ttfamily 2003.06223}}].

\bibitem{Burgess:2018dhc}
J.~M. Burgess, D.~B\'egu\'e, A.~Bacelj, D.~Giannios, F.~Berlato and J.~Greiner,
  \emph{{Gamma-ray bursts as cool synchrotron sources}},
  \href{https://doi.org/10.1038/s41550-019-0911-z}{\emph{Nature Astron.}
  {\bfseries 4} (2019) 174--179},
  [\href{https://arxiv.org/abs/1810.06965}{{\ttfamily 1810.06965}}].

\bibitem{Rees:1994nw}
M.~Rees and P.~M{\'e}sz{\'a}ros, \emph{{Unsteady outflow models for
  cosmological gamma-ray bursts}},
  \href{https://doi.org/10.1086/187446}{\emph{Astrophys. J. Lett.} {\bfseries
  430} (1994) L93--L96},
  [\href{https://arxiv.org/abs/astro-ph/9404038}{{\ttfamily
  astro-ph/9404038}}].

\bibitem{Toma:2010xw}
K.~Toma, X.-F. Wu and P.~M{\'e}sz{\'a}ros, \emph{{A Photosphere-Internal Shock
  Model of Gamma-Ray Bursts: Case Studies of Fermi/LAT Bursts}},
  \href{https://doi.org/10.1111/j.1365-2966.2011.18807.x}{\emph{Mon. Not. Roy.
  Astron. Soc.} {\bfseries 415} (2011) 1663--1680},
  [\href{https://arxiv.org/abs/1002.2634}{{\ttfamily 1002.2634}}].

\bibitem{Guiriec:2015ppa}
S.~Guiriec et~al., \emph{{Towards a Better Understanding of the GRB Phenomenon:
  a New Model for GRB Prompt Emission and its effects on the New
  L$_\mathrm{i}^\mathrm{NT}$-E$_\mathrm{peak,i}^\mathrm{rest,NT}$ relation}},
  \href{https://doi.org/10.1088/0004-637X/807/2/148}{\emph{Astrophys. J.}
  {\bfseries 807} (2015) 148},
  [\href{https://arxiv.org/abs/1501.07028}{{\ttfamily 1501.07028}}].

\bibitem{Zhang:2010jt}
B.~Zhang and H.~Yan, \emph{{The Internal-Collision-Induced Magnetic
  Reconnection and Turbulence (ICMART) Model of Gamma-Ray Bursts}},
  \href{https://doi.org/10.1088/0004-637X/726/2/90}{\emph{Astrophys. J.}
  {\bfseries 726} (2011) 90},
  [\href{https://arxiv.org/abs/1011.1197}{{\ttfamily 1011.1197}}].

\bibitem{Beniamini:2017fqh}
P.~Beniamini and D.~Giannios, \emph{{Prompt Gamma Ray Burst emission from
  gradual magnetic dissipation}},
  \href{https://doi.org/10.1093/mnras/stx717}{\emph{Mon. Not. Roy. Astron.
  Soc.} {\bfseries 468} (2017) 3202--3211},
  [\href{https://arxiv.org/abs/1703.07380}{{\ttfamily 1703.07380}}].

\bibitem{Gill:2020oon}
R.~Gill, J.~Granot and P.~Beniamini, \emph{{GRB Spectrum from Gradual
  Dissipation in a Magnetized Outflow}},
  \href{https://doi.org/10.1093/mnras/staa2870}{\emph{Mon. Not. Roy. Astron.
  Soc.} {\bfseries 499} (2020) 1356--1372},
  [\href{https://arxiv.org/abs/2008.10729}{{\ttfamily 2008.10729}}].

\bibitem{Ghisellini:2019lgz}
G.~Ghisellini, G.~Ghirlanda, G.~Oganesyan, S.~Ascenzi, L.~Nava, A.~Celotti
  et~al., \emph{{Proton\textendash{}synchrotron as the radiation mechanism of
  the prompt emission of gamma-ray bursts?}},
  \href{https://doi.org/10.1051/0004-6361/201937244}{\emph{Astron. Astrophys.}
  {\bfseries 636} (2020) A82},
  [\href{https://arxiv.org/abs/1912.02185}{{\ttfamily 1912.02185}}].

\bibitem{Florou2021}
I.~Florou, M.~Petropoulou and A.~Mastichiadis, \emph{{A marginally fast-cooling
  proton-synchrotron model for prompt GRBs}},
  \href{https://arxiv.org/abs/2102.02501}{{\ttfamily 2102.02501}}.

\bibitem{Bromberg:2011fg}
O.~Bromberg, E.~Nakar, T.~Piran and R.~Sari, \emph{{The propagation of
  relativistic jets in external media}},
  \href{https://doi.org/10.1088/0004-637X/740/2/100}{\emph{Astrophys. J.}
  {\bfseries 740} (2011) 100},
  [\href{https://arxiv.org/abs/1107.1326}{{\ttfamily 1107.1326}}].

\bibitem{1989Natur.340..126E}
D.~{Eichler}, M.~{Livio}, T.~{Piran} and D.~N. {Schramm},
  \emph{{Nucleosynthesis, neutrino bursts and {\ensuremath{\gamma}}-rays from
  coalescing neutron stars}},
  \href{https://doi.org/10.1038/340126a0}{\emph{Nature} {\bfseries 340} (July,
  1989) 126--128}.

\bibitem{1999ApJ...518..356P}
R.~{Popham}, S.~E. {Woosley} and C.~{Fryer}, \emph{{Hyperaccreting Black Holes
  and Gamma-Ray Bursts}},
  \href{https://doi.org/10.1086/307259}{\emph{Astrophys. J.} {\bfseries 518}
  (June, 1999) 356--374},
  [\href{https://arxiv.org/abs/astro-ph/9807028}{{\ttfamily
  astro-ph/9807028}}].

\bibitem{1997ApJ...482L..29M}
P.~{M{\'e}sz{\'a}ros} and M.~J. {Rees}, \emph{{Poynting Jets from Black Holes
  and Cosmological Gamma-Ray Bursts}},
  \href{https://doi.org/10.1086/310692}{\emph{Astrophys. J. Lett.} {\bfseries
  482} (June, 1997) L29--L32},
  [\href{https://arxiv.org/abs/astro-ph/9609065}{{\ttfamily
  astro-ph/9609065}}].

\bibitem{Goodman:1986az}
J.~Goodman, \emph{{Are gamma-ray bursts optically thick?}},
  \href{https://doi.org/10.1086/184741}{\emph{Astrophys. J. Lett.} {\bfseries
  308} (1986) L47--L50}.

\bibitem{Piran:1993jm}
T.~Piran, A.~Shemi and R.~Narayan, \emph{{Hydrodynamics of relativistic
  fireballs}}, \href{https://doi.org/10.1093/mnras/263.4.861}{\emph{Mon. Not.
  Roy. Astron. Soc.} {\bfseries 263} (1993) 861},
  [\href{https://arxiv.org/abs/astro-ph/9301004}{{\ttfamily
  astro-ph/9301004}}].

\bibitem{Lazzati:2013ym}
D.~Lazzati, B.~Morsony, R.~Margutti and M.~Begelman, \emph{{Photospheric
  emission as the dominant radiation mechanism in long-duration gamma-ray
  bursts}}, \href{https://doi.org/10.1088/0004-637X/765/2/103}{\emph{Astrophys.
  J.} {\bfseries 765} (2013) 103},
  [\href{https://arxiv.org/abs/1301.3920}{{\ttfamily 1301.3920}}].

\bibitem{Mizuta:2013yma}
A.~Mizuta and K.~Ioka, \emph{{Opening Angles of Collapsar Jets}},
  \href{https://doi.org/10.1088/0004-637X/777/2/162}{\emph{Astrophys. J.}
  {\bfseries 777} (2013) 162},
  [\href{https://arxiv.org/abs/1304.0163}{{\ttfamily 1304.0163}}].

\bibitem{Dermer:2009zz}
C.~D. Dermer and G.~Menon, \emph{{High energy radiation from black holes: Gamma
  rays, cosmic rays and neutrinos}}.
\newblock Princeton U. Pr., Princeton, USA, 2009.

\bibitem{Meszaros:1993cc}
P.~M{\'e}sz{\'a}ros, P.~Laguna and M.~Rees, \emph{{Gas dynamics of
  relativistically expanding gamma-ray burst sources: Kinematics, energetics,
  magnetic fields and efficiency}},
  \href{https://doi.org/10.1086/173154}{\emph{Astrophys. J.} {\bfseries 415}
  (1993) 181--190}, [\href{https://arxiv.org/abs/astro-ph/9301007}{{\ttfamily
  astro-ph/9301007}}].

\bibitem{Meszaros:1999gb}
P.~M{\'e}sz{\'a}ros and M.~Rees, \emph{{Steep slopes and preferred breaks in
  GRB spectra: The Role of photospheres and comptonization}},
  \href{https://doi.org/10.1086/308371}{\emph{Astrophys. J.} {\bfseries 530}
  (2000) 292--298}, [\href{https://arxiv.org/abs/astro-ph/9908126}{{\ttfamily
  astro-ph/9908126}}].

\bibitem{Piran:1999kx}
T.~Piran, \emph{{Gamma-ray bursts and the fireball model}},
  \href{https://doi.org/10.1016/S0370-1573(98)00127-6}{\emph{Phys. Rept.}
  {\bfseries 314} (1999) 575--667},
  [\href{https://arxiv.org/abs/astro-ph/9810256}{{\ttfamily
  astro-ph/9810256}}].

\bibitem{vanEerten:2018amz}
H.~van Eerten, \emph{{Gamma-ray burst afterglow blast waves}},
  \href{https://doi.org/10.1142/S0218271818420026}{\emph{Int. J. Mod. Phys. D}
  {\bfseries 27} (2018) 1842002},
  [\href{https://arxiv.org/abs/1801.01848}{{\ttfamily 1801.01848}}].

\bibitem{Panaitescu:2000bk}
A.~Panaitescu and P.~Kumar, \emph{{Analytic light-curves of gamma-ray burst
  afterglows: homogeneous versus wind external media}},
  \href{https://doi.org/10.1086/317090}{\emph{Astrophys. J.} {\bfseries 543}
  (2000) 66}, [\href{https://arxiv.org/abs/astro-ph/0003246}{{\ttfamily
  astro-ph/0003246}}].

\bibitem{Sari:1995cb}
R.~Sari and T.~Piran, \emph{{Hydrodynamic time scales and temporal structure in
  GRBs}}, \href{https://doi.org/10.1063/1.51659}{\emph{AIP Conf. Proc.}
  {\bfseries 384} (1996) 782--786},
  [\href{https://arxiv.org/abs/astro-ph/9512125}{{\ttfamily
  astro-ph/9512125}}].

\bibitem{Blandford:1976uq}
R.~Blandford and C.~McKee, \emph{{Fluid dynamics of relativistic blast waves}},
  \href{https://doi.org/10.1063/1.861619}{\emph{Phys. Fluids} {\bfseries 19}
  (1976) 1130--1138}.

\bibitem{Zhang:2018ond}
B.~Zhang, \emph{{The Physics of Gamma-Ray Bursts}}.
\newblock Cambridge University Press, 12, 2018.

\bibitem{Kobayashi:1997jk}
S.~Kobayashi, T.~Piran and R.~Sari, \emph{{Can internal shocks produce the
  variability in GRBs?}},
  \href{https://doi.org/10.1086/512791}{\emph{Astrophys. J.} {\bfseries 490}
  (1997) 92--98}, [\href{https://arxiv.org/abs/astro-ph/9705013}{{\ttfamily
  astro-ph/9705013}}].

\bibitem{Daigne:1998xc}
F.~Daigne and R.~Mochkovitch, \emph{{Gamma-ray bursts from internal shocks in a
  relativistic wind: temporal and spectral properties}},
  \href{https://doi.org/10.1046/j.1365-8711.1998.01305.x}{\emph{Mon. Not. Roy.
  Astron. Soc.} {\bfseries 296} (1998) 275},
  [\href{https://arxiv.org/abs/astro-ph/9801245}{{\ttfamily
  astro-ph/9801245}}].

\bibitem{Guetta:2000ye}
D.~Guetta, M.~Spada and E.~Waxman, \emph{{Efficiency and spectrum of internal
  gamma-ray burst shocks}},
  \href{https://doi.org/10.1086/321543}{\emph{Astrophys. J.} {\bfseries 557}
  (2001) 399}, [\href{https://arxiv.org/abs/astro-ph/0011170}{{\ttfamily
  astro-ph/0011170}}].

\bibitem{Bustamante:2016wpu}
M.~Bustamante, K.~Murase, W.~Winter and J.~Heinze, \emph{{Multi-messenger light
  curves from gamma-ray bursts in the internal shock model}},
  \href{https://doi.org/10.3847/1538-4357/837/1/33}{\emph{Astrophys. J.}
  {\bfseries 837} (2017) 33},
  [\href{https://arxiv.org/abs/1606.02325}{{\ttfamily 1606.02325}}].

\bibitem{Rudolph:2019ccl}
A.~Rudolph, J.~Heinze, A.~Fedynitch and W.~Winter, \emph{{Impact of the
  Collision Model on the Multi-messenger Emission from Gamma-Ray Burst Internal
  Shocks}}, \href{https://doi.org/10.3847/1538-4357/ab7ea7}{\emph{Astrophys.
  J.} {\bfseries 893} (2020) 72},
  [\href{https://arxiv.org/abs/1907.10633}{{\ttfamily 1907.10633}}].

\bibitem{Beloborodov:2017use}
A.~Beloborodov and P.~M\'esz\'aros, \emph{{Photospheric Emission of Gamma-Ray
  Bursts}}, \href{https://doi.org/10.1007/s11214-017-0348-6}{\emph{Space Sci.
  Rev.} {\bfseries 207} (2017) 87--110},
  [\href{https://arxiv.org/abs/1701.04523}{{\ttfamily 1701.04523}}].

\bibitem{Thompson:1994zh}
C.~Thompson, \emph{{A Model of gamma-ray bursts}}, {\emph{Mon. Not. Roy.
  Astron. Soc.} {\bfseries 270} (1994) 480}.

\bibitem{Giannios:2006mx}
D.~Giannios, \emph{{Prompt emission spectra from the photosphere of a GRB}},
  \href{https://doi.org/10.1051/0004-6361:20065000}{\emph{Astron. Astrophys.}
  {\bfseries 457} (2006) 763--770},
  [\href{https://arxiv.org/abs/astro-ph/0602397}{{\ttfamily
  astro-ph/0602397}}].

\bibitem{Gill:2014fwa}
R.~Gill and C.~Thompson, \emph{{Non-thermal Gamma-ray Emission from Delayed
  Pair Breakdown in a Magnetized and Photon-rich Outflow}},
  \href{https://doi.org/10.1088/0004-637X/796/2/81}{\emph{Astrophys. J.}
  {\bfseries 796} (2014) 81},
  [\href{https://arxiv.org/abs/1406.4774}{{\ttfamily 1406.4774}}].

\bibitem{Rees:2004gt}
M.~Rees and P.~M{\'e}sz{\'a}ros, \emph{{Dissipative photosphere models of
  gamma-ray bursts and x-ray flashes}},
  \href{https://doi.org/10.1086/430818}{\emph{Astrophys. J.} {\bfseries 628}
  (2005) 847--852}, [\href{https://arxiv.org/abs/astro-ph/0412702}{{\ttfamily
  astro-ph/0412702}}].

\bibitem{Vurm:2012be}
I.~Vurm, Y.~Lyubarsky and T.~Piran, \emph{{On thermalization in gamma-ray burst
  jets and the peak energies of photospheric spectra}},
  \href{https://doi.org/10.1088/0004-637X/764/2/143}{\emph{Astrophys. J.}
  {\bfseries 764} (2013) 143},
  [\href{https://arxiv.org/abs/1209.0763}{{\ttfamily 1209.0763}}].

\bibitem{Vurm:2015yfa}
I.~Vurm and A.~M. Beloborodov, \emph{{Radiative Transfer Models for Gamma-Ray
  Bursts}}, \href{https://doi.org/10.3847/0004-637X/831/2/175}{\emph{Astrophys.
  J.} {\bfseries 831} (2016) 175},
  [\href{https://arxiv.org/abs/1506.01107}{{\ttfamily 1506.01107}}].

\bibitem{Beloborodov:2012ys}
A.~M. Beloborodov, \emph{{Regulation of the spectral peak in gamma-ray
  bursts}}, \href{https://doi.org/10.1088/0004-637X/764/2/157}{\emph{Astrophys.
  J.} {\bfseries 764} (2013) 157},
  [\href{https://arxiv.org/abs/1207.2707}{{\ttfamily 1207.2707}}].

\bibitem{Giannios:2006jb}
D.~Giannios and H.~C. Spruit, \emph{{Spectral and timing properties of a
  dissipative GRB photosphere}},
  \href{https://doi.org/10.1051/0004-6361:20066739}{\emph{Astron. Astrophys.}
  {\bfseries 469} (2007) 1--9},
  [\href{https://arxiv.org/abs/astro-ph/0611385}{{\ttfamily
  astro-ph/0611385}}].

\bibitem{Thompson:2013yna}
C.~Thompson and R.~Gill, \emph{{Hot Electromagnetic Outflows. III. Displaced
  Fireball in a Strong Magnetic Field}},
  \href{https://doi.org/10.1088/0004-637X/791/1/46}{\emph{Astrophys. J.}
  {\bfseries 791} (2014) 46},
  [\href{https://arxiv.org/abs/1310.2480}{{\ttfamily 1310.2480}}].

\bibitem{Beloborodov:2009be}
A.~M. Beloborodov, \emph{{Collisional mechanism for GRB emission}},
  \href{https://doi.org/10.1111/j.1365-2966.2010.16770.x}{\emph{Mon. Not. Roy.
  Astron. Soc.} {\bfseries 407} (2010) 1033},
  [\href{https://arxiv.org/abs/0907.0732}{{\ttfamily 0907.0732}}].

\bibitem{Lazzati:2009xx}
D.~Lazzati, B.~J. Morsony and M.~Begelman, \emph{{Very high efficiency
  photospheric emission in long duration gamma-ray bursts}},
  \href{https://doi.org/10.1088/0004-637X/700/1/L47}{\emph{Astrophys. J. Lett.}
  {\bfseries 700} (2009) L47--L50},
  [\href{https://arxiv.org/abs/0904.2779}{{\ttfamily 0904.2779}}].

\bibitem{Gottlieb:2019aae}
O.~Gottlieb, A.~Levinson and E.~Nakar, \emph{{High efficiency photospheric
  emission entailed by formation of a collimation shock in gamma-ray bursts}},
  \href{https://doi.org/10.1093/mnras/stz1828}{\emph{Mon. Not. Roy. Astron.
  Soc.} {\bfseries 488} (2019) 1416--1426},
  [\href{https://arxiv.org/abs/1904.07244}{{\ttfamily 1904.07244}}].

\bibitem{Bahcall:2000sa}
J.~N. Bahcall and P.~M{\'e}sz{\'a}ros, \emph{{5-GeV to 10-GeV neutrinos from
  gamma-ray burst fireballs}},
  \href{https://doi.org/10.1103/PhysRevLett.85.1362}{\emph{Phys. Rev. Lett.}
  {\bfseries 85} (2000) 1362--1365},
  [\href{https://arxiv.org/abs/hep-ph/0004019}{{\ttfamily hep-ph/0004019}}].

\bibitem{Kashiyama:2013ata}
K.~Kashiyama, K.~Murase and P.~M\'esz\'aros, \emph{{Neutron-Proton-Converter
  Acceleration Mechanism at Subphotospheres of Relativistic Outflows}},
  \href{https://doi.org/10.1103/PhysRevLett.111.131103}{\emph{Phys. Rev. Lett.}
  {\bfseries 111} (2013) 131103},
  [\href{https://arxiv.org/abs/1304.1945}{{\ttfamily 1304.1945}}].

\bibitem{Wang:2008zm}
X.-Y. Wang and Z.-G. Dai, \emph{{Prompt TeV neutrinos from dissipative
  photospheres of gamma-ray bursts}},
  \href{https://doi.org/10.1088/0004-637X/691/2/L67}{\emph{Astrophys. J. Lett.}
  {\bfseries 691} (2009) L67--L71},
  [\href{https://arxiv.org/abs/0807.0290}{{\ttfamily 0807.0290}}].

\bibitem{Murase:2008sp}
K.~Murase, \emph{{Prompt High-Energy Neutrinos from Gamma-Ray Bursts in the
  Photospheric and Synchrotron Self-Compton Scenarios}},
  \href{https://doi.org/10.1103/PhysRevD.78.101302}{\emph{Phys. Rev. D}
  {\bfseries 78} (2008) 101302},
  [\href{https://arxiv.org/abs/0807.0919}{{\ttfamily 0807.0919}}].

\bibitem{Xiao:2017blv}
D.~Xiao, Z.-G. Dai and P.~M\'esz\'aros, \emph{{Prompt Neutrino Emission of
  Gamma-Ray Bursts in the Dissipative Photospheric Scenario Revisited: Possible
  Contributions from Cocoons}},
  \href{https://doi.org/10.3847/1538-4357/aa76e5}{\emph{Astrophys. J.}
  {\bfseries 843} (2017) 17},
  [\href{https://arxiv.org/abs/1706.01293}{{\ttfamily 1706.01293}}].

\bibitem{Murase:2013hh}
K.~Murase, K.~Kashiyama and P.~M\'esz\'aros, \emph{{Subphotospheric Neutrinos
  from Gamma-Ray Bursts: The Role of Neutrons}},
  \href{https://doi.org/10.1103/PhysRevLett.111.131102}{\emph{Phys. Rev. Lett.}
  {\bfseries 111} (2013) 131102},
  [\href{https://arxiv.org/abs/1301.4236}{{\ttfamily 1301.4236}}].

\bibitem{Beloborodov:2016jmz}
A.~M. Beloborodov, \emph{{Sub-photospheric shocks in relativistic explosions}},
  \href{https://doi.org/10.3847/1538-4357/aa5c8c}{\emph{Astrophys. J.}
  {\bfseries 838} (2017) 125},
  [\href{https://arxiv.org/abs/1604.02794}{{\ttfamily 1604.02794}}].

\bibitem{Zhang:2013ycn}
B.~Zhang and B.~Zhang, \emph{{Gamma-Ray Burst Prompt Emission Light Curves and
  Power Density Spectra in the ICMART Model}},
  \href{https://doi.org/10.1088/0004-637X/782/2/92}{\emph{Astrophys. J.}
  {\bfseries 782} (2014) 92},
  [\href{https://arxiv.org/abs/1312.7701}{{\ttfamily 1312.7701}}].

\bibitem{Drenkhahn:2001ue}
G.~Drenkhahn, \emph{{Acceleration of GRB outflows by Poynting flux
  dissipation}},
  \href{https://doi.org/10.1051/0004-6361:20020390}{\emph{Astron. Astrophys.}
  {\bfseries 387} (2002) 714},
  [\href{https://arxiv.org/abs/astro-ph/0112509}{{\ttfamily
  astro-ph/0112509}}].

\bibitem{Parfrey:2014wga}
K.~Parfrey, D.~Giannios and A.~M. Beloborodov, \emph{{Black-hole jets without
  large-scale net magnetic flux}},
  \href{https://doi.org/10.1093/mnrasl/slu162}{\emph{Mon. Not. Roy. Astron.
  Soc.} {\bfseries 446} (2015) L61--L65},
  [\href{https://arxiv.org/abs/1410.0374}{{\ttfamily 1410.0374}}].

\bibitem{2015MNRAS.450..183S}
L.~{Sironi}, M.~{Petropoulou} and D.~{Giannios}, \emph{{Relativistic jets shine
  through shocks or magnetic reconnection?}},
  \href{https://doi.org/10.1093/mnras/stv641}{\emph{Mon. Not. Roy. Astron.
  Soc.} {\bfseries 450} (2015) 183--191},
  [\href{https://arxiv.org/abs/1502.01021}{{\ttfamily 1502.01021}}].

\bibitem{2016ApJ...818L...9G}
F.~Guo, X.~Li, H.~Li, W.~Daughton, B.~Zhang, N.~Lloyd-Ronning et~al.,
  \emph{{Efficient Production of High-energy Nonthermal Particles During
  Magnetic Reconnection in a Magnetically Dominated Ion\textendash{}electron
  Plasma}}, \href{https://doi.org/10.3847/2041-8205/818/1/L9}{\emph{Astrophys.
  J. Lett.} {\bfseries 818} (2016) L9},
  [\href{https://arxiv.org/abs/1511.01434}{{\ttfamily 1511.01434}}].

\bibitem{Werner:2016fxe}
G.~Werner, D.~Uzdensky, M.~Begelman, B.~Cerutti and K.~Nalewajko,
  \emph{{Non-thermal particle acceleration in collisionless relativistic
  electron\textendash{}proton reconnection}},
  \href{https://doi.org/10.1093/mnras/stx2530}{\emph{Mon. Not. Roy. Astron.
  Soc.} {\bfseries 473} (2018) 4840--4861},
  [\href{https://arxiv.org/abs/1612.04493}{{\ttfamily 1612.04493}}].

\bibitem{2019ApJ...880...37P}
M.~{Petropoulou}, L.~{Sironi}, A.~{Spitkovsky} and D.~{Giannios},
  \emph{{Relativistic Magnetic Reconnection in Electron-Positron-Proton
  Plasmas: Implications for Jets of Active Galactic Nuclei}},
  \href{https://doi.org/10.3847/1538-4357/ab287a}{\emph{Astrophys. J.}
  {\bfseries 880} (July, 2019) 37},
  [\href{https://arxiv.org/abs/1906.03297}{{\ttfamily 1906.03297}}].

\bibitem{Sironi:2014jfa}
L.~Sironi and A.~Spitkovsky, \emph{{Relativistic Reconnection: an Efficient
  Source of Non-Thermal Particles}},
  \href{https://doi.org/10.1088/2041-8205/783/1/L21}{\emph{Astrophys. J. Lett.}
  {\bfseries 783} (2014) L21},
  [\href{https://arxiv.org/abs/1401.5471}{{\ttfamily 1401.5471}}].

\bibitem{Guo2014}
F.~Guo, H.~Li, W.~Daughton and Y.-H. Liu, \emph{Formation of hard power laws in
  the energetic particle spectra resulting from relativistic magnetic
  reconnection},
  \href{https://doi.org/10.1103/PhysRevLett.113.155005}{\emph{Phys. Rev. Lett.}
  {\bfseries 113} (Oct, 2014) 155005}.

\bibitem{2016ApJ...816L...8W}
G.~R. {Werner}, D.~A. {Uzdensky}, B.~{Cerutti}, K.~{Nalewajko} and M.~C.
  {Begelman}, \emph{{The Extent of Power-law Energy Spectra in Collisionless
  Relativistic Magnetic Reconnection in Pair Plasmas}},
  \href{https://doi.org/10.3847/2041-8205/816/1/L8}{\emph{Astrophys. J. Lett.}
  {\bfseries 816} (Jan., 2016) L8},
  [\href{https://arxiv.org/abs/1409.8262}{{\ttfamily 1409.8262}}].

\bibitem{Oganesyan:2017ork}
G.~Oganesyan, L.~Nava, G.~Ghirlanda and A.~Celotti, \emph{{Detection of
  Low-energy Breaks in Gamma-Ray Burst Prompt Emission Spectra}},
  \href{https://doi.org/10.3847/1538-4357/aa831e}{\emph{Astrophys. J.}
  {\bfseries 846} (2017) 137},
  [\href{https://arxiv.org/abs/1709.04689}{{\ttfamily 1709.04689}}].

\bibitem{Ravasio:2019kiw}
M.~Ravasio, G.~Ghirlanda, L.~Nava and G.~Ghisellini, \emph{{Evidence of two
  spectral breaks in the prompt emission of gamma ray bursts}},
  \href{https://doi.org/10.1051/0004-6361/201834987}{\emph{Astron. Astrophys.}
  {\bfseries 625} (2019) A60},
  [\href{https://arxiv.org/abs/1903.02555}{{\ttfamily 1903.02555}}].

\bibitem{Daigne:2010fb}
F.~Daigne, Z.~Bosnjak and G.~Dubus, \emph{{Reconciling observed GRB prompt
  spectra with synchrotron radiation ?}},
  \href{https://doi.org/10.1051/0004-6361/201015457}{\emph{Astron. Astrophys.}
  {\bfseries 526} (2011) A110},
  [\href{https://arxiv.org/abs/1009.2636}{{\ttfamily 1009.2636}}].

\bibitem{Crumley:2018kvf}
P.~Crumley, D.~Caprioli, S.~Markoff and A.~Spitkovsky, \emph{{Kinetic
  simulations of mildly relativistic shocks \textendash{} I. Particle
  acceleration in high Mach number shocks}},
  \href{https://doi.org/10.1093/mnras/stz232}{\emph{Mon. Not. Roy. Astron.
  Soc.} {\bfseries 485} (2019) 5105--5119},
  [\href{https://arxiv.org/abs/1809.10809}{{\ttfamily 1809.10809}}].

\bibitem{Sironi:2010rb}
L.~Sironi and A.~Spitkovsky, \emph{{Particle Acceleration in Relativistic
  Magnetized Collisionless Electron-Ion Shocks}},
  \href{https://doi.org/10.1088/0004-637X/726/2/75}{\emph{Astrophys. J.}
  {\bfseries 726} (2011) 75},
  [\href{https://arxiv.org/abs/1009.0024}{{\ttfamily 1009.0024}}].

\bibitem{Kumar:2009ps}
P.~Kumar and R.~B. Duran, \emph{{On the generation of high energy photons
  detected by the Fermi Satellite from gamma-ray bursts}},
  \href{https://doi.org/10.1111/j.1745-3933.2009.00766.x}{\emph{Mon. Not. Roy.
  Astron. Soc.} {\bfseries 400} (2009) 75},
  [\href{https://arxiv.org/abs/0905.2417}{{\ttfamily 0905.2417}}].

\bibitem{Santana:2013saa}
R.~Santana, R.~B. Duran and P.~Kumar, \emph{{Magnetic Fields In Relativistic
  Collisionless Shocks}},
  \href{https://doi.org/10.1088/0004-637X/785/1/29}{\emph{Astrophys. J.}
  {\bfseries 785} (2014) 29},
  [\href{https://arxiv.org/abs/1309.3277}{{\ttfamily 1309.3277}}].

\bibitem{Beniamini:2016hzc}
P.~Beniamini, L.~Nava and T.~Piran, \emph{{A revised analysis of gamma-ray
  bursts\textquoteright{} prompt efficiencies}},
  \href{https://doi.org/10.1093/mnras/stw1331}{\emph{Mon. Not. Roy. Astron.
  Soc.} {\bfseries 461} (2016) 51--59},
  [\href{https://arxiv.org/abs/1606.00311}{{\ttfamily 1606.00311}}].

\bibitem{Sironi:2013ri}
L.~Sironi, A.~Spitkovsky and J.~Arons, \emph{{The Maximum Energy of Accelerated
  Particles in Relativistic Collisionless Shocks}},
  \href{https://doi.org/10.1088/0004-637X/771/1/54}{\emph{Astrophys. J.}
  {\bfseries 771} (2013) 54},
  [\href{https://arxiv.org/abs/1301.5333}{{\ttfamily 1301.5333}}].

\bibitem{Wang:2015vpa}
X.-G. Wang et~al., \emph{{How bad or Good are the External Forward Shock
  Afterglow Models of Gamma-ray Bursts?}},
  \href{https://doi.org/10.1088/0067-0049/219/1/9}{\emph{Astrophys. J. Suppl.}
  {\bfseries 219} (2015) 9},
  [\href{https://arxiv.org/abs/1503.03193}{{\ttfamily 1503.03193}}].

\bibitem{Racusin:2011jf}
J.~L. Racusin et~al., \emph{{Fermi and Swift Gamma-Ray Burst Afterglow
  Population Studies}},
  \href{https://doi.org/10.1088/0004-637X/738/2/138}{\emph{Astrophys. J.}
  {\bfseries 738} (2011) 138},
  [\href{https://arxiv.org/abs/1106.2469}{{\ttfamily 1106.2469}}].

\bibitem{Ghirlanda:2017opl}
G.~Ghirlanda, F.~Nappo, G.~Ghisellini, A.~Melandri, G.~Marcarini, L.~Nava
  et~al., \emph{{Bulk Lorentz factors of Gamma-Ray Bursts}},
  \href{https://doi.org/10.1051/0004-6361/201731598}{\emph{Astron. Astrophys.}
  {\bfseries 609} (2018) A112},
  [\href{https://arxiv.org/abs/1711.06257}{{\ttfamily 1711.06257}}].

\bibitem{LloydRonning:2004gv}
N.~M. Lloyd-Ronning and B.~Zhang, \emph{{On the kinetic energy and radiative
  efficiency of gamma-ray bursts}},
  \href{https://doi.org/10.1086/423026}{\emph{Astrophys. J.} {\bfseries 613}
  (2004) 477--483}, [\href{https://arxiv.org/abs/astro-ph/0404107}{{\ttfamily
  astro-ph/0404107}}].

\bibitem{2015ApJ...815..101N}
K.~{Nalewajko}, D.~A. {Uzdensky}, B.~{Cerutti}, G.~R. {Werner} and M.~C.
  {Begelman}, \emph{{On the Distribution of Particle Acceleration Sites in
  Plasmoid-dominated Relativistic Magnetic Reconnection}},
  \href{https://doi.org/10.1088/0004-637X/815/2/101}{\emph{Astrophys. J.}
  {\bfseries 815} (Dec., 2015) 101},
  [\href{https://arxiv.org/abs/1508.02392}{{\ttfamily 1508.02392}}].

\bibitem{2018MNRAS.481.5687P}
M.~{Petropoulou} and L.~{Sironi}, \emph{{The steady growth of the high-energy
  spectral cut-off in relativistic magnetic reconnection}},
  \href{https://doi.org/10.1093/mnras/sty2702}{\emph{Mon. Not. Roy. Astron.
  Soc.} {\bfseries 481} (2018) 5687--5701},
  [\href{https://arxiv.org/abs/1808.00966}{{\ttfamily 1808.00966}}].

\bibitem{2020ApJ...899..151K}
P.~{Kilian}, X.~{Li}, F.~{Guo} and H.~{Li}, \emph{{Exploring the Acceleration
  Mechanisms for Particle Injection and Power-law Formation during
  Transrelativistic Magnetic Reconnection}},
  \href{https://doi.org/10.3847/1538-4357/aba1e9}{\emph{Astrophys. J.}
  {\bfseries 899} (Aug., 2020) 151},
  [\href{https://arxiv.org/abs/2001.02732}{{\ttfamily 2001.02732}}].

\bibitem{Lipari:2007su}
P.~Lipari, M.~Lusignoli and D.~Meloni, \emph{{Flavor Composition and Energy
  Spectrum of Astrophysical Neutrinos}},
  \href{https://doi.org/10.1103/PhysRevD.75.123005}{\emph{Phys. Rev. D}
  {\bfseries 75} (2007) 123005},
  [\href{https://arxiv.org/abs/0704.0718}{{\ttfamily 0704.0718}}].

\bibitem{Matthews:2020lig}
J.~Matthews, A.~Bell and K.~Blundell, \emph{{Particle acceleration in
  astrophysical jets}},
  \href{https://doi.org/10.1016/j.newar.2020.101543}{\emph{New Astron. Rev.}
  {\bfseries 89} (2020) 101543},
  [\href{https://arxiv.org/abs/2003.06587}{{\ttfamily 2003.06587}}].

\bibitem{Hummer:2010vx}
S.~H{\"u}mmer, M.~Ruger, F.~Spanier and W.~Winter, \emph{{Simplified models for
  photohadronic interactions in cosmic accelerators}},
  \href{https://doi.org/10.1088/0004-637X/721/1/630}{\emph{Astrophys. J.}
  {\bfseries 721} (2010) 630--652},
  [\href{https://arxiv.org/abs/1002.1310}{{\ttfamily 1002.1310}}].

\bibitem{Razzaque:2005bh}
S.~Razzaque, P.~M{\'e}sz{\'a}ros and E.~Waxman, \emph{{High energy neutrinos
  from a slow jet model of core collapse supernovae}},
  \href{https://doi.org/10.1142/S0217732305018414}{\emph{Mod. Phys. Lett. A}
  {\bfseries 20} (2005) 2351--2368},
  [\href{https://arxiv.org/abs/astro-ph/0509729}{{\ttfamily
  astro-ph/0509729}}].

\bibitem{Gao:2012ay}
S.~Gao, K.~Asano and P.~M{\'e}sz{\'a}ros, \emph{{High Energy Neutrinos from
  Dissipative Photospheric Models of Gamma Ray Bursts}},
  \href{https://doi.org/10.1088/1475-7516/2012/11/058}{\emph{JCAP} {\bfseries
  11} (2012) 058}, [\href{https://arxiv.org/abs/1210.1186}{{\ttfamily
  1210.1186}}].

\bibitem{Zyla:2020zbs}
{\scshape Particle Data Group} collaboration, P.~Zyla et~al., \emph{{Review of
  Particle Physics}}, \href{https://doi.org/10.1093/ptep/ptaa104}{\emph{PTEP}
  {\bfseries 2020} (2020) 083C01}.

\bibitem{PhysRev.137.B1306}
F.~C. Jones, \emph{{Inverse Compton Scattering of Cosmic-Ray Electrons}},
  \href{https://doi.org/10.1103/PhysRev.137.B1306}{\emph{Phys. Rev.} {\bfseries
  137} (Mar, 1965) B1306--B1311}.

\bibitem{Mucke:1999yb}
A.~M{\"u}cke, R.~Engel, J.~Rachen, R.~Protheroe and T.~Stanev, \emph{{SOPHIA:
  Monte Carlo simulations of photohadronic processes in astrophysics}},
  \href{https://doi.org/10.1016/S0010-4655(99)00446-4}{\emph{Comput. Phys.
  Commun.} {\bfseries 124} (2000) 290--314},
  [\href{https://arxiv.org/abs/astro-ph/9903478}{{\ttfamily
  astro-ph/9903478}}].

\bibitem{Asano:2006zzb}
K.~Asano and S.~Nagataki, \emph{{Very high energy neutrinos originating from
  kaons in gamma-ray bursts}},
  \href{https://doi.org/10.1086/503291}{\emph{Astrophys. J. Lett.} {\bfseries
  640} (2006) L9--L12},
  [\href{https://arxiv.org/abs/astro-ph/0603107}{{\ttfamily
  astro-ph/0603107}}].

\bibitem{Petropoulou:2014lja}
M.~Petropoulou, D.~Giannios and S.~Dimitrakoudis, \emph{{Implications of a PeV
  neutrino spectral cutoff in GRB models}},
  \href{https://doi.org/10.1093/mnras/stu1757}{\emph{Mon. Not. Roy. Astron.
  Soc.} {\bfseries 445} (2014) 570--580},
  [\href{https://arxiv.org/abs/1405.2091}{{\ttfamily 1405.2091}}].

\bibitem{Esteban:2020cvm}
I.~Esteban, M.~Gonzalez-Garcia, M.~Maltoni, T.~Schwetz and A.~Zhou, \emph{{The
  fate of hints: updated global analysis of three-flavor neutrino
  oscillations}}, \href{https://doi.org/10.1007/JHEP09(2020)178}{\emph{JHEP}
  {\bfseries 09} (2020) 178},
  [\href{https://arxiv.org/abs/2007.14792}{{\ttfamily 2007.14792}}].

\bibitem{Baerwald:2011ee}
P.~Baerwald, S.~H{\"u}mmer and W.~Winter, \emph{{Systematics in the
  Interpretation of Aggregated Neutrino Flux Limits and Flavor Ratios from
  Gamma-Ray Bursts}},
  \href{https://doi.org/10.1016/j.astropartphys.2011.11.005}{\emph{Astropart.
  Phys.} {\bfseries 35} (2012) 508--529},
  [\href{https://arxiv.org/abs/1107.5583}{{\ttfamily 1107.5583}}].

\bibitem{Aghanim:2018eyx}
{\scshape Planck} collaboration, N.~Aghanim et~al., \emph{{Planck 2018 results.
  VI. Cosmological parameters}},
  \href{https://doi.org/10.1051/0004-6361/201833910}{\emph{Astron. Astrophys.}
  {\bfseries 641} (2020) A6},
  [\href{https://arxiv.org/abs/1807.06209}{{\ttfamily 1807.06209}}].

\bibitem{Baerwald:2013pu}
P.~Baerwald, M.~Bustamante and W.~Winter, \emph{{UHECR escape mechanisms for
  protons and neutrons from GRBs, and the cosmic ray-neutrino connection}},
  \href{https://doi.org/10.1088/0004-637X/768/2/186}{\emph{Astrophys. J.}
  {\bfseries 768} (2013) 186},
  [\href{https://arxiv.org/abs/1301.6163}{{\ttfamily 1301.6163}}].

\bibitem{Samuelsson:2018fan}
F.~Samuelsson, D.~B{\'e}gu{\'e}, F.~Ryde and A.~Pe'er, \emph{{The Limited
  Contribution of Low- and High-Luminosity Gamma-Ray Bursts to Ultra-High
  Energy Cosmic Rays}},
  \href{https://doi.org/10.3847/1538-4357/ab153c}{\emph{Astrophys. J.}
  {\bfseries 876} (2019) 93},
  [\href{https://arxiv.org/abs/1810.06579}{{\ttfamily 1810.06579}}].

\bibitem{FermiLAT:2018ksx}
{\scshape Fermi-LAT} collaboration, M.~Ajello et~al., \emph{{Investigating the
  Nature of Late-time High-energy GRB Emission through Joint Fermi/Swift
  Observations}},
  \href{https://doi.org/10.3847/1538-4357/aad000}{\emph{Astrophys. J.}
  {\bfseries 863} (2018) 138},
  [\href{https://arxiv.org/abs/1808.01683}{{\ttfamily 1808.01683}}].

\bibitem{Zhang:2012qy}
B.~Zhang and P.~Kumar, \emph{{Model-dependent high-energy neutrino flux from
  Gamma-Ray Bursts}},
  \href{https://doi.org/10.1103/PhysRevLett.110.121101}{\emph{Phys. Rev. Lett.}
  {\bfseries 110} (2013) 121101},
  [\href{https://arxiv.org/abs/1210.0647}{{\ttfamily 1210.0647}}].

\bibitem{Guiriec:2016bae}
S.~Guiriec, C.~Kouveliotou, D.~Hartmann, J.~Granot, K.~Asano,
  P.~M{\'e}sz{\'a}ros et~al., \emph{{A Unified Model for GRB Prompt Emission
  from Optical to $\gamma$-Rays; Exploring GRBs as Standard Candle}},
  \href{https://doi.org/10.3847/2041-8205/831/1/L8}{\emph{Astrophys. J. Lett.}
  {\bfseries 831} (2016) L8},
  [\href{https://arxiv.org/abs/1606.07193}{{\ttfamily 1606.07193}}].

\bibitem{Arimoto:2016vhy}
M.~Arimoto, K.~Asano, M.~Ohno, P.~Veres, M.~Axelsson, E.~Bissaldi et~al.,
  \emph{{High-energy Non-thermal and Thermal Emission From GRB 141207a Detected
  by Fermi}},
  \href{https://doi.org/10.3847/1538-4357/833/2/139}{\emph{Astrophys. J.}
  {\bfseries 833} (2016) 139},
  [\href{https://arxiv.org/abs/1610.04867}{{\ttfamily 1610.04867}}].

\bibitem{Beniamini:2015eaa}
P.~Beniamini, L.~Nava, R.~Barniol~Duran and T.~Piran, \emph{{Energies of GRB
  blast waves and prompt efficiencies as implied by modelling of X-ray and GeV
  afterglows}}, \href{https://doi.org/10.1093/mnras/stv2033}{\emph{Mon. Not.
  Roy. Astron. Soc.} {\bfseries 454} (2015) 1073--1085},
  [\href{https://arxiv.org/abs/1504.04833}{{\ttfamily 1504.04833}}].

\bibitem{Razzaque:2013dsa}
S.~Razzaque, \emph{{Long-lived PeV\textendash{}EeV neutrinos from gamma-ray
  burst blastwave}},
  \href{https://doi.org/10.1103/PhysRevD.88.103003}{\emph{Phys. Rev. D}
  {\bfseries 88} (2013) 103003},
  [\href{https://arxiv.org/abs/1307.7596}{{\ttfamily 1307.7596}}].

\bibitem{Ajello:2019avs}
M.~Ajello et~al., \emph{{Fermi and Swift Observations of GRB 190114C: Tracing
  the Evolution of High-Energy Emission from Prompt to Afterglow}},
  \href{https://doi.org/10.3847/1538-4357/ab5b05}{\emph{Astrophys. J.}
  {\bfseries 890} (2020) 9},
  [\href{https://arxiv.org/abs/1909.10605}{{\ttfamily 1909.10605}}].

\bibitem{Deng:2015xea}
W.~Deng, H.~Li, B.~Zhang and S.~Li, \emph{{Relativistic MHD simulations of
  collision-induced magnetic dissipation in Poynting-flux-dominated
  jets/outflows}},
  \href{https://doi.org/10.1088/0004-637X/805/2/163}{\emph{Astrophys. J.}
  {\bfseries 805} (2015) 163},
  [\href{https://arxiv.org/abs/1501.07595}{{\ttfamily 1501.07595}}].

\bibitem{Liu:2020cvk}
K.~Liu, D.-B. Lin, K.~Wang, L.~Zhou, X.-G. Wang and E.-W. Liang,
  \emph{{Gamma-ray Burst Spectrum with a Time-dependent Injection Rate of
  High-energy Electrons}},
  \href{https://doi.org/10.3847/2041-8213/ab838e}{\emph{Astrophys. J. Lett.}
  {\bfseries 893} (2020) L14},
  [\href{https://arxiv.org/abs/2003.12231}{{\ttfamily 2003.12231}}].

\bibitem{GRBweb}
{GRB web catalog,}. \url{https://icecube.wisc.edu/~grbweb_public/}.

\bibitem{Gupta:2007yb}
N.~Gupta and B.~Zhang, \emph{{Prompt Emission of High Energy Photons from Gamma
  Ray Bursts}},
  \href{https://doi.org/10.1111/j.1365-2966.2007.12051.x}{\emph{Mon. Not. Roy.
  Astron. Soc.} {\bfseries 380} (2007) 78},
  [\href{https://arxiv.org/abs/0704.1329}{{\ttfamily 0704.1329}}].

\bibitem{Aartsen:2020fgd}
{\scshape IceCube Gen2} collaboration, M.~G. Aartsen et~al.,
  \emph{{IceCube-Gen2: The Window to the Extreme Universe}},
  \href{https://arxiv.org/abs/2008.04323}{{\ttfamily 2008.04323}}.

\bibitem{Adrian-Martinez:2016fdl}
{\scshape KM3Net} collaboration, S.~Adrian-Martinez et~al., \emph{{Letter of
  intent for KM3NeT 2.0}},
  \href{https://doi.org/10.1088/0954-3899/43/8/084001}{\emph{J. Phys. G}
  {\bfseries 43} (2016) 084001},
  [\href{https://arxiv.org/abs/1601.07459}{{\ttfamily 1601.07459}}].

\bibitem{Petropoulou:2014awa}
M.~Petropoulou, \emph{{The role of hadronic cascades in GRB models of efficient
  neutrino production}},
  \href{https://doi.org/10.1093/mnras/stu1079}{\emph{Mon. Not. Roy. Astron.
  Soc.} {\bfseries 442} (2014) 3026--3036},
  [\href{https://arxiv.org/abs/1405.7669}{{\ttfamily 1405.7669}}].

\bibitem{Petropoulou:2014sja}
M.~Petropoulou, S.~Dimitrakoudis, A.~Mastichiadis and D.~Giannios,
  \emph{{Hadronic supercriticality as a trigger for \ensuremath{\gamma}-ray
  burst emission}}, \href{https://doi.org/10.1093/mnras/stu1362}{\emph{Mon.
  Not. Roy. Astron. Soc.} {\bfseries 444} (2014) 2186--2199},
  [\href{https://arxiv.org/abs/1407.2915}{{\ttfamily 1407.2915}}].

\bibitem{Murase:2011cx}
K.~Murase, K.~Asano, T.~Terasawa and P.~Meszaros, \emph{{The Role of Stochastic
  Acceleration in the Prompt Emission of Gamma-Ray Bursts: Application to
  Hadronic Injection}},
  \href{https://doi.org/10.1088/0004-637X/746/2/164}{\emph{Astrophys. J.}
  {\bfseries 746} (2012) 164},
  [\href{https://arxiv.org/abs/1107.5575}{{\ttfamily 1107.5575}}].

\bibitem{Asano:2013jea}
K.~Asano and P.~M{\'e}sz{\'a}ros, \emph{{Photon and Neutrino Spectra of
  Time-Dependent Photospheric Models of Gamma-Ray Bursts}},
  \href{https://doi.org/10.1088/1475-7516/2013/09/008}{\emph{JCAP} {\bfseries
  09} (2013) 008}, [\href{https://arxiv.org/abs/1308.3563}{{\ttfamily
  1308.3563}}].

\bibitem{Mastichiadis:2020yjo}
A.~Mastichiadis, I.~Florou, E.~Kefala, S.~S. Boula and M.~Petropoulou, \emph{{A
  roadmap to hadronic supercriticalities: a comprehensive study of the
  parameter space for high-energy astrophysical sources}},
  \href{https://doi.org/10.1093/mnras/staa1308}{\emph{Mon. Not. Roy. Astron.
  Soc.} {\bfseries 495} (2020) 2458--2474},
  [\href{https://arxiv.org/abs/2003.06956}{{\ttfamily 2003.06956}}].

\bibitem{Band:1993eg}
D.~Band et~al., \emph{{BATSE observations of gamma-ray burst spectra. 1.
  Spectral diversity.}}, \href{https://doi.org/10.1086/172995}{\emph{Astrophys.
  J.} {\bfseries 413} (1993) 281--292}.

\bibitem{Amati:2006ky}
L.~Amati, \emph{{The E(p,i) - E(iso) correlation in grbs: updated observational
  status, re-analysis and main implications}},
  \href{https://doi.org/10.1111/j.1365-2966.2006.10840.x}{\emph{Mon. Not. Roy.
  Astron. Soc.} {\bfseries 372} (2006) 233--245},
  [\href{https://arxiv.org/abs/astro-ph/0601553}{{\ttfamily
  astro-ph/0601553}}].

\bibitem{Yu:2016epf}
H.-F. Yu et~al., \emph{{The Fermi GBM gamma-ray burst time-resolved spectral
  catalog: brightest bursts in the first four years}},
  \href{https://doi.org/10.1051/0004-6361/201527509}{\emph{Astron. Astrophys.}
  {\bfseries 588} (2016) A135},
  [\href{https://arxiv.org/abs/1601.05206}{{\ttfamily 1601.05206}}].

\bibitem{Yu:2018vfp}
H.-F. {Yu}, H.~{Dereli-B{\'e}gu{\'e}} and F.~{Ryde}, \emph{{Bayesian
  Time-resolved Spectroscopy of GRB Pulses}},
  \href{https://doi.org/10.3847/1538-4357/ab488a}{\emph{Astrophys. J.}
  {\bfseries 886} (Nov., 2019) 20},
  [\href{https://arxiv.org/abs/1810.07313}{{\ttfamily 1810.07313}}].

\bibitem{Asano:2015oia}
K.~Asano and T.~Terasawa, \emph{{Stochastic Acceleration Model of Gamma-Ray
  Burst with Decaying Turbulence}},
  \href{https://doi.org/10.1093/mnras/stv2152}{\emph{Mon. Not. Roy. Astron.
  Soc.} {\bfseries 454} (2015) 2242--2248},
  [\href{https://arxiv.org/abs/1509.04477}{{\ttfamily 1509.04477}}].

\bibitem{Sakamoto:2007yi}
T.~Sakamoto et~al., \emph{{The First Swift BAT Gamma-Ray Burst Catalog}},
  \href{https://doi.org/10.1086/523646}{\emph{Astrophys. J. Suppl.} {\bfseries
  175} (2008) 179--190}, [\href{https://arxiv.org/abs/0707.4626}{{\ttfamily
  0707.4626}}].

\end{thebibliography}\endgroup

\end{document}